\documentclass[aps,pre,reprint,superscriptaddress]{revtex4-2}

\usepackage{graphicx}
\usepackage{subfigure}
\usepackage{color}
\usepackage[dvipsnames]{xcolor}
\usepackage{amsmath,amssymb}
\usepackage[linktocpage,colorlinks=true,linkcolor=blue,citecolor=blue,breaklinks=true,urlcolor=blue]{hyperref}
\usepackage{verbatim}
\usepackage{breakcites}
\usepackage{wrapfig}
\usepackage{soul}
\usepackage{xparse}
\usepackage{multirow}
\usepackage{bm}
\usepackage[shortlabels]{enumitem}
\usepackage{bm}
\usepackage{float}

\usepackage{xr}
\makeatletter

\newcommand*{\addFileDependency}[1]{
\typeout{(#1)}
\@addtofilelist{#1}
\IfFileExists{#1}{}{\typeout{No file #1.}}
}\makeatother

\newcommand{\fig}[1]{Fig.~\ref{#1}}

\newcommand{\eq}[1]{Eq.~\ref{#1}}

\newcommand{\sctn}[1]{\S~\ref{#1}}
\newcommand{\movie}[1]{Movie~\ref{#1}}

\newcommand{\figsi}[1]{Fig.~\ref{#1}}
\newcommand{\tabsi}[1]{Table~\ref{#1}}
\renewcommand{\vec}[1]{\boldsymbol{#1}}
\newcommand{\tens}[1]{\underline{\underline{#1}}}

\newcommand{\grad}{\vec{\nabla}}

\newcommand{\kb}{k_\textmd{B}}
\newcommand{\kbt}{\kb T}

\newcommand{\av}[1]{\left\langle #1 \right\rangle}

\newcommand{\abs}[1]{\left| #1 \right|}

\newcommand{\ie}{\textit{i.e.}}

\newcommand{\dt}{\delta t}
\newcommand{\act}{\alpha}
\newcommand{\actMIN}{\act_\mathrm{eq}}
\newcommand{\actMin}{\act_\mathrm{turb}}
\newcommand{\actMax}{\act_\dagger}
\DeclareDocumentCommand\length{ g }{%
    {\ell%
        \IfNoValueF {#1} { _{#1} }%
    }%
}

\newcommand{\lenSys}{\length{\text{sys}}}

\newcommand{\lenDef}{\length{d}}
\newcommand{\lenVel}{\length{v}}

\newcommand{\lenDens}{\length{\rho}}

\DeclareDocumentCommand\corr{ m g }{%
    {C_{#1,#1}%
        \IfNoValueF {#2} { \left(#2\right)}%
    }%
}
\DeclareDocumentCommand\xCorr{ m m g }{%
    {C_{#1,#2}%
        \IfNoValueF {#3} { \left(#3\right)}%
    }%
}
\DeclareDocumentCommand\xCovar{ m m g }{%
    {G_{#1,#2}%
        \IfNoValueF {#3} { \left(#3\right)}%
    }%
}
\DeclareDocumentCommand\pcc{ m m g }{%
    {\mathcal{P}_{#1,#2}%
        \IfNoValueF {#3} { \left(#3\right)}%
    }%
}
\newcommand{\NCell}{N_\text{av}}  
\newcommand{\CellDens}{\rho_{C}}

\newcommand{\SigF}{\mathcal{S}_{C}}
\newcommand{\SigFn}[3]{\SigF \left( #3; #1, #2 \right)}
\newcommand{\Sigpos}{\sigma_\mathrm{p}}
\newcommand{\Sigwid}{\sigma_\mathrm{w}}
\newcommand{\ActSum}{\act_\mathrm{C}^\mathrm{P}} 
\newcommand{\ActAv}{\act_\mathrm{C}^\mathrm{C}} 
\newcommand{\SigSum}{\act_\mathrm{C}^\mathrm{MP}} 
\newcommand{\SigAv}{\act_\mathrm{C}^\mathrm{MC}} 

\newcommand{\tocite}[1]{\textcolor{cyan}{\textsuperscript{*}}}

\newcounter{siequation}
\newcounter{sifigure}
\newcounter{sitable}
\newcounter{sisection}
\newcounter{simovie}
\begin{document}


\title{Mitigating Density Fluctuations in Particle-based Active Nematic Simulations}

\author{Timofey Kozhukhov}
\affiliation{School of Physics and Astronomy, The University of Edinburgh, Peter Guthrie Tait Road, Edinburgh, EH9 3FD, UK.}

\author{Benjamin Loewe}
\affiliation{School of Physics and Astronomy, The University of Edinburgh, Peter Guthrie Tait Road, Edinburgh, EH9 3FD, UK.}
\affiliation{
Facultad de Física, Pontificia Universidad Católica de Chile, Santiago 7820436, Chile.}

\author{Tyler N. Shendruk}
\affiliation{School of Physics and Astronomy, The University of Edinburgh, Peter Guthrie Tait Road, Edinburgh, EH9 3FD, UK.}

\date{\today}

\begin{abstract}
\noindent
Understanding active matter has led to new perspectives on biophysics and non-equilibrium dynamics. 
However, the development of numerical tools for simulating active fluids capable of incorporating non-trivial boundaries or inclusions has lagged behind.
Active particle-based methods, which typically excel at this, suffer from large density fluctuations that affect the dynamics of inclusions.
To this end, we advance the Active-Nematic Multi-Particle Collision Dynamics algorithm, a particle-based method for simulating active nematics, by addressing the large density fluctuations that arise from activity.
This paper introduces three novel activity formulations that mitigate the coupling between activity and local density.
Local density fluctuations are decreased to a level comparable to the passive limit while retaining the phenomenology of active nematics and increasing the active turbulence regime four-fold. 
These developments extend the technique into a flexible tool for modeling active systems, including solutes and inclusions, with broad applications for the study of biophysical systems.
\end{abstract}


\maketitle

\section{Introduction}\label{sctn:intro} 

A common motif in biomaterials is that their elongated constituent elements often align over length-scales much larger than themselves~\cite{needleman2017, yaman2019, doostmohammadi2022}. 
This quasi-long-range orientational order has led to the development of the theory of active nematic liquid crystals~\cite{Alert2022AnnRevCondMat}. 
The study of active nematics, spontaneously flowing systems that exhibit apolar orientational order, has made great strides~\cite{Giomi2015PRX, Shankar2019, Beller2020, hardouin2022} and found wide-ranging applicability, especially in biological settings~\cite{li2019, liu2021, Aranson2022, Ladoux2022}. 
The development of idealized biophysical experimental systems~\cite{Dogic2012Nature-MicrotubuleActiveNematic, zhang2021} and subsequent continuum models~\cite{Giomi2015PRX, Yeomans2016NatComm} has led to a general understanding of the bulk behavior of these systems~\cite{Doostmohammadi2018-Review, Alert2022AnnRevCondMat}. 
Consequently, the focus has expanded to studying active nematics confined within complicated boundaries~\cite{Loewe2021NJP, Thampi2022, Sattvic2023, Alexander2023}, and the development of strategies to control active flows~\cite{Selinger2021PRE, zhang2021, Fraden2023-LightAN}.

\begin{figure*}[tb]
    \centering 
    \includegraphics[width=\linewidth]{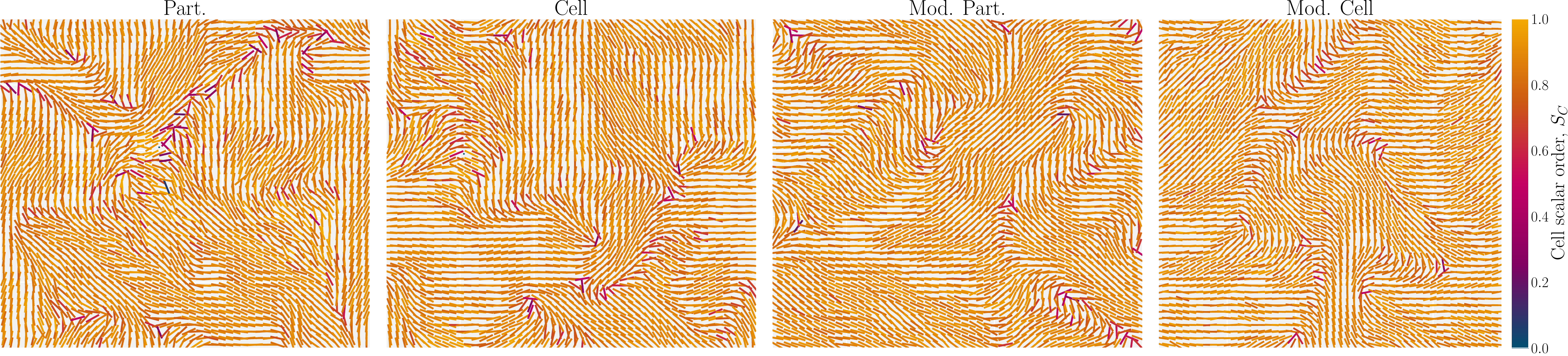}
    \caption{
        \textbf{Alternative formulations of cellular activity do not affect bulk active nematic turbulence.}
        Snapshots of local director field $\vec{n}_C$ colored by the scalar order parameter $S_C$ for the four different formulations of cellular activity, in a square periodic domain of size $\lenSys=50$.
        Corresponds to frame 75 of \movie{simov:DirSnapshots}.
    }
    \label{fig:DirSnapshots}
\end{figure*}

Sustained progress into both of these avenues relies on the development of numerical tools capable of flexibly incorporating nontrivial boundaries and inclusions. 
One such method is Active-Nematic Multi-Particle Collision Dynamics (AN-MPCD)~\cite{Kozhukhov2022-ANMPCD}, which builds on nematic MPCD~\cite{Shendruk2015SoftMatter-NMPCD} and can reproduce active turbulence in bulk (\fig{fig:DirSnapshots}). 
Active-Nematic MPCD has proven useful in simulating active turbulence in 3D~\cite{Hijar2023-ANMPCD}. 
Unlike other commonly used active-nematic solvers~\cite{Marenduzzo2007PRE-ANLB, Beller2020}, the particle-based nature of MPCD allows for relatively straightforward incorporation of surfaces representing boundaries or inclusions --- a feature that has been exploited to study the dynamics of active nematics in porous media~\cite{Keogh2023-Darcy}. 

While the particle-based nature of MPCD is advantageous for handling complex geometries, it comes with its own set of challenges:
As with any particle-based simulation technique~\cite{Chate2006PRL, PeshkovChate2014PRL, Bertin2009JPhysA, Bar2020AnnRev, Valeriani2022}, passive MPCD has local density fluctuations that can affect the dynamics of inclusions~\cite{Zantop2021JCP-MPCDState}. 
In the passive case, these are relatively small; however, in the active case these are exacerbated, as exemplified from active Brownian particle simulations and the associated motility-induced phase separation~\cite{Cates2015-MIPS}. 
The active component of AN-MPCD introduces large number fluctuations that effectively limit the regime over which the algorithm reproduces continuous fields~\cite{Kozhukhov2022-ANMPCD}. 
In the originally proposed AN-MPCD algorithm, each individual MPCD particle is considered to be an active unit. 
While sensible from a particle-based perspective, from the continuum perspective this means that the strength of the local active dipole force is proportional to the local fluid density. 
This implementation ties the local degree of activity to the local density, resulting in a positive feedback: activity leads to regions of high density, which in turn further increases local activity. 

With the aim of increasing the operational regime of AN-MPCD and improving its suitability for simulating active suspensions, this article introduces three new variant formulations to AN-MPCD that mitigate the density-activity feedback by either: 
    \emph{(i)} completely untying the strength of activity from the local density, making it uniform for the entire system, 
    \emph{(ii)} modulating the dependence of activity on density, or 
    \emph{(iii)} a combination of the previous two methods. 
By exploring the interplay between density and the strength of activity, our study sheds light on how the particular process of local energy injection affects collective dynamics, which may be useful to understand other methods of active control, such as light activated active nematics~\cite{zhang2021, bate2022, Fraden2023-LightAN} or spatially varying fuel concentration~\cite{Selinger2021PRE, ruske2022, coelho2022}.

Each of the proposed methods quantitatively reproduces the same phenomenology of active nematics (\fig{fig:DirSnapshots}, \movie{simov:DirSnapshots}). 
However, the three new formulations greatly reduce the degree of density fluctuations. 
In particular, the modulated activities decrease and cap the degree of pressure-induced advection of inclusions to a level only slightly greater than passive MPCD. 
Furthermore, the effective turbulence regimes are shown to increase by roughly four times with respect to the original particle-based activity implementation. 
As such, these three new activity formulations address the primary short-coming of AN-MPCD for simulating continuous active nematic fields. 
They not only expand the operational regime of the method, but also improve its suitability for modeling solutes and inclusions, transforming it into a flexible numerical tool for modeling composite active systems. 
This could include passive colloidal particles or polymers suspended in spontaneously flowing biofluids.  

\section{MPCD Method}\label{sctn:algorithm}

Multi-Particle Collision Dynamics (MPCD) discretizes a continuous fluid into point-like particles, which propagate through streaming and collision steps.
Each MPCD particle $i$ of mass $m_i$, position $\vec{r}_i$, and velocity $\vec{v}_i$ follows a simple Eulerian ballistic streaming step for a time $\delta t$ to a new position
\begin{equation}
    \label{eq:MPCD-Stream}
    \vec{r}_i (t + \delta t) = \vec{r}_i (t) + \vec{v}_i (t) \delta t
\end{equation}
before undergoing a collision event. 
Collisions between particles are encoded by the collision operation $\vec{\Xi}_{i,C}$, which stochastically exchanges momenta and other particle properties within a given MPCD cell $C$ of size $a$, while ensuring that all appropriate conservation laws are satisfied locally within each cell. 
The collision operation is applied to each particle within a cell as 
\begin{equation}
    \label{eq:MPCD-Collision}
    \vec{v}_i (t + \delta t) = \av{\vec{v}}_C(t) + \vec{\Xi}_{i,C} (t), 
\end{equation}
where the cellular average $\av{\vec{v}}_C(t) = \sum_j^{\CellDens} m_i \vec{v}_i (t) / \sum_j^{\CellDens} m_i$ is over the $\CellDens(t)$ particles currently in cell $C$. 
A widely used passive MPCD collision operation for thermalized isotropic fluids is the Andersen collision operator~\cite{GompperIhle2009Book-MPCD, Gompper2007EPL} 
\begin{equation}
    \label{eq:MPCD-Andersen}
    \vec{\Xi}_{i, C}^\mathrm{0} = 
    \vec{\xi}_i -
    \av{\vec{\xi}}_C +
    \left( \tens{I}_C^{-1} \cdot \delta \vec{\mathcal{L}}_\mathrm{vel}\right) \times \vec{r}_i' ,
\end{equation}
in which the first two terms represent a stochastic exchange of translational momentum with $\vec{\xi}_i$ being a random velocity drawn from the Maxwell-Boltzmann distribution for mass $m_i$ and thermal energy $\kbt$ and $\av{\vec{\xi}}_C$ being the cellular average of the random velocities. 
The third term imposes conservation of momentum.
The moment of inertia $\tens{I}_C$ is computed from the position of point particles in cell $C$ relative to the cell center of mass $\av{\vec{r}}_C$, where $\vec{r}_i' = \vec{r}_i - \av{\vec{r}}_C$.
The change in angular momentum due to the collision is $\delta\vec{\mathcal{L}}_\mathrm{vel} = \sum_j^{\CellDens} m_j \left[ \vec{r}_j' \times \left( \vec{v}_j-\vec{\xi}_j \right) \right]$.

The Andersen collision operation (\eq{eq:MPCD-Andersen}) highlights the key features of an MPCD operator with each term, namely: 
It reproduces hydrodynamics by satisfying momentum conservation laws by removing residuals, and it introduces viscous dissipation through stochastic collisions. 
The collision operation $\vec{\Xi}_{i,C}$ governs the fluid's material properties and transport coefficients~\cite{GompperIhle2009Book-MPCD}. 
Additional material properties can be simulated through more complicated collision operations~\cite{tao2008,Stark2013-ViscoMPCD,Toneian2019}. 
The most relevant example for this study is the nematic collision operation~\cite{Shendruk2015SoftMatter-NMPCD}. 
By prescribing an orientation $\vec{u}_i$ to each particle, and a cell-average director $\vec{n}_C$ (calculated via the tensor order parameter), it is possible to devise a nematic collision operations $\vec{\Xi}_{i,C}^\mathrm{N}$ and $\tens{R}_{i,C}$ to reproduce stochastic linearized nematohydrodynamics.

During each cellular collision, all particle's orientations are altered by 
\begin{equation}
    \label{eq:NMPCD-Orientation}
    \vec{u}_i(t+\delta t) = \tens{R}_{i,C} \cdot \vec{u}_i(t) ,
\end{equation}
in which $\tens{R}_{i,C}$ is a rotation matrix representing a nematic orientation collision operation composed of multiple contributions.
Inspired by the Andersen collision operation (\eq{eq:MPCD-Andersen}), 
$\tens{R}_{i,C}$ possesses a component that re-orients particle $i$ by a random orientation drawn from a locally equilibrated Maier-Saupe distribution $\sim e^{US_C (\vec{u}_i\cdot \vec{n}_C)^2}$ about the cellular director $\vec{n}_C$ with scalar order $S_C$ for an interaction potential $U$~\cite{Shendruk2015SoftMatter-NMPCD}. 
Additionally, the rotation of the particles is coupled to their flow because
the strain rate and vorticity rotate the orientation of the particles, modeled as slender rods, through the Jeffery's equation with tumbling parameter $\lambda$ and hydrodynamic susceptibility $\chi$. 
The stochastic operation and Jeffery's equation generate angular momentum $\delta \vec{\mathcal{L}}_\mathrm{N} = \gamma_\mathrm{R} \sum_j^{\CellDens} \vec{u}_j \times \dot{\vec{u}}_j$ for a rotational friction coefficient $\gamma_\mathrm{R}$. 
This is balanced by adding $\vec{\Xi}_{i,C}^\mathrm{N} = - \left( \tens{I}_C^{-1} \cdot \delta \vec{\mathcal{L}}_\mathrm{N}\right) \times \vec{r}_i' $ to the translational collision operation (\eq{eq:MPCD-Andersen}). 
Conservation of angular momentum induces nematic backflow. 

Activity is introduced through a local cell-based active force dipole, which is applied by adding an active contribution to the collision operation
\begin{equation}
    \vec{\Xi}_i^\mathrm{AN} = 
    \act_C \delta t 
    \left(
        \frac{\kappa_i}{m_i} - \frac{\av{\kappa}_C}{\av{m}_C} 
    \right) \vec{n}_C, \label{eq:AN-CollisionOp}
\end{equation}
in which $\act_C$ denotes the local cell dipole strength~\cite{Kozhukhov2022-ANMPCD}. 
To determine the direction of the active force acting on particle $i$, cell $C$ is dissected by a plane through the center of mass with normal $\vec{n}_C$, and $\kappa_i(\vec{r}_i', \vec{n}_c)=\pm1$ depending on whether $\vec{r}_i'$ is above or below the plane~\cite{Kozhukhov2022-ANMPCD}. 
The two terms in \eq{eq:AN-CollisionOp} represent the activity that provides individual impulses per unit mass for each particle and a term that removes any residual linear momentum change from the active force term.
The induced local force dipole reproduces the bending instability~\cite{SimhaRamaswamy2002PhysicaA}, unbinding of topological defects~\cite{Thampi2014EPL}, and active turbulence~\cite{Kozhukhov2022-ANMPCD}. 
Since the collision algorithm injects energy but conserves momentum, the algorithm simulates wet active nematics~\cite{Marchetti2013RevModPhys-Hydrodynamics, Kozhukhov2022-ANMPCD}.

In each of the four algorithms considered in this article, the activity is included through \eq{eq:AN-CollisionOp}. 
The differences enter through the form chosen for the local cell dipole strength $\act_C\left(\CellDens\right)$. 
The local strength $\act_C(\rho_C)$ can be chosen to be linearly proportional to $\CellDens$ (\emph{particle-carried activity}; \sctn{sctn:PCAct}, \cite{Kozhukhov2022-ANMPCD}), constant with respect to $\CellDens$ (\emph{cell-carried activity}; \sctn{sctn:CCAct}), or modulated by a density kernel (\sctn{sctn:ModAct}).

\subsection{Simulation Units and Parameters}
Results are reported in MPCD units. 
The unit of length is the MPCD cell size $a \equiv 1$, the unit of energy is thermal energy $\kbt \equiv 1$, and the unit of mass is particles' mass $m$. 
Since only a single species of fluid particles is considered here, $m_i = m \equiv 1$. 
As a result, the unit of time is $a\sqrt{m/\kbt} = 1$.
The time step size is set to $\dt = 0.1$. 
The average number of particles per cell is set to $\NCell = \av{\CellDens} = 20$ throughout, where $\av{\cdot}$ is the average over the entire system.
Due to the selection of $a \equiv 1$, and $m \equiv 1$, the number density of a cell $\CellDens$ is equivalent to the mass density.

The mean field potential is set to $U=10$, which puts the system deep in the nematic phase~\cite{Shendruk2015SoftMatter-NMPCD}. 
The rotational friction coefficient is $\gamma_\mathrm{R}=0.01$, hydrodynamic susceptibility is $\chi=0.5$ and tumbling parameter is $\lambda=2$ for a flow-aligning nematic. 

Simulations run in two dimensions within a periodic box of size $\lenSys \times \lenSys = 200\times 200$, unless stated otherwise. 
Particles are initialized with random positions, thermally sampled velocities and uniform orientation aligned along the $\hat{y}$ axis. 
All simulations include $10^4$ warmup steps and run for an additional $5\times10^4$ data collection steps.
A summary of all simulation parameters is provided in \tabsi{sitab:UnitsRef}.

\section{Active-Nematic MPCD} \label{sctn:ANSigMPCD}

\begin{figure}[tb]
    \centering
    \begin{subfigure}
        \centering
        \includegraphics[width=0.95\linewidth]{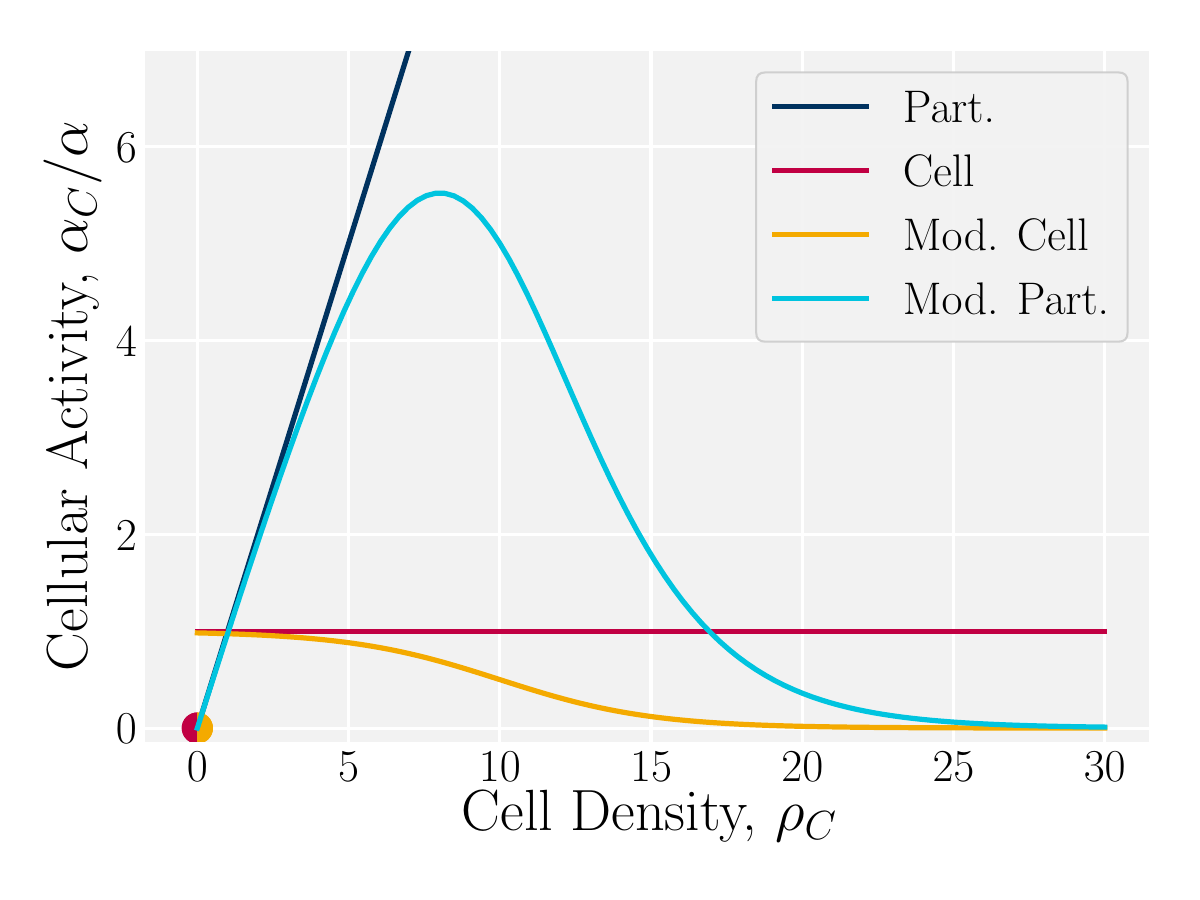}
    \end{subfigure}
    \caption{\textbf{Formulations of cellular activity $\act_\mathrm{C}$ as a function of local density $\CellDens$. }
    A schematic of the alternative formulations for cellular activity, normalized by the particle activity $\act$.
    ($\av{\CellDens}=5$, $\Sigpos=1$, and $\Sigwid=1$). 
    At $\CellDens=0$, cell- and modulated cell-based activity take a value of 0, indicated by the marker.
    }
    \label{fig:ActivitySchematic}
\end{figure}

In order to employ AN-MPCD as described in \sctn{sctn:algorithm}, one must choose a form for the activity. 
In the following section, four formulations are described.  

\subsection{Particle-Carried Activity}
\label{sctn:PCAct}

The original AN-MPCD algorithm models activity as though each MPCD particle carries activity~\cite{Kozhukhov2022-ANMPCD}. 
Therefore, each cell's activity is set to 
\begin{equation}
    \ActSum = \sum_{i=1}^{\CellDens} \act_i \label{eq:ActSumDef} ,
\end{equation}
in which $\act_i$ is the constant activity prescribed to each particle $i$. 
In this definition, the cellular activity is simply the sum of the activities of the particles. 
Each particle contributes linearly to the cellular activity and so this definition is labelled \emph{particle-carried activity}. 
For a system in which every particle carries the same activity ($\act_i=\act\; \forall\; i$), the cellular activity is simply
\begin{equation}
    \ActSum = \CellDens \act \label{eq:ActSumFixed} ,
\end{equation}
which is directly proportional to the number of particles in the cell (\fig{fig:ActivitySchematic} (Part.); dark blue).

The $\ActSum$ activity formulation (\eq{eq:ActSumFixed}) is effective in simulating active nematic liquid crystals~\cite{Kozhukhov2022-ANMPCD}, satisfying the theoretically expected scalings of active nematics for a range of parameters~\cite{Thampi2014EPL, Hemingway2016SoftMatter}. 
However, unlike continuum solvers of the active nematic equations of motion~\cite{Beller2020, Keogh2022, Head2024-DTensor} which typically assume incompressibility, this particle-based approach possesses a large degree of density fluctuations~\cite{Kozhukhov2022-ANMPCD}. 
In some cases, such as in modeling bacteria suspension that do exhibit large variations in swimmer density~\cite{Zottl2016, liu2021}, the fluctuations may be desired, while in others they may hamper studies of systems in which density-induced pressure gradients may affect the physics, such as suspensions of colloid~\cite{Wani2022} or polymer solutes~\cite{Jain2022, chen2019}. 
While it was shown that density fluctuations can be mitigated by limiting the activity regime (i.e., providing an upper bound for $\alpha$)~\cite{Kozhukhov2022-ANMPCD}, this might not always be desirable when reproducing experimental data. 
The fluctuations result from a positive feedback, in which high density regions result in large local activities, which then create more higher density regions. 

\subsection{Cell-Carried Activity}
\label{sctn:CCAct}

To break the positive feedback, activity should not increase linearly with density. 
In this section, the activity is made nearly constant. 
The formulation considered here normalizes $\ActSum$ by $\CellDens$ to remove the direct density dependence from \eq{eq:ActSumDef} via the redefinition of each cell's activity as
\begin{align}
    \alpha_C^C &= 
    \begin{cases}
        \av{\act_i}_C  & \text{if } \CellDens > 0 \\
        0 & \text{if } \CellDens = 0 
    \end{cases}, \label{eq:ActAvDef} \\
\end{align}
where $\av{\act}_C = \left( \sum_i^{\CellDens} \act_i \right)/ \CellDens = \ActSum/\CellDens$. 
For a single population of particles with constant particle activity ($\act_i=\act\; \forall\; i$), cell activities are constant and 
\begin{equation}
    \alpha_C^C = \act
\end{equation}
everywhere, except in the case of empty cells since there can be no force dipole in an empty cell.
In contrast to $\ActSum$, this homogenizes the cellular activity throughout the simulation: every non-empty MPCD cell has the same activity, regardless of the local density. 

In this version, the activity is carried by cells rather than by the particles, and so this choice is referred to as \emph{cell-carried activity}  (\fig{fig:ActivitySchematic} (Cell.); red).

\subsection{Modulated activities}
\label{sctn:ModAct}

While \emph{cell-carried activity} $\ActAv$ breaks the positive feedback, it does not proactively discourage high densities. 
To further mitigate the activity-induced density variations, activity could be formulated to impose a negative feedback at high densities. 
To do this, a modulation function $\SigF$ is introduced as a multiplicative factor on $\ActSum$ or $\ActAv$. 
This modulation function depends on the local density $\CellDens$ and is chosen to decrease to zero as $\CellDens\to\infty$ but only activating at high densities, leaving the small density behavior intact. 
Thus, a sigmoidal modulation function 
\begin{equation}
    \SigF = 
    \frac{1}{2}
    \left(
        1 - \tanh \left(
            \frac{\CellDens - \av{\CellDens} \left( 1 + \Sigpos \right)}{\av{\CellDens} \Sigwid}
        \right)
    \right)
    \label{eq:SigFnDef}
\end{equation}
is selected.
The modulation function $\SigFn{\Sigpos}{\Sigwid}{\CellDens}$ compares a cell's instantaneous density $\CellDens(t)$ to the system-wide mean density $\av{\CellDens}$ and returns a value in the interval $(0,1)$. 
The two parameters $\Sigpos$ and $\Sigwid$ set the position and width of the sigmoidal drop. 
In particular, $\Sigpos$ sets the position of the sigmoid midpoint, with a value of $\Sigpos=0$ placing the midpoint at $\av{\CellDens}$, while $\Sigpos>0$ ($\Sigpos<0$) shifts the midpoint to higher (lower) densities. 

The modulation can be applied to either particle- or cell-carried activity. 
In the case of the \emph{modulated cell-carried activity}, 
\begin{align}
    \SigAv
    &= \ActAv \SigFn{\Sigpos}{\Sigwid}{\CellDens} \label{eq:SigAvDef}
\end{align}
Because $\ActAv$ is independent of density, the modulation acts like a switch lowering the activity at densities above $\Sigpos\av{\CellDens}$ (\fig{fig:ActivitySchematic} (Mod. Cell); yellow). 

\begin{figure}[tb]
    \centering
    \begin{subfigure}
        \centering
        \includegraphics[width=0.95\linewidth]{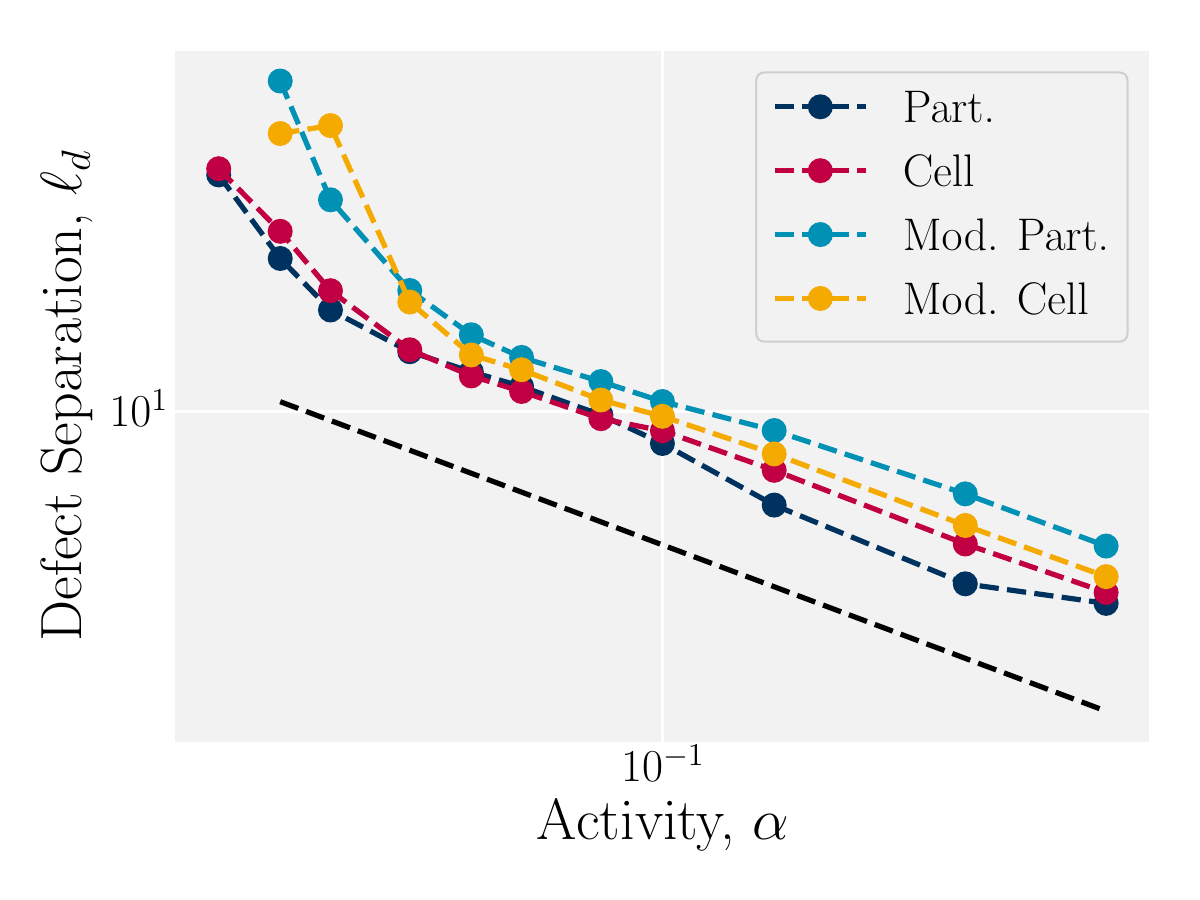}
    \end{subfigure}
    \caption{\textbf{Defect separation.} 
    Steady-state defect density $\rho_d$ is used to compute the root-mean square defect separation $\lenDef=\rho_d^{-1/2}$ for all four formulations of cellular activity. 
    ``Part.'' corresponds to \emph{particle-carried activity} (\sctn{sctn:PCAct}; \eq{eq:ActSumDef}); 
    ``Cell'' to \emph{cell-carried activity} (\sctn{sctn:CCAct}; \eq{eq:ActAvDef});
    ``Mod. Part.'' to \emph{modulated particle-carried activity} (\sctn{sctn:ModAct}; \eq{eq:SigSumDef}); 
    ``Mod. Cell'' to \emph{modulated cell-carried activity} (\sctn{sctn:ModAct}; \eq{eq:SigAvDef}). 
    The dashed line indicates a scaling of $\lenDef \sim \act^{-1/2}$, which is the theoretically expected scaling for active nematic turbulence. 
    }
    \label{fig:Turb-Defects}
\end{figure}

Similarly, the modulation can be applied to create a \emph{modulated particle-carried activity}
\begin{align}
    \SigSum
    &= \ActSum \SigFn{\Sigpos}{\Sigwid}{\CellDens} \label{eq:SigSumDef}. 
\end{align}
The activity formulation $\SigSum$ appears qualitatively similar to a Maxwell-Boltzmann distribution, mimicking the linear behavior of $\ActSum$ for small $\CellDens$, but decaying to zero at large $\CellDens$ (\fig{fig:ActivitySchematic} (Mod. Part.); cyan).

\subsection{Activity Parameters}

In this study, we limit our consideration to a single species of active fluid particles by setting $\act_i = \act \;\forall \; i$ and choosing to consider extensile activity ($\act>0$). 
The sigmoidal modulation parameters are set to $\Sigpos=0.4$ and $\Sigwid=0.5$, which are found to give optimal density behavior without lowering effective activity (see \figsi{sifig:PhaseDiagrams}).

\section{Results}\label{sctn:results}

Previous work has verified that \emph{particle-carried activity} ($\ActSum$) reproduces active turbulence~\cite{Kozhukhov2022-ANMPCD}. 
It was demonstrated that the \emph{particle-carried activity} algorithm possesses four distinct regimes as a function of activity:
for small activities, $\act \lesssim \actMIN \approx 10^{-3}$, the thermostat dominates over activity.
In the second regime, $\actMIN \lesssim \act \lesssim \actMin \approx 0.02$, there is enough active stress to form bend walls within finite sized systems. 
In the third regime, $\actMin \lesssim \act \lesssim \actMax$, the active stress is sufficient for pair creation events to occur, driving the system into active turbulence.
Finally, for $\act \gtrsim \actMax$ density fluctuations become large. 
The following sections will demonstrate that each of these regimes is still present in the modulated methods. 

Qualitatively, all four activity formulations reproduce active turbulence (\fig{fig:DirSnapshots}; \movie{simov:DirSnapshots}). 
In each, the transient dynamics of active turbulence onset is the same (\movie{simov:BendInstability}):
The activity-induced hydrodynamic instability~\cite{Ramaswamy2007Science} generates narrow kink walls~\cite{Chate2019PRL, Kumar2022SoftMatter} that are perpendicular to the initially uniform director. 
Along the kink walls, active forces drive viscometric flows~\cite{Head2024-DTensor}. 
At sufficiently high activity, the free energy cost of the bend walls is large enough that topological defect pairs unbind, with defect core sizes comparable to the MPCD cell size $a$. 
The $+1/2$ defects are self-motile, moving rapidly through the system, while the $-1/2$ defects are only passively advected by the flows. 
This description holds for all four activity formulations, with the distance separating defects and the speed of the $+1/2$ qualitatively similar in each (\fig{fig:DirSnapshots}).

\subsection{Scaling of Active Turbulence}

\begin{figure}[tb]
    \centering
    \begin{subfigure}
        \centering
        \includegraphics[width=0.95\linewidth]{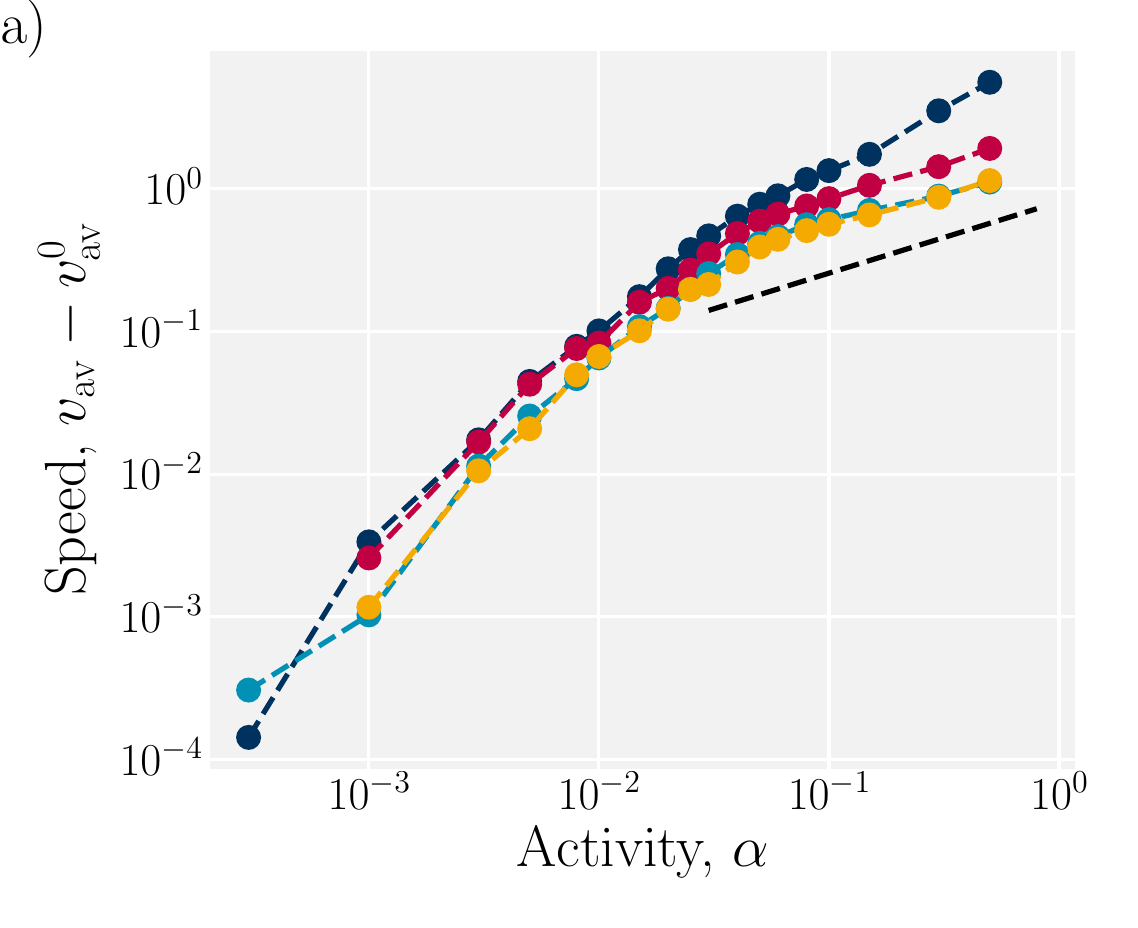}
    \end{subfigure}
    \begin{subfigure}
        \centering
        \includegraphics[width=0.95\linewidth]{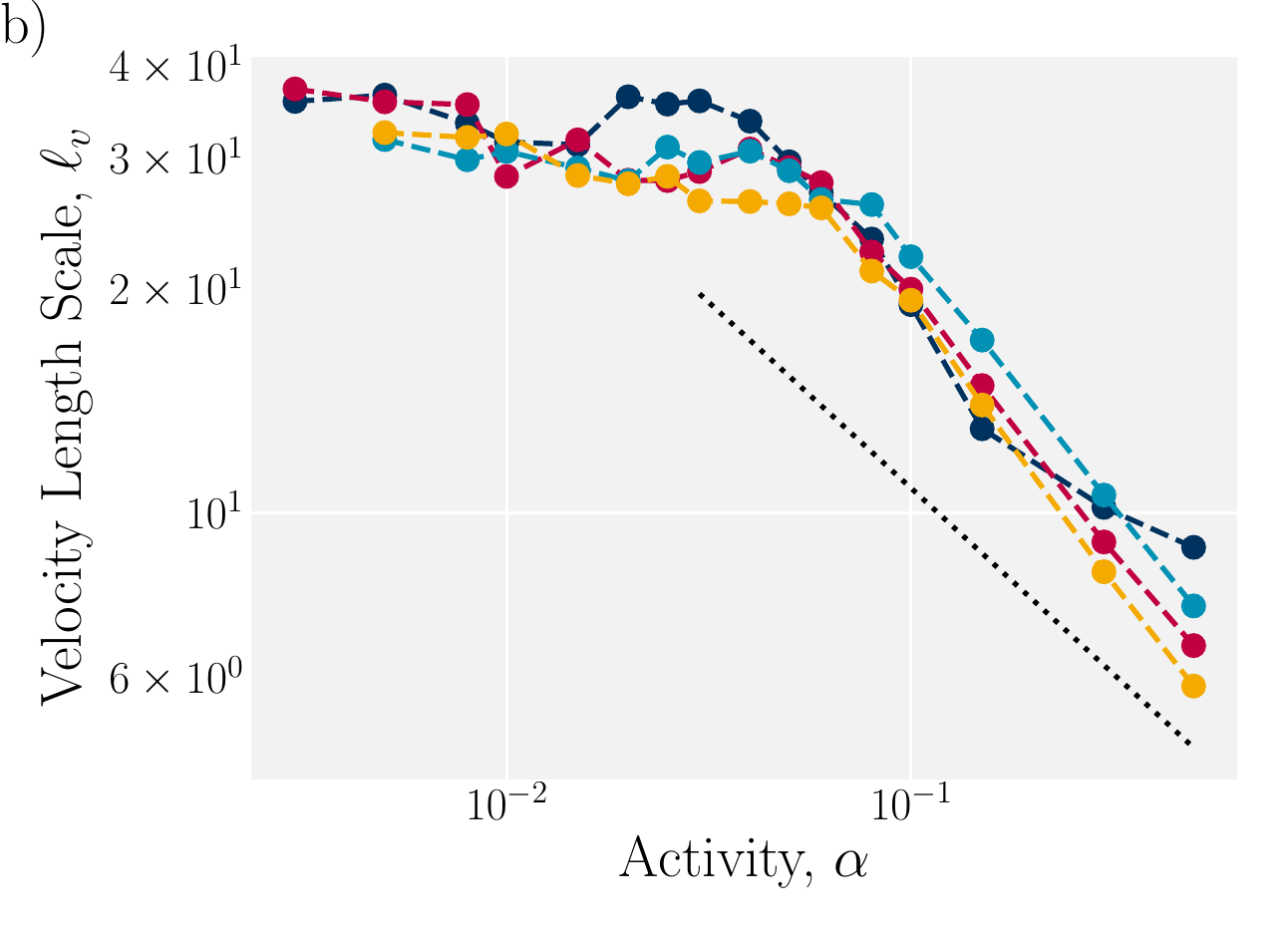}
    \end{subfigure}
    \caption{\textbf{Velocity scalings of active turbulence.}
    \textbf{(a)}
    The velocity contribution due to active flows, $v_\mathrm{av}-v_\mathrm{av}^0$, for all four formulations of cellular activity (same as \fig{fig:Turb-Defects}).
    The dashed line indicates the theoretically expected scaling of $v_\mathrm{av} \sim \act^{1/2}$.
    \textbf{(b)}
    The velocity length scale of active turbulence $\lenVel$ in the turbulence regime for all four activity formulations.
    The dotted line indicates the theoretically expected scaling of $\lenVel \sim \act^{-1/2}$. 
    }
    \label{fig:Turb-FlowSpeed}
\end{figure}

\begin{figure*}[tb]
    \centering 
    \includegraphics[width=\linewidth]{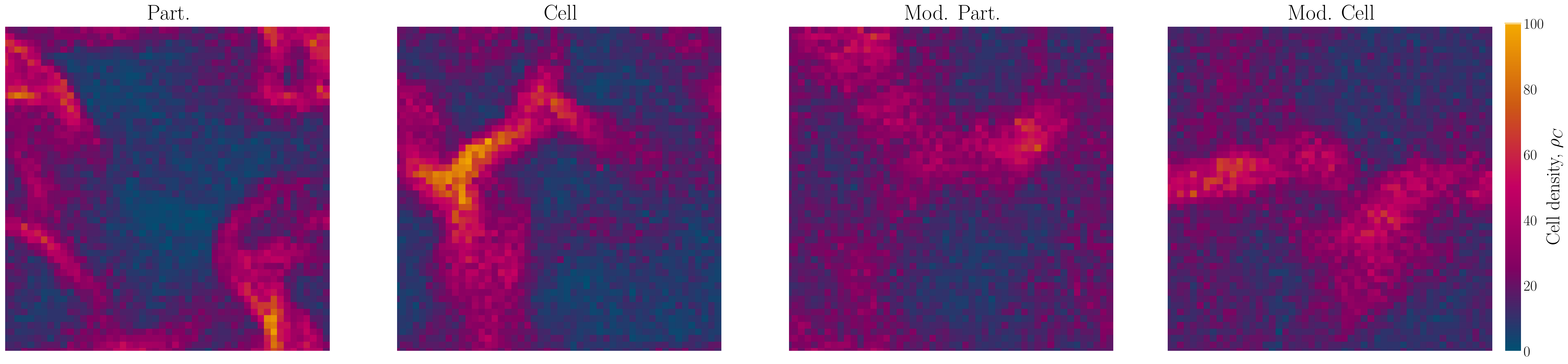}
    \caption{
        \textbf{Alternative formulations of cellular activity mitigate density fluctuations.}
        Snapshots of local density field $\rho_C$ for the four different formulations of cellular activity, in a square periodic domain of size $\lenSys=50$. 
        Corresponds to frame 75 of \movie{simov:DensSnapshots}.
    }
    \label{fig:DensSnapshots}
\end{figure*}

A common feature of active nematic turbulence is the presence of unbound $\pm 1/2$ nematic defects~\cite{Dogic2015NatureMat}.
In 2D, the defects with the lowest energetic cost compatible with nematic symmetry are the $\pm 1/2$ defects, corresponding to windings of $\pm \pi/2$~\cite{Selinger2018SoftMatter}. 
The fact that active turbulence coincides with the unbinding of topological defects is not a coincidence, as defect dynamics drive active turbulence~\cite{Giomi2015PRX}. 
\emph{Particle-carried} AN-MPCD successfully exhibits defect pair creation and annihilation events, producing an active length scale (\fig{fig:Turb-Defects}). 
The active length scale is found by computing the root-mean square mean defect separation $\lenDef=\rho_d^{-1/2}$ from the mean number density of defects $\rho_d$. 
Likewise, the \emph{cell-carried activity} and both modulated formulations successfully reproduce defects and the expected scaling behavior.

For all activity formulations, the defect density is zero until a critical activity ($\actMin$) above which turbulence occurs. 
The $\actMin$ value varies slightly between formulations (\fig{fig:Turb-Defects}), with the onset for the modulated methods occurring only at a slightly higher activity.
Once the critical activity is reached, the defect density increases with activity, and the defect separation length decreases as $\lenDef\sim \act^{-1/2}$ (\fig{fig:Turb-Defects}; dashed line). 
Measured exponents for each formulation (\tabsi{sitab:ScalingRef}) demonstrate that the modulated methods reproduce the theoretical expectations for the nematic length scale in active turbulence~\cite{Thampi2014EPL, Hemingway2016SoftMatter}. 
This demonstrates the scaling of active turbulence is not affected by the choice of cell activity formulation. 

\begin{figure*}[tb]
    \centering
    \begin{subfigure}
        \centering
        \includegraphics[width=0.49\linewidth]{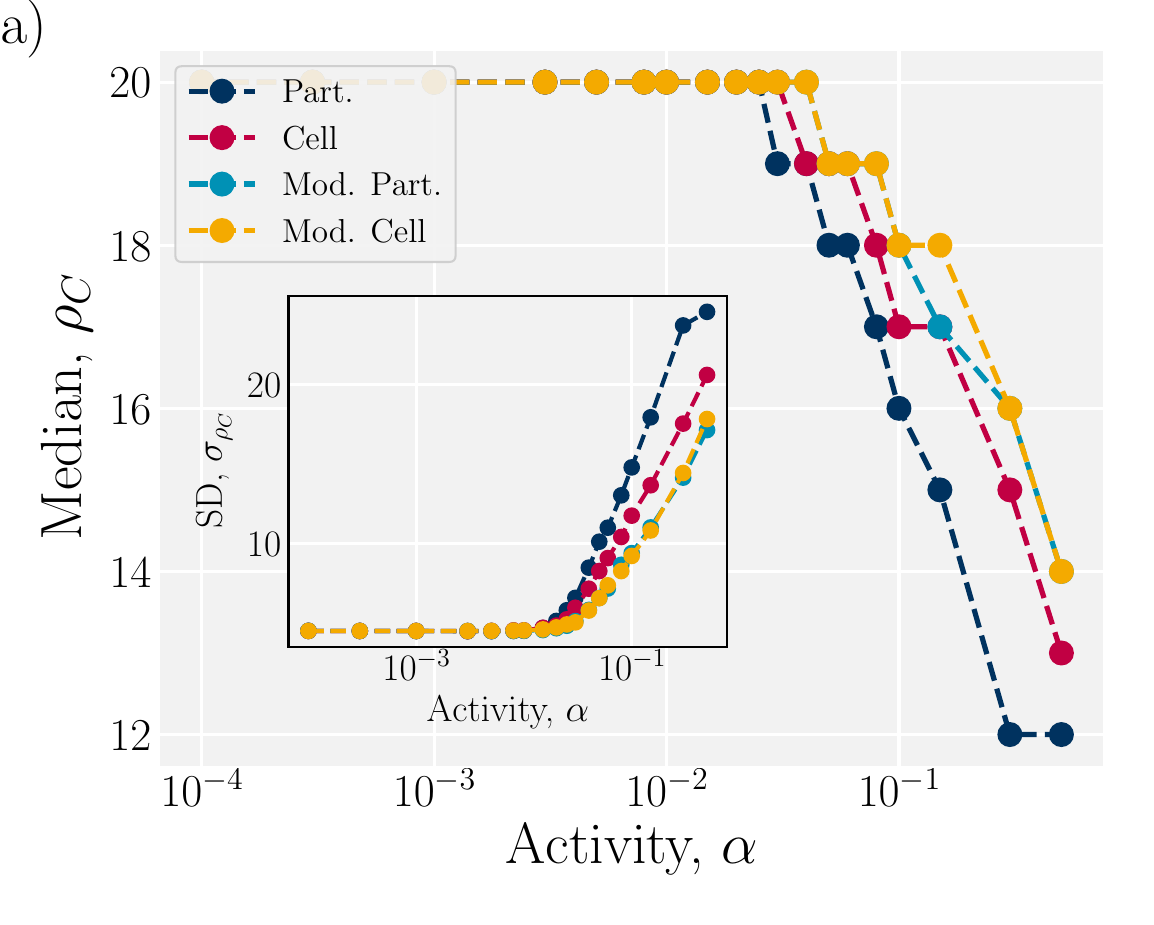}
    \end{subfigure}
    \begin{subfigure}
        \centering
        \includegraphics[width=0.49\linewidth]{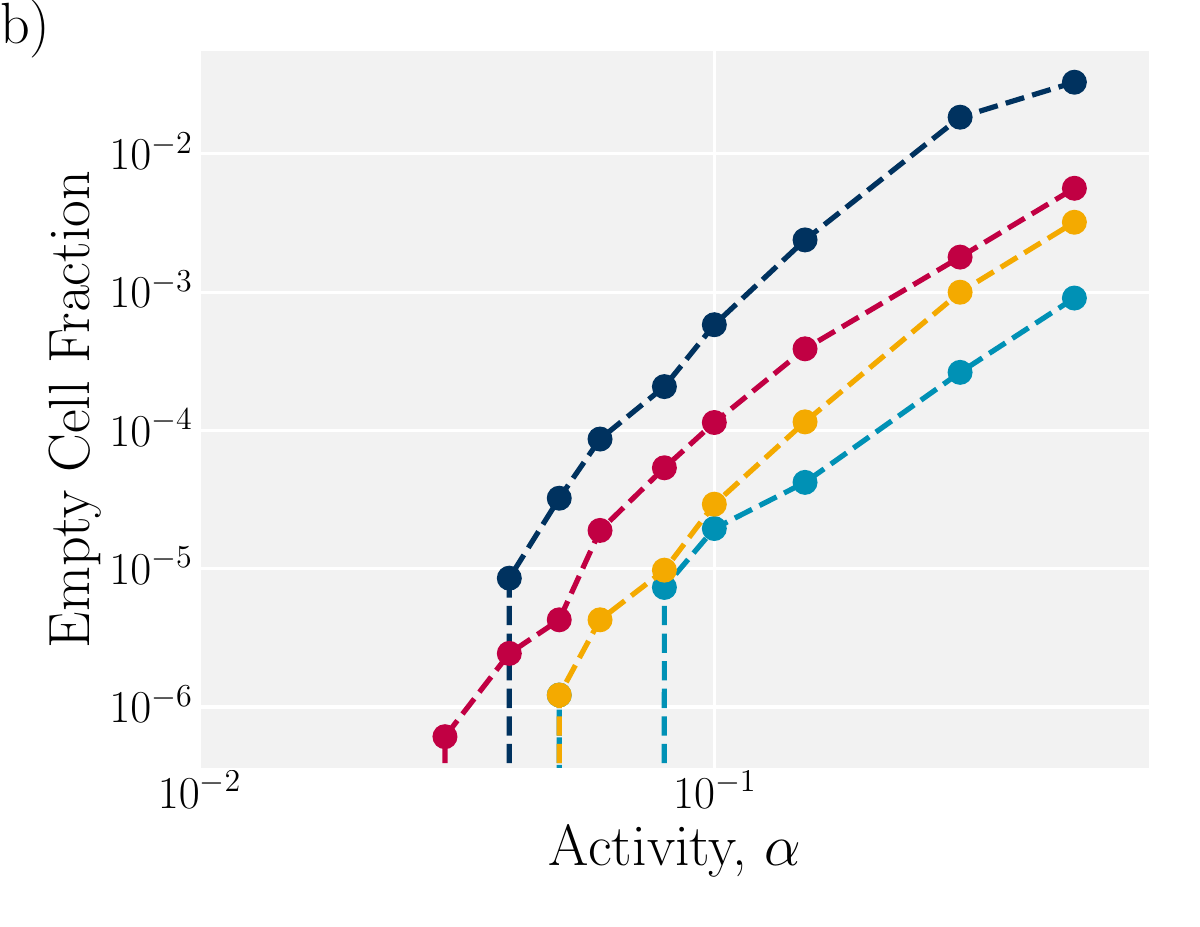}
    \end{subfigure}
    \begin{subfigure}
        \centering
        \includegraphics[width=0.49\linewidth]{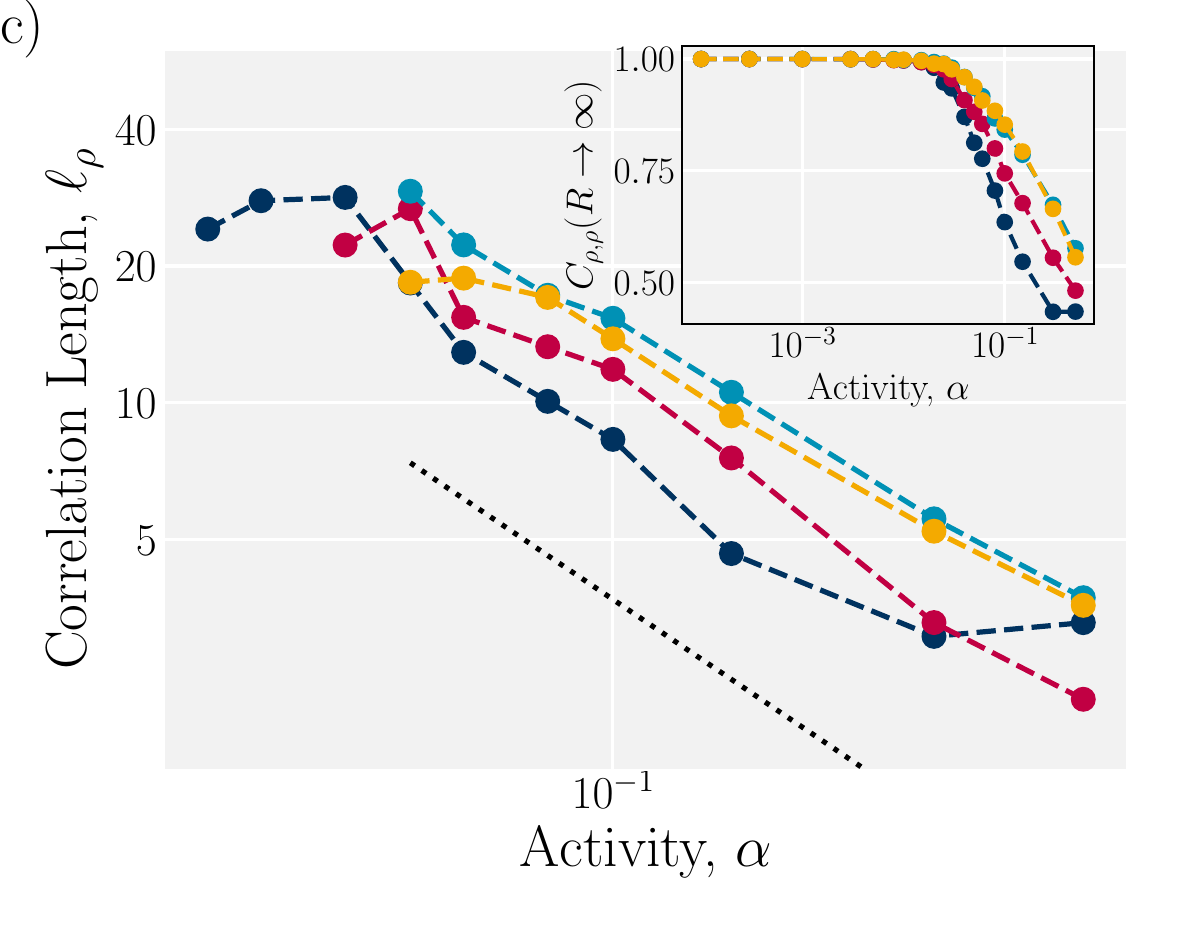}
    \end{subfigure}
    \begin{subfigure}
        \centering
        \includegraphics[width=0.49\linewidth]{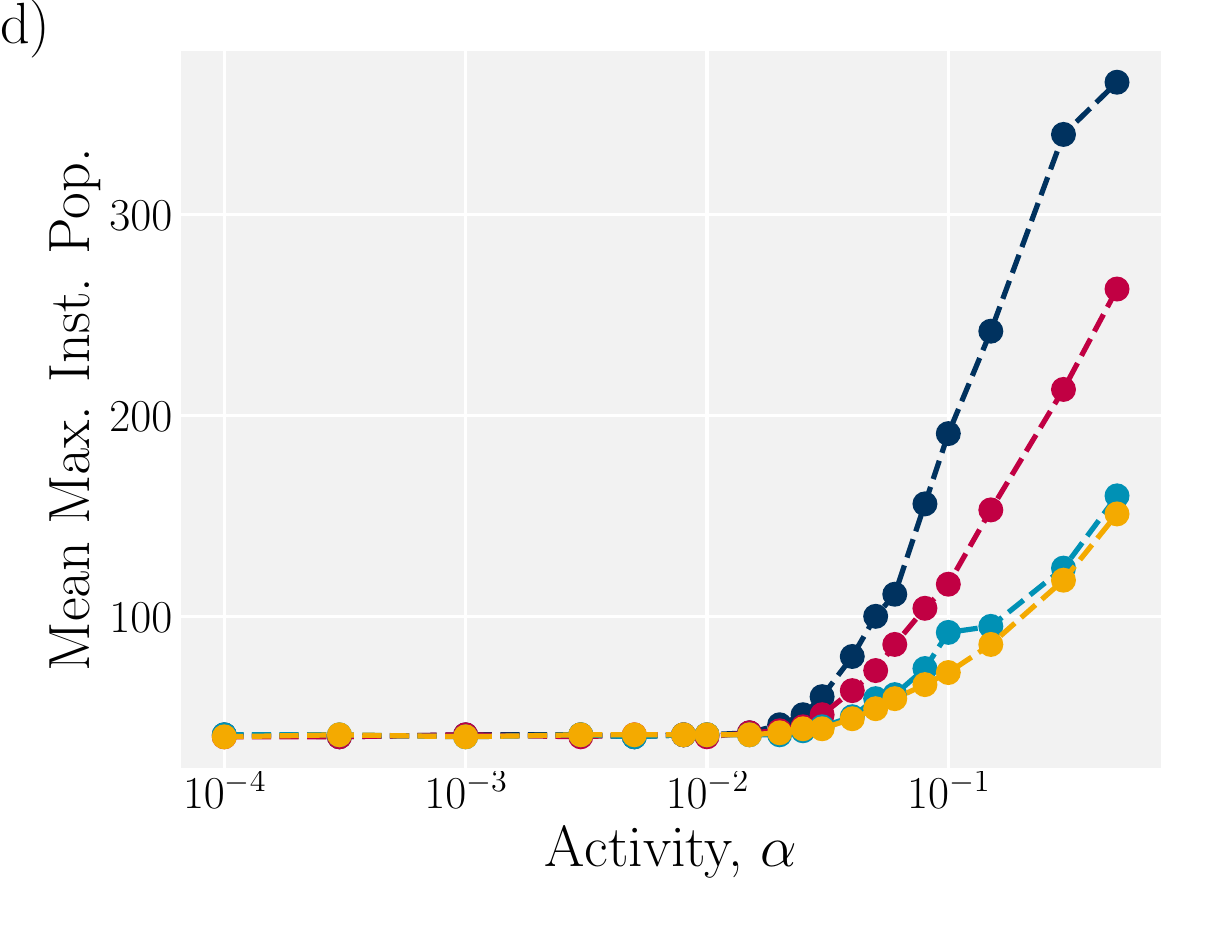}
    \end{subfigure}
    \caption{\textbf{Modulated formulations for cellular activity mitigate density fluctuations. } 
    \textbf{(a) }
    The median cell population $\CellDens$ for all four formulations of cellular activity.
    Modulated methods result the system median dropping below the mean density within the turbulence regime.
    \textbf{Inset: }
    The standard deviation of cell population $\sigma_{\rho}$ for all four formulations of cellular activity.
    \textbf{(b) }
    The fraction of empty cells with $\CellDens=0$ for all four formulations of cellular activity.
    \emph{particle-carried activity} in particular has a large fraction of empty cells at the highest activities, while alternative formulations result in a more homogeneous system.
    \textbf{(c) }
    The density correlation length $\lenDens$ for all four formulations of cellular activity, representative of the width of high density nematic bands throughout the system.
    Density bands become more dense and thinner as activity is raised, but alternative formulations mitigate this effect.
    The dotted line indicates a scaling of $\lenDens \sim \act^{-1}$.
    \textbf{Inset: }
    The far-field value of the density correlation function $\corr{\rho} (R\to\infty)$, indicating the degree of density homogeneity.
    Modulated methods, in particular, result in a more homogeneous system. 
    \textbf{(d) }
    The temporal average of the spatial maxima of local density, $\av{ \max\left[ \rho_C(\vec{r}, t) \right]_{\vec r} }_t$.
    }
    \label{fig:Dens-DensMetrics}
\end{figure*}

Likewise, the spontaneous active flows scale as $v_\mathrm{av} \sim \act^{1/2}$~\cite{Giomi2014PhilTransactionsA}.
The speed resulting from active turbulence for each formulation (\fig{fig:Turb-FlowSpeed}a) is obtained by measuring the average root-mean square flow speed $v_\mathrm{av}$ and subtracting the residual thermal speed $v_\mathrm{av}^0 = \lim_{\act\to 0}v_\mathrm{av} \neq 0$. 
This yields two distinct scaling regimes:
At lower activities, $\act \lesssim \actMin$, system-spanning bend walls drive slow but coherent flows in the system. 
This results in a scaling that is consistent with $v_\mathrm{av} \sim \act$ (\tabsi{sitab:ScalingRef}).
At higher activities when the system enters into active turbulence ($\act \gtrsim \actMin$), all formulations match the theoretical scaling, as indicated by the dashed line in \fig{fig:Turb-FlowSpeed}a and \tabsi{sitab:ScalingRef}, with the exception of \emph{particle-carried activity} at very large activities. 
Both regimes exist for all four activity formulations. 
All activity formulations reproduce the expected flow speed scaling laws in the active turbulence regime and within their operational regime.

\begin{figure}[tb]
    \centering
    \begin{subfigure}
        \centering
        \includegraphics[width=0.9\linewidth]{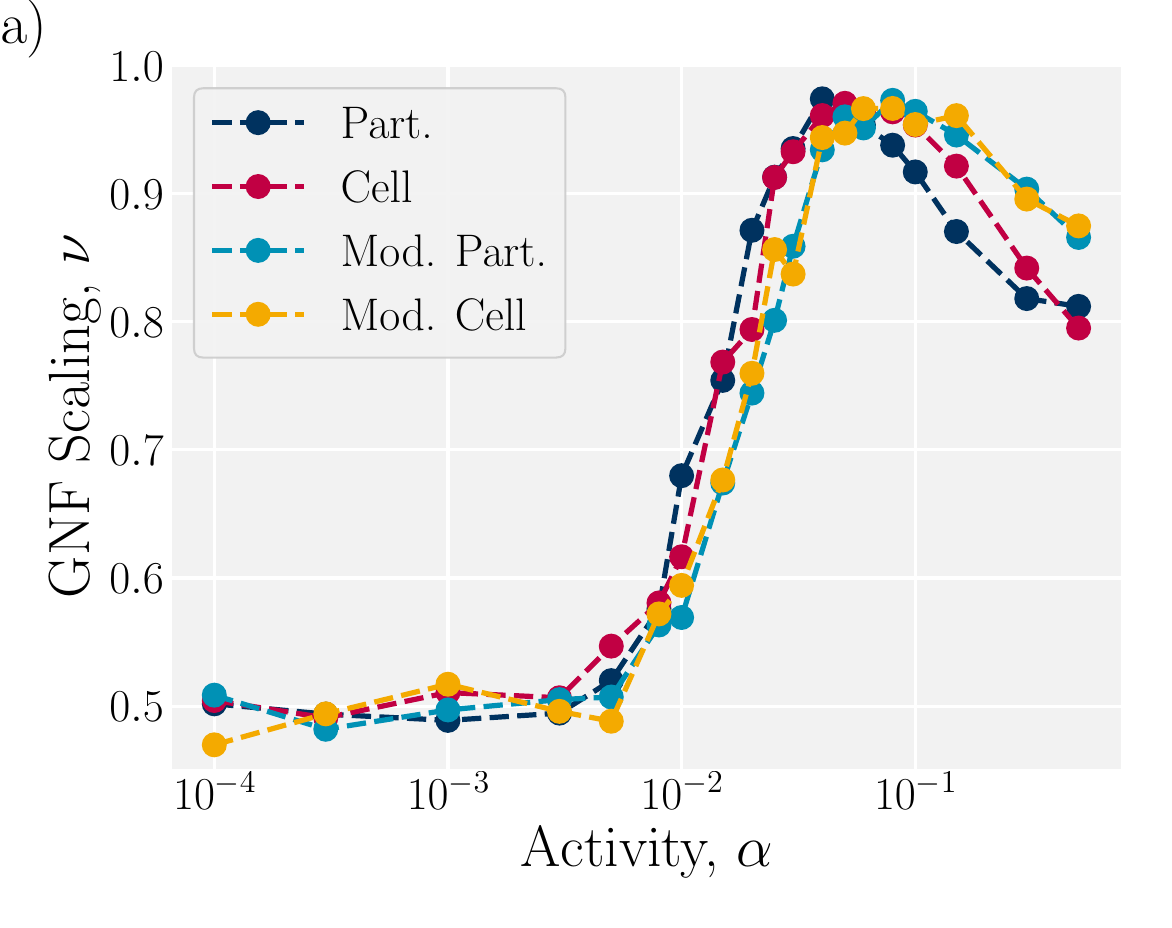}
    \end{subfigure}
    \begin{subfigure}
        \centering
        \includegraphics[width=0.9\linewidth]{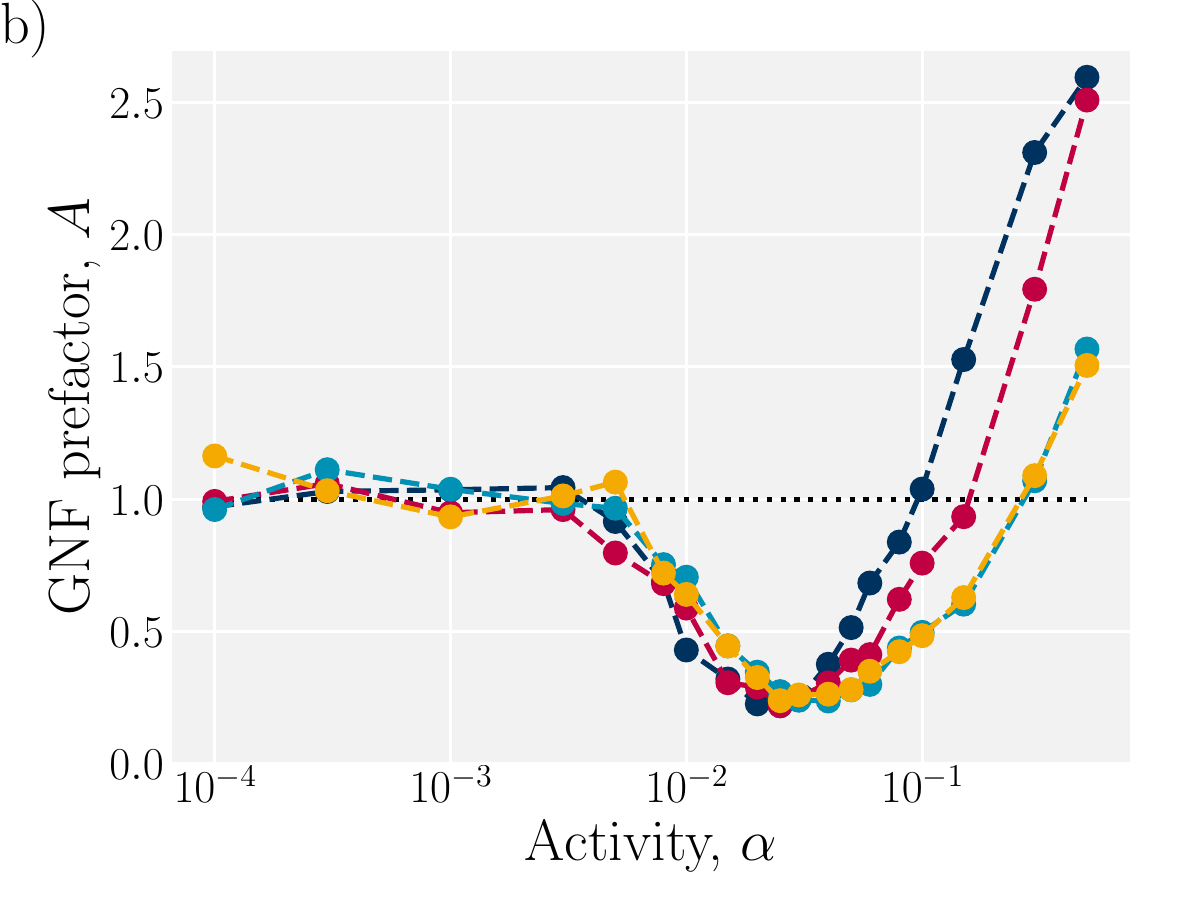}
    \end{subfigure}
    \begin{subfigure}
        \centering
        \includegraphics[width=0.9\linewidth]{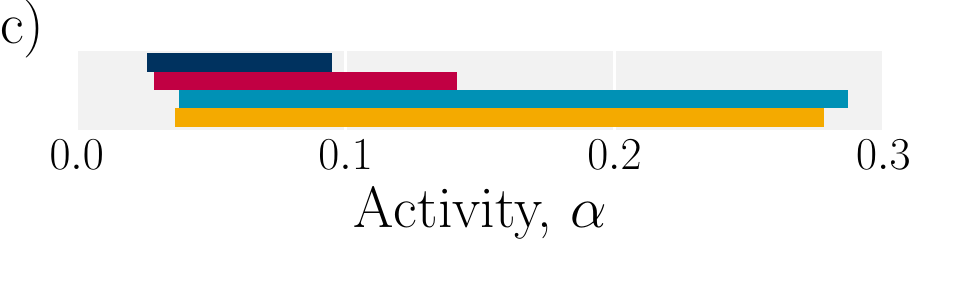}
    \end{subfigure}
    \caption{\textbf{Giant number fluctuations (GNF) analysis reveals algorithm regimes. } 
    \textbf{(a) }
    The scaling $\nu$ of giant number fluctuations, $\sigma_\rho = A\rho^\nu$.
    Each of these methods has similar scaling, with a shift for the modulated formulations.
    \textbf{(b) }
    The prefactor $A$ of fits to the giant number fluctuations, $\sigma_\rho = A\rho^\nu$.
    In the passive regime $\act\lesssim\actMIN$ these remain at a unit value. 
    Once in the nematic bend regime $\actMIN\lesssim\act\lesssim\actMin$ these begin to drop, with a minimum when turbulence forms at $\act\simeq\actMin$.
    In the turbulence regime $\actMin\lesssim\act$ the prefactor rises until it exceeds unit value when $\act\gtrsim\actMax$.
    \textbf{(c) }
    The effective turbulence regime $\actMin\lesssim\actMax$ for the four formulations of cellular activity, obtained from (b).
    }
    \label{fig:Dens-GNF}
\end{figure}

\begin{figure*}[tb]
    \centering
    \begin{subfigure}
        \centering
        \includegraphics[width=0.32\linewidth]{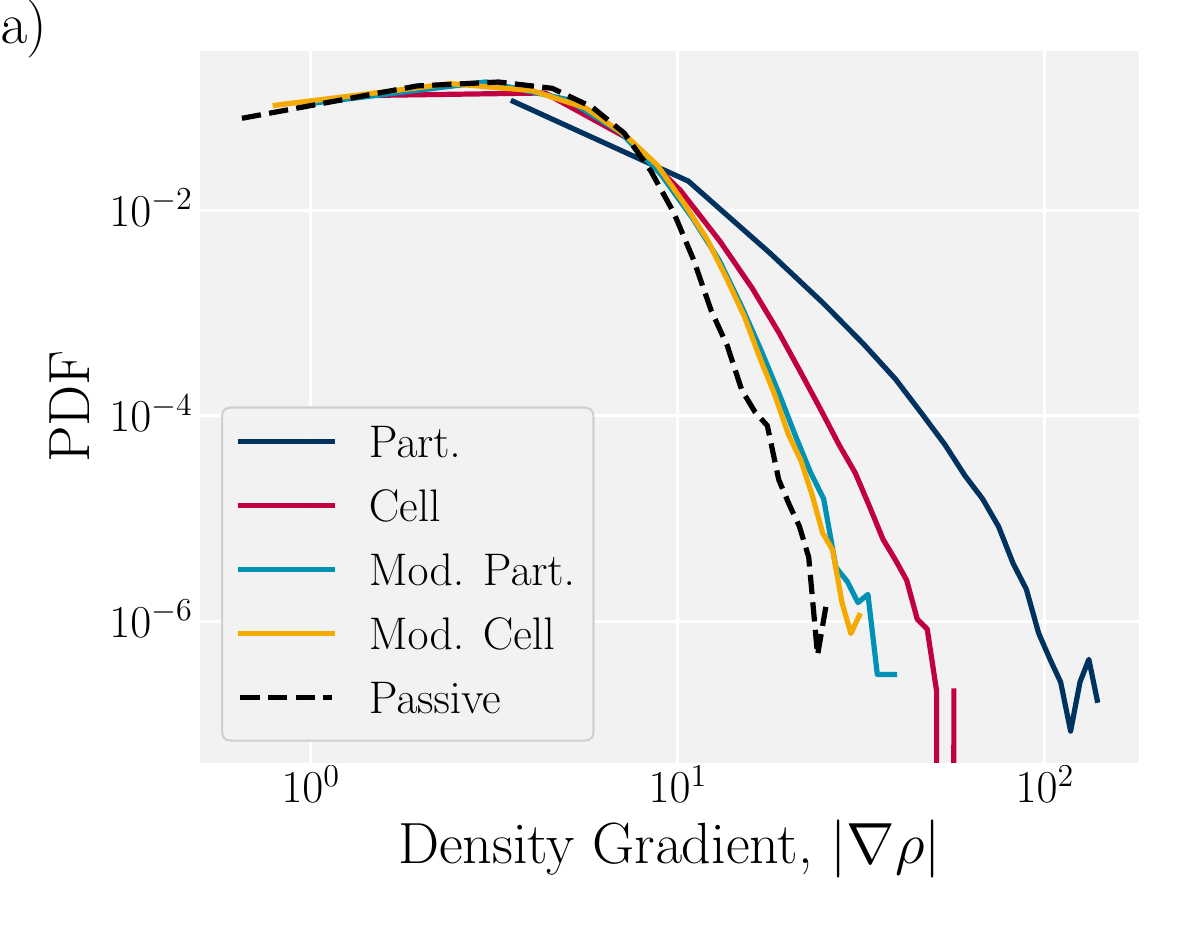}
    \end{subfigure}
    \begin{subfigure}
        \centering
        \includegraphics[width=0.32\linewidth]{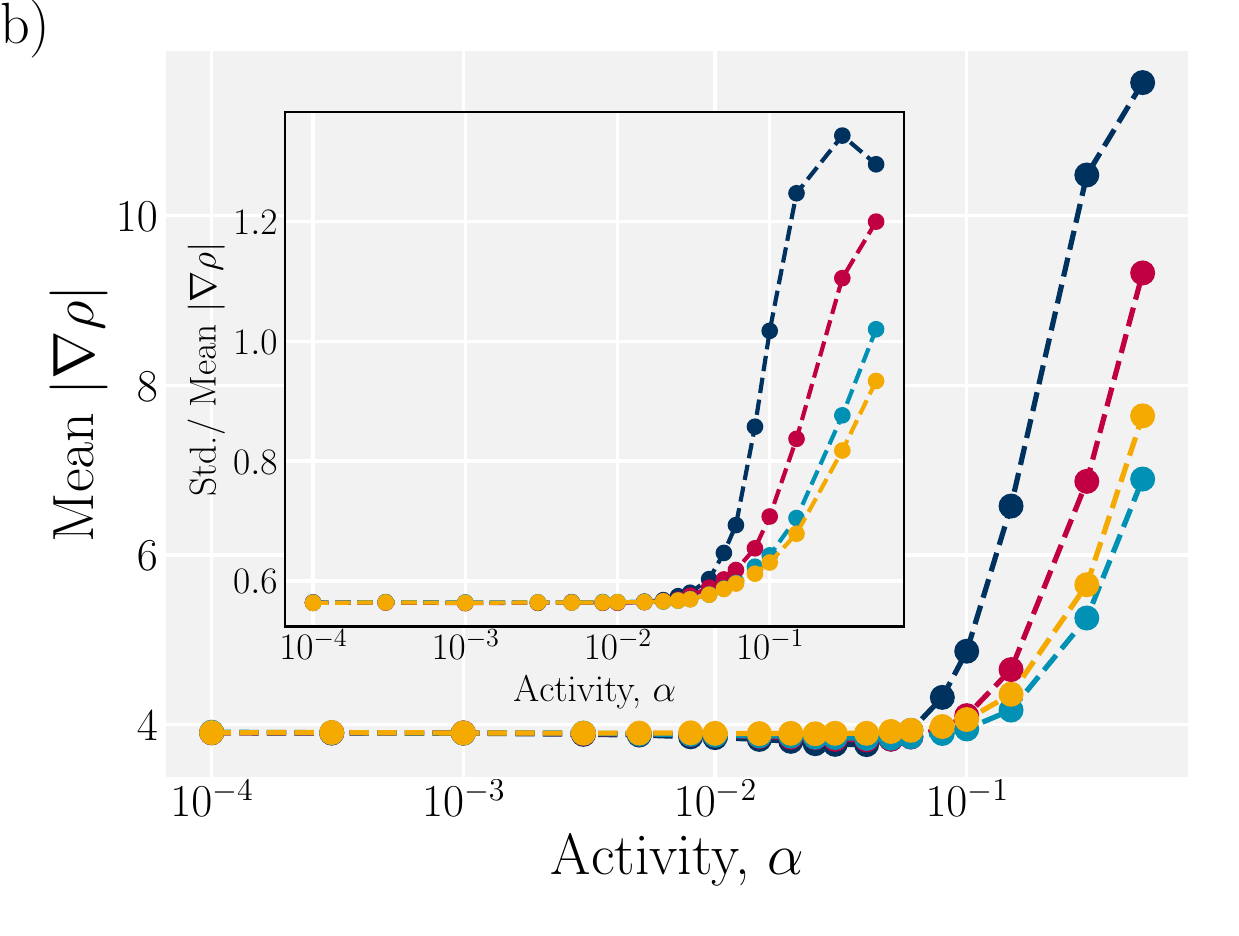}
    \end{subfigure}
    \begin{subfigure}
        \centering
        \includegraphics[width=0.32\linewidth]{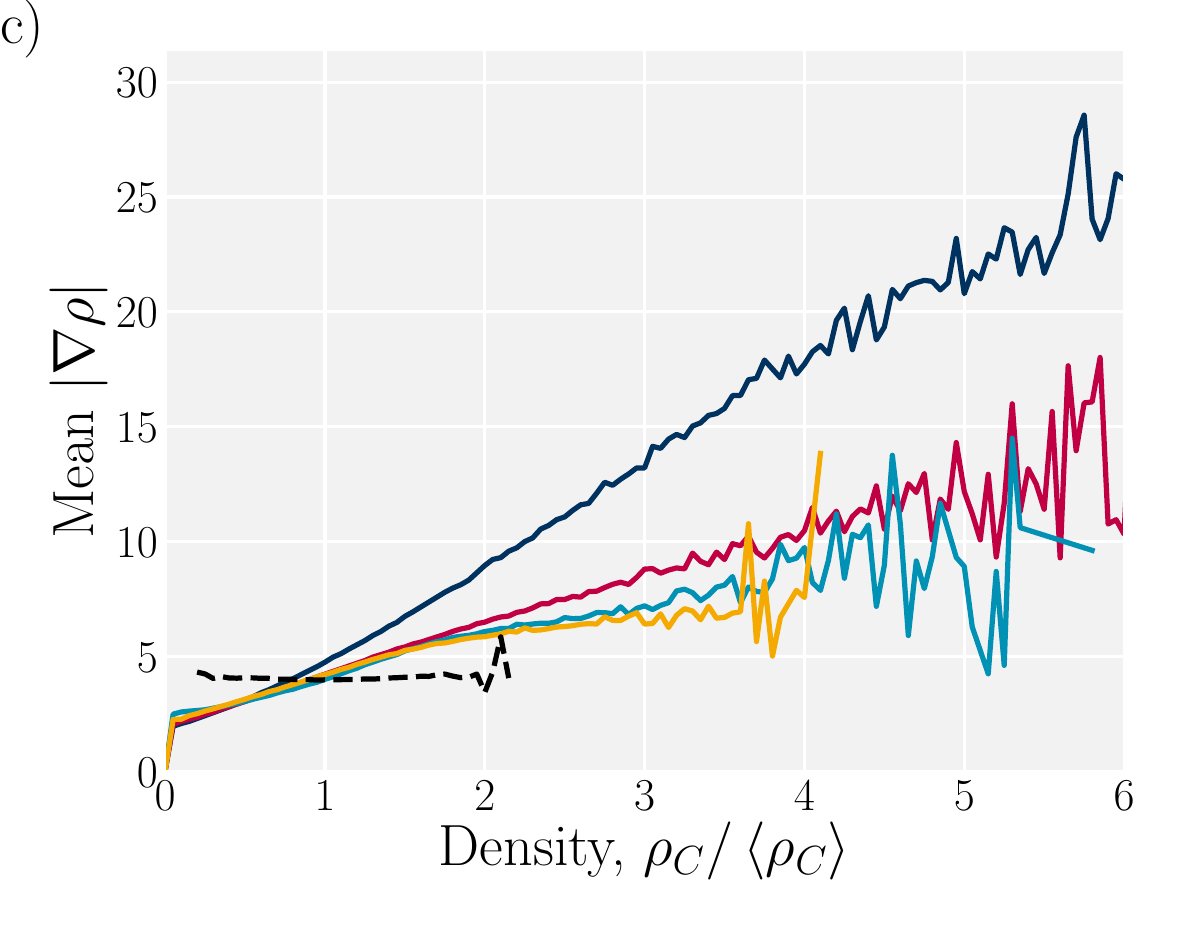}
    \end{subfigure}
    \caption{\textbf{Modulated formulations for cellular activity mitigate density induced drift. }
    \textbf{(a) }
    Distributions of the magnitude of pressure induced drift, $\abs{\grad\rho}$, for all four formulations of cellular activity at a fixed activity of $\act=0.1$ and for passive N-MPCD. 
    \textbf{(b) }
    The mean magnitude of pressure induced drift, $\av{\abs{\grad\rho}}$, for all four formulations of cellular activity.
    \textbf{Inset: } 
    The standard deviation of pressure induced drift normalized by the mean for all four formulations of cellular activity.
    \textbf{(c) }
    The magnitude of pressure induced drift as a function of local density, $\abs{\grad\rho}(\CellDens)$, for all four formulations of cellular activity at a fixed activity of $\act=0.1$ and for passive N-MPCD.
    The beginnings of plateaus of maximum pressure induced drift are seen for the modulated methods, capping the effect density fluctuations can have on solutes. 
    }
    \label{fig:Dens-Ficks}
\end{figure*}

The radial velocity auto-correlation functions, $\corr{v}(R) = \av{\vec{v}(\vec r)\cdot \vec{v}(\vec r + R \vec{\hat{r}})}_{\vec{r},t} / \av{v^2(\vec r)}_{\vec{r},t}$ (\figsi{sifig:Vel-Corr}) are fit by an exponential decay to measure the velocity length scale of active turbulence, $\lenVel$ (\fig{fig:Turb-FlowSpeed}b). 
For $\actMIN \lesssim \act \lesssim \actMin$, the active flow driven by the nascent nematic deformations of the bend walls are largely cohesive, leading to high $\lenVel$. 
Furthermore, these flows are parallel to the bend walls, which increases their amplitude, but crucially, not their wavelength. 
The characteristic decorrelation length scale is largely independent of the particle activity $\act$ in this regime. 
On the other hand in the turbulent regime, the length scale decays as $\lenDef\sim\act^{-1/2}$ (\tabsi{sitab:ScalingRef}), which matches the theoretical prediction~\cite{Thampi2014EPL, Hemingway2016SoftMatter}. 

In conclusion, regardless of the chosen activity formalism, AN-MPCD successfully reproduces the theoretically expected scaling laws for active turbulence within its operational regime. 
However, the onset of turbulence varies depending on formalism, indicating that the activity operational regimes depend on the implementation of activity.

\subsection{Density Variation}

As a particle-based model of active matter, AN-MPCD exhibits density fluctuations~\cite{Chate2006PRL, PeshkovChate2014PRL, Bertin2009JPhysA, Bar2020AnnRev}.
Furthermore, since its equation of state corresponds to an ideal gas~\cite{GompperIhle2009Book-MPCD, Zantop2021JCP-MPCDState}, MPCD simulates a compressible fluid~\cite{theers2014, Akhter2014}. 
Together these produce density fluctuations, which in turn induce pressure gradients and resulting flows. 
If significant, these flows may affect the dynamics of suspended solutes, such as colloids, tracers and polymers. 
This section will show that the modulated activities greatly reduce density fluctuations when compared to \emph{particle-carried activity} by characterizing the behavior of density fluctuations and the corresponding pressure induced gradients, demonstrated in \fig{fig:DensSnapshots}; \movie{simov:DensSnapshots}.

In the low activity limit, the density distributions are Gaussian about $\av{\CellDens}$ (\figsi{sifig:Dens-Prob}). 
As the activity is increased, the distributions skew, producing high density tails (\figsi{sifig:Dens-Prob}). 
While qualitatively all density distributions behave similarly, a quantitative analysis reveals differences. 
While the mean of the distributions is fixed by the number of particles in the simulation, the median is constantly $\CellDens=20$ at low activity for all activity formulations (\fig{fig:Dens-DensMetrics}a). 
In each formulation, the cell population median starts to decrease in the turbulent regime, $\act \gtrsim \actMin$ (\fig{fig:Dens-DensMetrics}a). 
While all four formulations exhibit this behavior, \emph{particle-carried activity} $\ActSum$ exhibits the largest drop. 
Since particle number is conserved, the smaller cell population median in the active turbulent regime arises because most cells lose particles to a relatively few cells that drastically increase their local density. 

This results in an associated increase in the standard deviations (\fig{fig:Dens-DensMetrics}a inset). 
The standard deviations are constant at low activities but rise in the active-turbulence regime. 
This is true in all formulations of activity; however, it is once again most pronounced for the \emph{particle-carried activity} $\ActSum$ (\fig{fig:Dens-DensMetrics}a inset). 
The second largest standard deviation is for \emph{cell-carried activity} $\ActAv$, while the two modulated versions, $\SigSum$ and $\SigAv$, are the smallest and relatively indistinguishable. 
Thus, the modulation decreases density variation, as intended. 

It is useful to consider the extrema of low and high densities. 
At the lowest of densities, the fraction of completely empty cells ($\CellDens=0$) raises sharply above $\act \gtrsim \actMin$ (\fig{fig:Dens-DensMetrics}b).
Because observables are computed from cell averages, a large fraction of empty cells marks the break down of the MPCD algorithm. 
In this regard, \emph{particle-carried activity} has an order of magnitude larger fraction of empty cells when compared to all other formalisms at $\act \gtrsim \actMin$ activity values.  
In particular, the \emph{modulated particle-carried activity} $\SigSum$ has the lowest fraction of empty cells --- each more than an order of magnitude smaller than \emph{particle-carried activity}. 

In the other extreme, in the turbulent regime high density cells form thin, high-density \emph{nematic bands}~\cite{Chate2019} (\movie{simov:DensSnapshots}). 
The radial density auto-correlation function $\corr{\rho}(R) = \av{\CellDens(\vec r) \CellDens(\vec r + R \vec{\hat{r}})}_{\vec{r},t} / \av{{\CellDens}^2(\vec r)}_{\vec{r},t}$ characterizes the width of these bands (\figsi{sifig:Dens-Corr}) through the decorrelation length $\ell_{\rho}$ (\fig{fig:Dens-DensMetrics}c). 
As the system enters its turbulent regime and bands form, increasing activity decreases the density decorrelation lengths (\fig{fig:Dens-DensMetrics}c).
The density decorrelation length scales roughly inverse to activity as $\lenDens\sim \act^{-1}$ (see \tabsi{sitab:ScalingRef}), indicating nematic bands become increasingly thinner. 
While this behavior is shared by all activity formulations, \emph{particle-carried activity} leads to the thinnest bands, whereas the modulated methods provide the widest bands (with roughly twice the decorrelation length when compared to \emph{particle-carried activity}). 

Complementing the decorrelation length is the far-field value of the correlation function, $\corr{\rho}(R\to\infty)$ (\fig{fig:Dens-DensMetrics}c; inset). 
The far-field decorrelation describes how pronounced the density fluctuations are, providing an estimate of how dense bands are compared to the diluted surroundings. 
Not only does \emph{particle-carried activity} have the thinnest bands (\fig{fig:Dens-DensMetrics}c), it also has the largest density difference. 
On the other hand, the modulated methods $\SigSum$ and $\SigAv$ have the least particle accumulation within bands. 
This behavior is further verified by considering the time average of the instantaneous maximum cell density, $\av{ \max\left[ \rho_C(\vec{r}, t) \right]_{\vec r} }_t$, which measures how dense the peak of the nematic bands are for each activity formulation (\fig{fig:Dens-DensMetrics}d). 
Consistent with all other metrics of density variation, \emph{particle-carried activity} $\ActSum$ has significantly larger maximum instantaneous densities compared to the modulated methods.
This indicates that, while density bands are present in all activity formulations, the modulated activities lead to attenuated bands that are wider and less concentrated, and thus to more homogeneous systems. 
Overall, these results illustrate how \emph{cell-carried activity} $\ActAv$ reduces density variation compared to the original implementation of \emph{particle-carried activity} $\ActSum$; however, it is the \emph{modulated activities} $\SigSum$ and $\SigAv$ that produce the most homogeneous solvents.

The activity-driven variation of density can be further quantified by measuring the giant number density fluctuations~\cite{Ramaswamy2003EPL, Shi2013NatComm, liu2021}. 
The standard deviation of density in local regions of increasing size is measured as a function of the average number of particles $\sigma_\rho(\rho_C)$ (\figsi{sifig:GNF}). 
The resulting standard deviations are fit to a power law, $\sigma_\rho(\rho_C) = A\rho_C^\nu$, which in equilibrium scales as $\sigma_\rho \sim \rho_C^{1/2}$ according to the central limit theorem. 
However, as for many non-equilibrium active particle systems~\cite{Ramaswamy2007Science, Henkes2011-Jamming, Henkes2018-NematicSphere, Toner2019, liu2021}, $\nu\to 1$ in the turbulent regime irrespective of the activity formulation (\fig{fig:Dens-GNF}a). 

While the phrase ``giant number fluctuations'' refers to the increasing of $\nu$ above $1/2$, it does not necessarily refer to the actual magnitude of the variations. 
Indeed, what matters for solute dynamics is not necessarily the exponent but rather the local amplitude of the variations, which is quantified by the prefactor $A$ (\fig{fig:Dens-GNF}b).
This prefactor can be interpreted as $A=\sigma_\rho(1)$ is therefore a measure of local density fluctuations in the limit of a single particle in a cell. 
The behavior of $A$ is found to be a particularly clear descriptor of the different regimes of activity for each formulation. 
At low activities ($\act < \actMIN$), the source of fluctuations is the thermostat and the prefactor is identical for all formulations of activity (\fig{fig:Dens-GNF}b). 
At higher activities ($\actMIN \lesssim \act \lesssim \actMin$), the active force overcomes the thermostat and the system begins to exhibit bend walls on scales comparable to the system size with their associated deterministic flows (\fig{fig:Dens-GNF}b). 
As flows become more deterministic, individual particles generally travel along the direction of the active forcing, surpressing thermal fluctuations ($A<1$). 
As the activity is further increased ($\actMin \lesssim \act \lesssim \actMax$), the bend instability is triggered and the active turbulence competes against coherent flows increasing fluctuations, and causing $A$ to rise (\fig{fig:Dens-GNF}b).
At even greater activities ($\act \gtrsim \actMax$), the amplitude $A>1$ indicates substantial non-equilibrium effects to the density (\fig{fig:Dens-GNF}b;). 
At these high activities, the activity-induced density effects are strong, and the algorithm no longer reproduces a continuum fluid on large length scales. 
This occurs first in \emph{particle-carried activity} and last in the \emph{modulated activities}. 
From this we can infer the critical activities for the bend instability $\actMIN$, the onset of active turbulence $\actMin$, and the onset of substantial density fluctuations $\actMax$. 
In particular, $\actMin$ and $\actMax$ serve as bounds for each activity formulation's operational turbulence regime (\fig{fig:Dens-GNF}c and \tabsi{sitab:OperationalRegimeRef}). 
The two \emph{modulated activities} $\SigSum$ and $\SigAv$ are seen to have the widest operational regime, roughly four times wider than the \emph{particle-carried activity} $\ActSum$ (\fig{fig:Dens-GNF}c). 

\subsection{Density Gradients}

While the attenuation of giant-number fluctuations by \emph{modulated activities} is promising, density bands may still induce local fluxes, which affect the behavior of solutes or suspended particles. 
As the MPCD equation of state corresponds to an ideal gas, Fick's first law states that density-induced diffusive fluxes are driven by density gradients, $\grad \rho$.
Hence, the magnitude of density gradients, $\abs{\grad \rho}$, measures the contribution of local density variations to the flux. 
The distribution of $\abs{\grad \rho}$ for a fixed activity $\act=0.1$ (\fig{fig:Dens-Ficks}a) reveals that all activity formulations present Maxwell-Boltzmann-like behavior. 
The \emph{particle-carried activity} has the longest tail, while the two \emph{modulated activities} have the shortest tails. 
The modulated activity formulations are very close to the passive case, which is a Maxwell-Boltzmann distribution with standard deviation $\sigma = 0.85\pm0.05$. 

Lower activities ($\act=0.08$) have narrower distributions, with less pronounced differences between the \emph{modulated activities} and \emph{cell-carried activity}, although \emph{particle-carried activity} maintains its significantly wider tail (\figsi{sifig:Ficks-Prob}a). 
Meanwhile higher activities ($\act=0.3$) have much wider distributions, although the \emph{modulated activities} still have the shortest tails (\figsi{sifig:Ficks-Prob}b).
As activity increases, the mean and width of the distributions increase (\fig{fig:Dens-Ficks}b, b inset), indicating that density-induced drift should be expected to become larger and more widely distributed. 
\emph{Particle-carried activity} consistently presents significantly wider tails, and therefore, larger expected fluxes than the other activity choices. 
The \emph{modulated activities} generally have the lowest gradients, barely increasing beyond the equilibrium limit of $\act\to 0$ for activities $\act \lesssim 0.1$ (\fig{fig:Dens-Ficks}b).

To further investigate why this is the case, the magnitude of density gradients as a function of local density is measured for a fixed activity ($\act=0.1$) and compared with the passive case (\fig{fig:Dens-Ficks}c).
While the passive case remains constant for all densities, both \emph{particle-} and \emph{cell-carried activities} exhibit monotonically increasing $\abs{\grad\rho}$ as a function of local density. 
However, the \emph{modulated activities} begin to plateau to a value only slightly beyond the passive limit as $\CellDens\to 1.5\av{\CellDens}$.
As with the distributions of $\abs{\grad\rho}$, this behaviour still occurs for lower activities (\figsi{sifig:Ficks-Dens}a), but is less pronounced. 
Meanwhile for higher activities (\figsi{sifig:Ficks-Dens}b), the \emph{modulated activties} exhibit a fully formed plateau. 
This indicates that the \emph{modulated activities} $\SigSum$ and $\SigAv$ have comparable Fickian fluxes to the passive limit.


\begin{figure}[tb]
    \centering
    \begin{subfigure}
        \centering
        \includegraphics[width=0.95\linewidth]{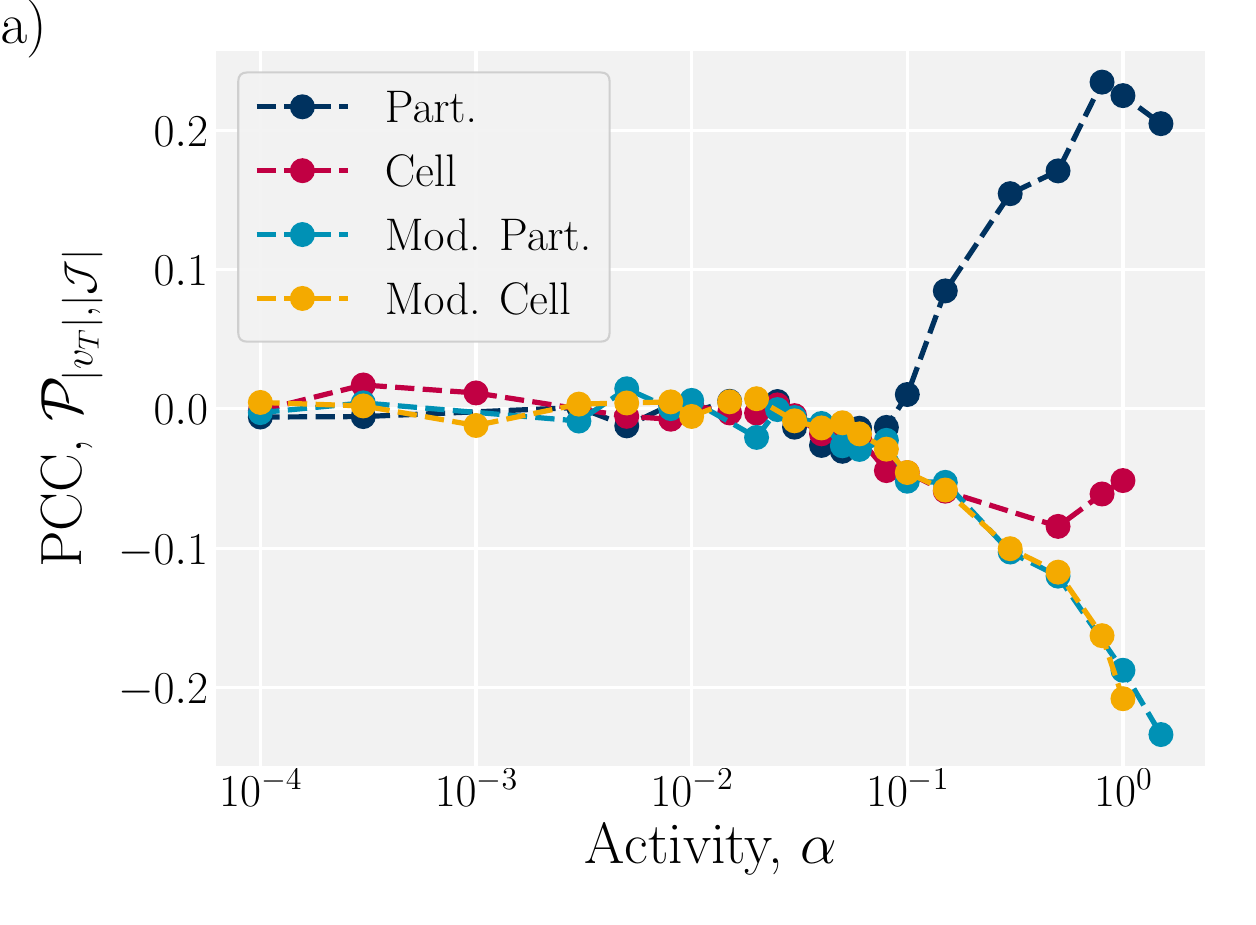}
    \end{subfigure}
    \begin{subfigure}
        \centering
        \includegraphics[width=0.95\linewidth]{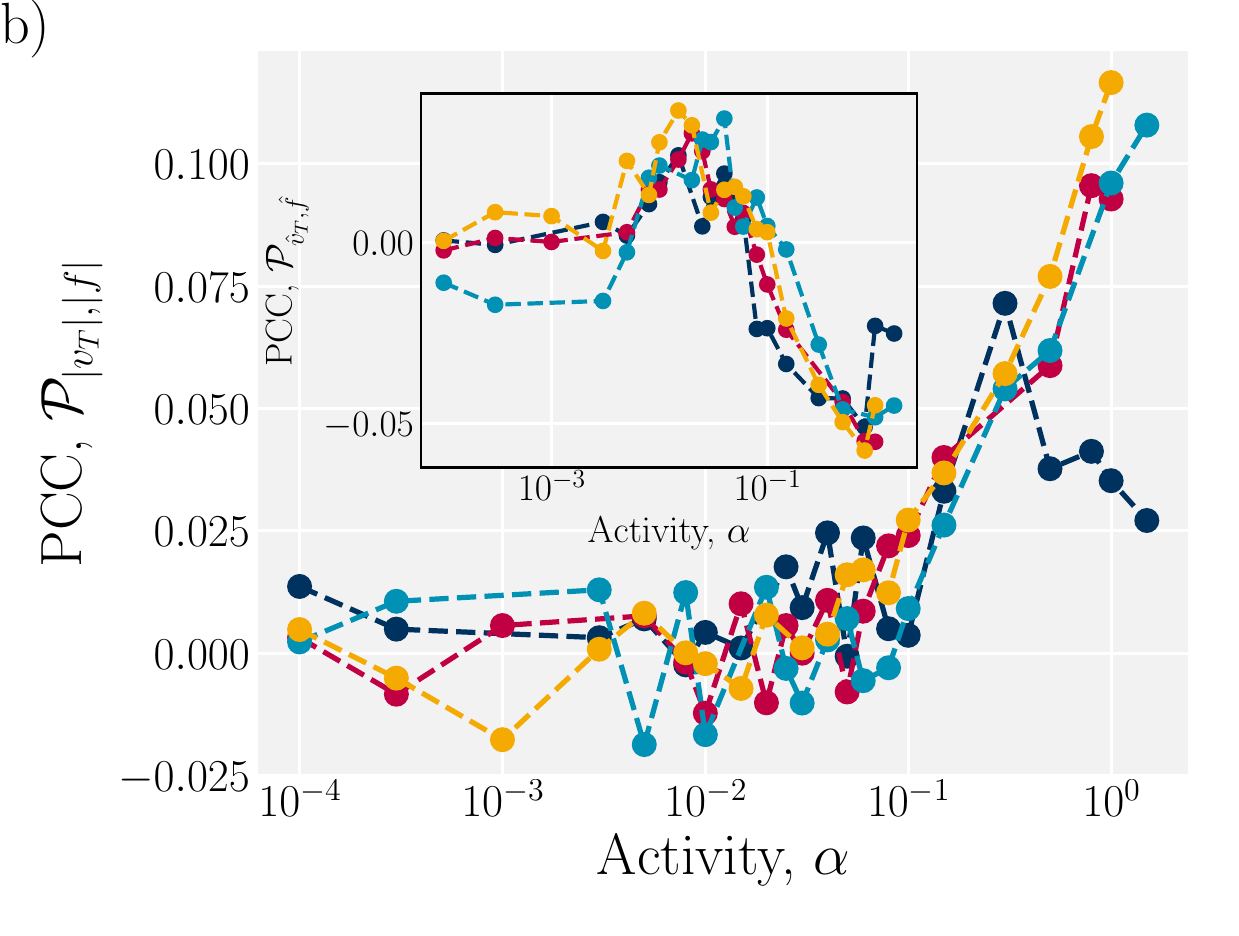}
    \end{subfigure}
    \caption{\textbf{Pearson correlation coefficients (PCC) of tracer particle velocity to cell quantities have non-trivial dependency on activity formulation. } 
    \textbf{(a) }
    PCC $\pcc{\abs{v_T}}{\abs{\mathcal{J}}}$ of tracer particle speeds $\abs{v_T}$ compared to the magnitude of pressure induced drift $\abs{\mathcal{J}}=\abs{\grad \rho}$.
    \textbf{(b) }
    PCC $\pcc{\abs{v_T}}{\abs{f}}$ of tracer particle speeds $\abs{v_T}$ compared to the magnitude of active force $\abs{f}=\abs{\grad \cdot \vec{n}_C}$.
    \textbf{Inset: }
    PCC $\pcc{\hat{v}_T}{\hat{f}}$ of tracer particle velocity directions $\hat{v}_T$ compared to the direction of active force $\hat{f}=(\grad \cdot \vec{n}_\mathrm{C}) \vec{n}_\mathrm{C}/\abs{f}$.
    }
    \label{fig:Dens-PCC}
\end{figure}

The density gradients suggest that Fickian fluxes in the \emph{modulated activities} should not affect solutes substantially more than in passive MPCD. 
To confirm this, point-like tracer particles are suspended within the active nematic solvent, whose trajectories can be recorded and directly analyzed. 
Tracer particles have the same mass $m_T=m$ as the solvent, but are not active $\act_T=0$.
The velocity of tracer particles $\vec{v}_T$ is correlated to the local fluid fields at the level of MPCD cells. 
In particular, the correlation between tracer velocity and terms proportional to the active forcing $\vec{f}=(\grad \cdot \vec{n}_\mathrm{C})\vec{n}_\mathrm{C}$ and proportional to the local density gradients $\vec{\mathcal{J}}=\grad \rho$ are measured. 
This is done using the Pearson correlation coefficient (PCC), $\pcc{X}{Y}=\av{(X-\av{X}(Y-\av{Y}} / \sigma_X \sigma_Y$ between variables $X$ and $Y$ with standard deviations $\sigma_X=\av{(X-\av{X}}^{1/2}$.
The PCC is used to reveal whether the magnitudes of the local density gradient $\abs{\mathcal{J}}=\abs{\vec\nabla\rho}$ or active forcing $\abs{f}=\abs{\vec\nabla\cdot\vec{n}_C}$ correlate to tracer speed $\abs{v_T}$. 
Likewise, angular PCC determines whether tracer particles are more biased in the direction of the density gradients $\hat{\mathcal{J}}=\vec\nabla\rho / \abs{\mathcal{J}}$ or active forcing $\hat{f}=(\vec\nabla\cdot\vec{n}_C)\vec{n}_C / \abs{f}$.

The angular correlation between the density gradients and tracer velocity, $\pcc{\hat{v}_T}{\hat{\mathcal{J}}}$, indicates complete decorrelation, irrespective of activity (\figsi{sifig:PCC-Angular}a). 
This indicates that tracers are not biased in the direction of density gradients. 
On the other hand, the correlation of the magnitude between the tracer particle velocity and the density gradient, $\pcc{\abs{v_T}}{\abs{\mathcal{J}}}$, is decorrelated at low activities but, within the active turbulence regime, it is not (\fig{fig:Dens-PCC}a; $\act \gtrsim \actMin$). 
Within the turbulence regime, the various activity formulations behave quite differently. 
The Pearson correlation coefficient of the magnitude for \emph{particle-carried activity} strongly increases with activity. 
This positive correlation indicates that tracer particles are more likely to travel faster when the density gradient is large, which is expected due to the formulation of \emph{particle-carried activity}. 
In contrast, all other activity choices result in negative Pearson correlation coefficients that decrease with activity, indicating that tracer particles move less quickly in regions with greater density gradients. 
This is particularly striking for the \emph{modulated activity} variants. 
This is because density-induced fluxes tend to occur in regions of high density, whereas the modulation strongly caps the active forcing at those high densities, leading to an anti-correlation with density gradients. 
While the strength of correlation and anti-correlation can reach $\pm0.2$ at the highest activities, the magnitude of correlation coefficients remains $\approx0.05$ within the operational regimes identified in \fig{fig:Dens-GNF}c. 
This is comparable to the uncertainty on the correlation coefficients of the passive MPCD case. 
Taking both the angular and magnitude PCC into account, the \emph{modulated activities} are indeed counteracting the activity-density feedback:
There is no directional bias of the tracer particle velocity with respect to the density gradient, and the tracer particles slow down in regions of high density gradients.

The Pearson correlation coefficient between the direction of the active force and the tracer velocity $\pcc{\hat{v}_T}{\hat{f}}$ is similar for all formulations, with a small peak for $\actMIN \lesssim \act \lesssim \actMin$ (\fig{fig:Dens-PCC}b inset). 
This is the activity regime where the active length scale is comparable to the system size and bend walls form. 
The positive correlation results from tracers traveling with the bend walls. 
In the turbulence regime, the dynamics become slightly anti-correlated at a similar scale of $\pcc{\hat{v}_{T}}{\hat{f}}\approx -0.05$. 
This contrasts with $\pcc{\abs{v_{T}}}{\abs{f}}$ (\fig{fig:Dens-PCC}b), which is decorrelated until the turbulence regime and then rises, indicating that active forcing correlates with the velocity of the tracer. 
The PCC $\pcc{\abs{v_{T}}}{\abs{f}}$ of each activity formulation continues to rise for all activities, except \emph{particle-carried activity}, which falls off once $\act \gtrsim \actMax$, due to the algorithm being overcome by density fluctuations. 
The increasing correlation of the active-force magnitude with the tracer velocity is expected in the turbulence regime, as the active force becomes the primary driver of the tracer particle motion.
This correlation analysis shows that the modulated methods do not produce density gradients that dominate over direct active forcing.

\subsection{Net Active Forcing}

\begin{figure}[tb]
    \centering
    \begin{subfigure}
        \centering
        \includegraphics[width=0.95\linewidth]{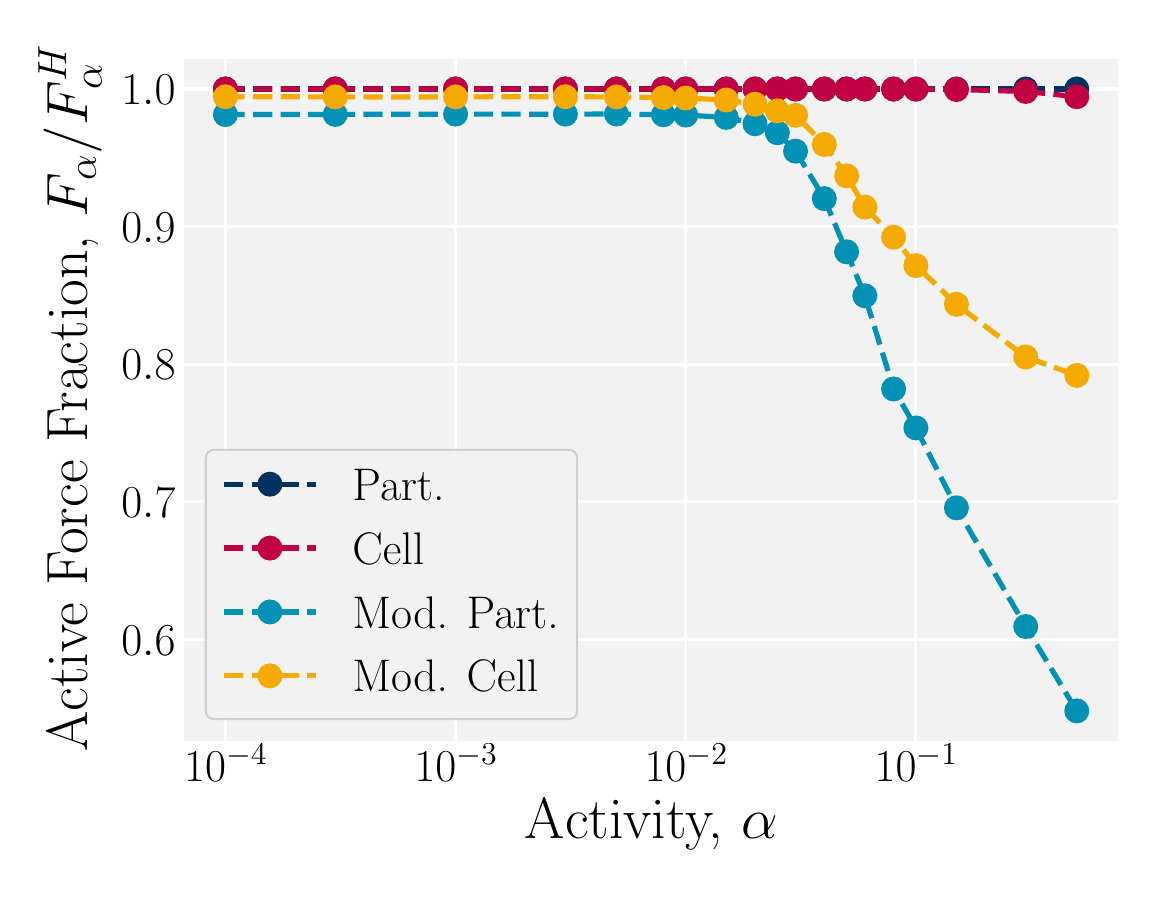}
    \end{subfigure}
    \caption{\textbf{Modulation of cellular activity results in active force fluctuations. }
    The temporally averaged active force input into the system, $F_\act$, is computed for each formulation of cellular activity and normalized with the active force input at average density $F^H_\act$. 
    While the active force from \emph{particle-carried activity} remains constant, all other methods exhibit some degree of active force fluctuations.
    }
    \label{fig:Energy-Global}
\end{figure}

Having characterized the fluctuations associated with each of the definitions of the local activity, it is worth returning to the definitions to better understand how the various activity formulations produce these variations. 
Since the \emph{particle-carried activity} formulation establishes a linear relationship between the cell's activity and particle density, $\ActSum=\CellDens\act$, and since the total number of particles is conserved throughout the simulation, the net activity $F_\act = \sum_C \act_C$ is constant. 
Comparing the net activity with the net activity for the homogeneous system, $F^H_\act=\sum_C \act_C(\NCell)$ (\ie, the active force input for average density), acts as a measure of energy input per collision operation.
For \emph{particle-carried activity}, this is definitionally unity (\fig{fig:Energy-Global}). 

This is not the case for the other activity formulations, since they are not linear with local density (\fig{fig:ActivitySchematic}). 
While \fig{fig:Energy-Global} appears to indicate that the \emph{cell-carried activity} is also independent of $\act$, there is in fact a marginal deviation from a homogeneous system at the highest activities. 
This is due to entirely empty cells in the highest activity regime, which results in a small drop in the net active forcing $F_\act$ input (\fig{fig:ActivitySchematic}; \eq{eq:ActAvDef}). 
In contrast, the \emph{modulated activities} never reach a net active forcing equivalent to an homogeneous system, even for the smallest activities (\fig{fig:Energy-Global}).
This is because inherent fluctuations within MPCD make it exceedingly unlikely that the system will remain at the average density. 
In the \emph{modulated activity} formalism, this causes the net active forcing to be slightly attenuated even in the low activity limit. 
As the activity increases, density fluctuations increase and the attenuation also rises, reducing the net active forcing. 
Between the two modulated formulations, $\SigSum$ has a greater loss in net active forcing because the cell activity at average density $\SigSum(\av{\CellDens})$ is much larger than $\SigAv(\av{\CellDens})$ (\fig{fig:ActivitySchematic}). 
Overall, this shows that these new formulations for cell activity trade off reduced density fluctuations for an increase in active forcing fluctuations.

\section{Discussion}

Active-nematic MPCD offers many advantages for simulating active fluids in complex environments or with suspended mesoscale inclusions, such as polymers and filaments. 
However, like any coarse-grained approach, AN-MPCD also possesses limitations and an operational regime. 
Here, we have considered how activity induces density variations in AN-MPCD, as it does in any particle-based approach~\cite{Chate2006PRL, PeshkovChate2014PRL, Bertin2009JPhysA, Bar2020AnnRev, Valeriani2022}. 
Any modifications that can curb density variations will extend the applicability of AN-MPCD for simulations of homogeneous fluids. 

This manuscript considered four activity formulations as a function of local density: 
\emph{(i)} The original \emph{particle-carried activity} $\ActSum$, which is proportional to the local density, 
\emph{(ii)} a modified \emph{cell-carried activity} $\ActAv$, which ensures activity is homogeneously applied throughout the system domain, 
\emph{(iii)} \emph{modulated particle-carried activity} $\SigSum$,  
\emph{(iv)} and \emph{modulated cell-carried activity}  $\SigAv$.
The two modulated formulations introduce a negative feedback by effectively decreasing cellular activity whenever the instantaneous local density is too large.
A sigmoidal modulation is chosen to smoothly switch activity ``on/ off''.
All four formulations are found to reproduce the theoretically expected scaling laws of active nematic turbulence, albeit with slightly differing turbulence regimes.
The \emph{modulated activities} $\SigSum$ and $\SigAv$ are found to have the widest effective turbulence regime, roughly four times wider than the \emph{particle-carried activity} $\ActSum$.

All four formulations are found to exhibit density fluctuations, common to active particle based methods, but the new formulations ($\ActAv$, $\SigSum$, $\SigAv$) exhibit reduced variations over the original \emph{particle-carried activity} $\ActSum$ formulation.
The \emph{modulated activities}, in particular, are found to have significantly reduced density fluctuations. 
The two modulated approaches are comparable, though \emph{modulated particle-carried activity} results in less empty MPCD cells.
To verify the suitability of AN-MPCD for the study of solutes suspended within an active fluid, the density gradients characterized density-induced Fickian fluxes. 
Within reasonable limits, the \emph{modulated activities} are found to have comparable density gradients to passive MPCD.
Finally, the Pearson correlation coefficients between tracer particle velocities and local density gradients or active forces confirm that density gradients have minimal effect on the velocity of solutes, which are in fact driven by the active forcing.

Both human crowds~\cite{Helbing2011} and quorum sensing bacteria~\cite{Huang2011, Lenz2012} have been shown to exhibit slowdown effects in dense environments. 
These result in a local velocity-density profile akin to the modulation function $\SigF$ applied to the active forcing of \emph{modulated activities} in this work, suggesting that modulation of active velocity can mitigate density fluctuations in the general particle case.
Based on the results reported here, future research studying other active particle systems, such as dry nematics~\cite{Chate2006PRL, Henkes2018-NematicSphere} and active Brownian particles~\cite{Cates2015-MIPS}, should consider the effects of activity modulation.

Although simple in concept and implementation, the AN-MPCD algorithm has much potential for simulating ``hypercomplex'' active materials~\cite{Dogic2014} --- suspensions of mesoscale solutes embedded in an active background. 
Both composite materials and activity are ubiquitous in biology. 
Biopolymer composites, such as cytoskeletal components or cells embedded in extra-cellular polymeric substances exemplify mesoscale active/passive mixtures. 
While many studies consider active particles in passive environments, AN-MPCD opens a pathway for simulating passive particles dispersed in an active host.

\vspace{\baselineskip}

\paragraph{Acknowledgments}
This research has received funding from the European Research Council (ERC) under the European Union’s Horizon 2020 research and innovation programme (Grant agreement No. 851196). 
For the purpose of open access, the author has applied a Creative Commons Attribution (CC BY) licence to any Author Accepted Manuscript version arising from this submission.

    
    

\bibliography{biblio}

\begin{thebibliography}{70}%
\makeatletter
\providecommand \@ifxundefined [1]{%
 \@ifx{#1\undefined}
}%
\providecommand \@ifnum [1]{%
 \ifnum #1\expandafter \@firstoftwo
 \else \expandafter \@secondoftwo
 \fi
}%
\providecommand \@ifx [1]{%
 \ifx #1\expandafter \@firstoftwo
 \else \expandafter \@secondoftwo
 \fi
}%
\providecommand \natexlab [1]{#1}%
\providecommand \enquote  [1]{``#1''}%
\providecommand \bibnamefont  [1]{#1}%
\providecommand \bibfnamefont [1]{#1}%
\providecommand \citenamefont [1]{#1}%
\providecommand \href@noop [0]{\@secondoftwo}%
\providecommand \href [0]{\begingroup \@sanitize@url \@href}%
\providecommand \@href[1]{\@@startlink{#1}\@@href}%
\providecommand \@@href[1]{\endgroup#1\@@endlink}%
\providecommand \@sanitize@url [0]{\catcode `\\12\catcode `\$12\catcode
  `\&12\catcode `\#12\catcode `\^12\catcode `\_12\catcode `\%12\relax}%
\providecommand \@@startlink[1]{}%
\providecommand \@@endlink[0]{}%
\providecommand \url  [0]{\begingroup\@sanitize@url \@url }%
\providecommand \@url [1]{\endgroup\@href {#1}{\urlprefix }}%
\providecommand \urlprefix  [0]{URL }%
\providecommand \Eprint [0]{\href }%
\providecommand \doibase [0]{https://doi.org/}%
\providecommand \selectlanguage [0]{\@gobble}%
\providecommand \bibinfo  [0]{\@secondoftwo}%
\providecommand \bibfield  [0]{\@secondoftwo}%
\providecommand \translation [1]{[#1]}%
\providecommand \BibitemOpen [0]{}%
\providecommand \bibitemStop [0]{}%
\providecommand \bibitemNoStop [0]{.\EOS\space}%
\providecommand \EOS [0]{\spacefactor3000\relax}%
\providecommand \BibitemShut  [1]{\csname bibitem#1\endcsname}%
\let\auto@bib@innerbib\@empty
\bibitem [{\citenamefont {Needleman}\ and\ \citenamefont
  {Dogic}(2017)}]{needleman2017}%
  \BibitemOpen
  \bibfield  {author} {\bibinfo {author} {\bibfnamefont {D.}~\bibnamefont
  {Needleman}}\ and\ \bibinfo {author} {\bibfnamefont {Z.}~\bibnamefont
  {Dogic}},\ }\bibfield  {title} {\bibinfo {title} {Active matter at the
  interface between materials science and cell biology},\ }\href@noop {}
  {\bibfield  {journal} {\bibinfo  {journal} {Nature Reviews Materials}\
  }\textbf {\bibinfo {volume} {2}},\ \bibinfo {pages} {1} (\bibinfo {year}
  {2017})}\BibitemShut {NoStop}%
\bibitem [{\citenamefont {Yaman}\ \emph {et~al.}(2019)\citenamefont {Yaman},
  \citenamefont {Demir}, \citenamefont {Vetter},\ and\ \citenamefont
  {Kocabas}}]{yaman2019}%
  \BibitemOpen
  \bibfield  {author} {\bibinfo {author} {\bibfnamefont {Y.~I.}\ \bibnamefont
  {Yaman}}, \bibinfo {author} {\bibfnamefont {E.}~\bibnamefont {Demir}},
  \bibinfo {author} {\bibfnamefont {R.}~\bibnamefont {Vetter}},\ and\ \bibinfo
  {author} {\bibfnamefont {A.}~\bibnamefont {Kocabas}},\ }\bibfield  {title}
  {\bibinfo {title} {Emergence of active nematics in chaining bacterial
  biofilms},\ }\href@noop {} {\bibfield  {journal} {\bibinfo  {journal} {Nature
  Communications}\ }\textbf {\bibinfo {volume} {10}},\ \bibinfo {pages} {2285}
  (\bibinfo {year} {2019})}\BibitemShut {NoStop}%
\bibitem [{\citenamefont {Doostmohammadi}\ and\ \citenamefont
  {Ladoux}(2022)}]{doostmohammadi2022}%
  \BibitemOpen
  \bibfield  {author} {\bibinfo {author} {\bibfnamefont {A.}~\bibnamefont
  {Doostmohammadi}}\ and\ \bibinfo {author} {\bibfnamefont {B.}~\bibnamefont
  {Ladoux}},\ }\bibfield  {title} {\bibinfo {title} {Physics of liquid crystals
  in cell biology},\ }\href@noop {} {\bibfield  {journal} {\bibinfo  {journal}
  {Trends in Cell Biology}\ }\textbf {\bibinfo {volume} {32}},\ \bibinfo
  {pages} {140} (\bibinfo {year} {2022})}\BibitemShut {NoStop}%
\bibitem [{\citenamefont {Alert}\ \emph {et~al.}(2022)\citenamefont {Alert},
  \citenamefont {Casademunt},\ and\ \citenamefont
  {Joanny}}]{Alert2022AnnRevCondMat}%
  \BibitemOpen
  \bibfield  {author} {\bibinfo {author} {\bibfnamefont {R.}~\bibnamefont
  {Alert}}, \bibinfo {author} {\bibfnamefont {J.}~\bibnamefont {Casademunt}},\
  and\ \bibinfo {author} {\bibfnamefont {J.-F.}\ \bibnamefont {Joanny}},\
  }\bibfield  {title} {\bibinfo {title} {Active turbulence},\ }\href@noop {}
  {\bibfield  {journal} {\bibinfo  {journal} {Annual Review of Condensed Matter
  Physics}\ }\textbf {\bibinfo {volume} {13}},\ \bibinfo {pages} {null}
  (\bibinfo {year} {2022})}\BibitemShut {NoStop}%
\bibitem [{\citenamefont {Giomi}(2015)}]{Giomi2015PRX}%
  \BibitemOpen
  \bibfield  {author} {\bibinfo {author} {\bibfnamefont {L.}~\bibnamefont
  {Giomi}},\ }\bibfield  {title} {\bibinfo {title} {Geometry and topology of
  turbulence in active nematics},\ }\href@noop {} {\bibfield  {journal}
  {\bibinfo  {journal} {Physical Review X}\ }\textbf {\bibinfo {volume} {5}},\
  \bibinfo {pages} {031003} (\bibinfo {year} {2015})}\BibitemShut {NoStop}%
\bibitem [{\citenamefont {Shankar}\ and\ \citenamefont
  {Marchetti}(2019)}]{Shankar2019}%
  \BibitemOpen
  \bibfield  {author} {\bibinfo {author} {\bibfnamefont {S.}~\bibnamefont
  {Shankar}}\ and\ \bibinfo {author} {\bibfnamefont {M.~C.}\ \bibnamefont
  {Marchetti}},\ }\bibfield  {title} {\bibinfo {title} {Hydrodynamics of active
  defects: From order to chaos to defect ordering},\ }\href@noop {} {\bibfield
  {journal} {\bibinfo  {journal} {Phys. Rev. X}\ }\textbf {\bibinfo {volume}
  {9}},\ \bibinfo {pages} {041047} (\bibinfo {year} {2019})}\BibitemShut
  {NoStop}%
\bibitem [{\citenamefont {Duclos}\ \emph {et~al.}(2020)\citenamefont {Duclos},
  \citenamefont {Adkins}, \citenamefont {Banerjee}, \citenamefont {Peterson},
  \citenamefont {Varghese}, \citenamefont {Kolvin}, \citenamefont {Baskaran},
  \citenamefont {Pelcovits}, \citenamefont {Powers}, \citenamefont {Baskaran}
  \emph {et~al.}}]{Beller2020}%
  \BibitemOpen
  \bibfield  {author} {\bibinfo {author} {\bibfnamefont {G.}~\bibnamefont
  {Duclos}}, \bibinfo {author} {\bibfnamefont {R.}~\bibnamefont {Adkins}},
  \bibinfo {author} {\bibfnamefont {D.}~\bibnamefont {Banerjee}}, \bibinfo
  {author} {\bibfnamefont {M.~S.}\ \bibnamefont {Peterson}}, \bibinfo {author}
  {\bibfnamefont {M.}~\bibnamefont {Varghese}}, \bibinfo {author}
  {\bibfnamefont {I.}~\bibnamefont {Kolvin}}, \bibinfo {author} {\bibfnamefont
  {A.}~\bibnamefont {Baskaran}}, \bibinfo {author} {\bibfnamefont {R.~A.}\
  \bibnamefont {Pelcovits}}, \bibinfo {author} {\bibfnamefont {T.~R.}\
  \bibnamefont {Powers}}, \bibinfo {author} {\bibfnamefont {A.}~\bibnamefont
  {Baskaran}}, \emph {et~al.},\ }\bibfield  {title} {\bibinfo {title}
  {Topological structure and dynamics of three-dimensional active nematics},\
  }\href@noop {} {\bibfield  {journal} {\bibinfo  {journal} {Science}\ }\textbf
  {\bibinfo {volume} {367}},\ \bibinfo {pages} {1120} (\bibinfo {year}
  {2020})}\BibitemShut {NoStop}%
\bibitem [{\citenamefont {Hardo{\"u}in}\ \emph {et~al.}(2022)\citenamefont
  {Hardo{\"u}in}, \citenamefont {Dor{\'e}}, \citenamefont {Laurent},
  \citenamefont {Lopez-Leon}, \citenamefont {Ign{\'e}s-Mullol},\ and\
  \citenamefont {Sagu{\'e}s}}]{hardouin2022}%
  \BibitemOpen
  \bibfield  {author} {\bibinfo {author} {\bibfnamefont {J.}~\bibnamefont
  {Hardo{\"u}in}}, \bibinfo {author} {\bibfnamefont {C.}~\bibnamefont
  {Dor{\'e}}}, \bibinfo {author} {\bibfnamefont {J.}~\bibnamefont {Laurent}},
  \bibinfo {author} {\bibfnamefont {T.}~\bibnamefont {Lopez-Leon}}, \bibinfo
  {author} {\bibfnamefont {J.}~\bibnamefont {Ign{\'e}s-Mullol}},\ and\ \bibinfo
  {author} {\bibfnamefont {F.}~\bibnamefont {Sagu{\'e}s}},\ }\bibfield  {title}
  {\bibinfo {title} {Active boundary layers in confined active nematics},\
  }\href@noop {} {\bibfield  {journal} {\bibinfo  {journal} {Nature
  Communications}\ }\textbf {\bibinfo {volume} {13}},\ \bibinfo {pages} {6675}
  (\bibinfo {year} {2022})}\BibitemShut {NoStop}%
\bibitem [{\citenamefont {Li}\ \emph {et~al.}(2019)\citenamefont {Li},
  \citenamefont {Shi}, \citenamefont {Huang}, \citenamefont {Chen},
  \citenamefont {Xiao}, \citenamefont {Liu}, \citenamefont {Chat{\'e}},\ and\
  \citenamefont {Zhang}}]{li2019}%
  \BibitemOpen
  \bibfield  {author} {\bibinfo {author} {\bibfnamefont {H.}~\bibnamefont
  {Li}}, \bibinfo {author} {\bibfnamefont {X.-q.}\ \bibnamefont {Shi}},
  \bibinfo {author} {\bibfnamefont {M.}~\bibnamefont {Huang}}, \bibinfo
  {author} {\bibfnamefont {X.}~\bibnamefont {Chen}}, \bibinfo {author}
  {\bibfnamefont {M.}~\bibnamefont {Xiao}}, \bibinfo {author} {\bibfnamefont
  {C.}~\bibnamefont {Liu}}, \bibinfo {author} {\bibfnamefont {H.}~\bibnamefont
  {Chat{\'e}}},\ and\ \bibinfo {author} {\bibfnamefont {H.}~\bibnamefont
  {Zhang}},\ }\bibfield  {title} {\bibinfo {title} {Data-driven quantitative
  modeling of bacterial active nematics},\ }\href@noop {} {\bibfield  {journal}
  {\bibinfo  {journal} {Proceedings of the National Academy of Sciences}\
  }\textbf {\bibinfo {volume} {116}},\ \bibinfo {pages} {777} (\bibinfo {year}
  {2019})}\BibitemShut {NoStop}%
\bibitem [{\citenamefont {Liu}\ \emph {et~al.}(2021)\citenamefont {Liu},
  \citenamefont {Zeng}, \citenamefont {Ma},\ and\ \citenamefont
  {Cheng}}]{liu2021}%
  \BibitemOpen
  \bibfield  {author} {\bibinfo {author} {\bibfnamefont {Z.}~\bibnamefont
  {Liu}}, \bibinfo {author} {\bibfnamefont {W.}~\bibnamefont {Zeng}}, \bibinfo
  {author} {\bibfnamefont {X.}~\bibnamefont {Ma}},\ and\ \bibinfo {author}
  {\bibfnamefont {X.}~\bibnamefont {Cheng}},\ }\bibfield  {title} {\bibinfo
  {title} {Density fluctuations and energy spectra of {3D} bacterial
  suspensions},\ }\href@noop {} {\bibfield  {journal} {\bibinfo  {journal}
  {Soft Matter}\ }\textbf {\bibinfo {volume} {17}},\ \bibinfo {pages} {10806}
  (\bibinfo {year} {2021})}\BibitemShut {NoStop}%
\bibitem [{\citenamefont {Aranson}(2022)}]{Aranson2022}%
  \BibitemOpen
  \bibfield  {author} {\bibinfo {author} {\bibfnamefont {I.~S.}\ \bibnamefont
  {Aranson}},\ }\bibfield  {title} {\bibinfo {title} {Bacterial active
  matter},\ }\href@noop {} {\bibfield  {journal} {\bibinfo  {journal} {Reports
  on Progress in Physics}\ }\textbf {\bibinfo {volume} {85}},\ \bibinfo {pages}
  {076601} (\bibinfo {year} {2022})}\BibitemShut {NoStop}%
\bibitem [{\citenamefont {Balasubramaniam}\ \emph {et~al.}(2022)\citenamefont
  {Balasubramaniam}, \citenamefont {Mège},\ and\ \citenamefont
  {Ladoux}}]{Ladoux2022}%
  \BibitemOpen
  \bibfield  {author} {\bibinfo {author} {\bibfnamefont {L.}~\bibnamefont
  {Balasubramaniam}}, \bibinfo {author} {\bibfnamefont {R.-M.}\ \bibnamefont
  {Mège}},\ and\ \bibinfo {author} {\bibfnamefont {B.}~\bibnamefont
  {Ladoux}},\ }\bibfield  {title} {\bibinfo {title} {Active nematics across
  scales from cytoskeleton organization to tissue morphogenesis},\ }\href@noop
  {} {\bibfield  {journal} {\bibinfo  {journal} {Current Opinion in Genetics \&
  Development}\ }\textbf {\bibinfo {volume} {73}},\ \bibinfo {pages} {101897}
  (\bibinfo {year} {2022})}\BibitemShut {NoStop}%
\bibitem [{\citenamefont {Sanchez}\ \emph {et~al.}(2012)\citenamefont
  {Sanchez}, \citenamefont {Chen}, \citenamefont {DeCamp}, \citenamefont
  {Heymann},\ and\ \citenamefont
  {Dogic}}]{Dogic2012Nature-MicrotubuleActiveNematic}%
  \BibitemOpen
  \bibfield  {author} {\bibinfo {author} {\bibfnamefont {T.}~\bibnamefont
  {Sanchez}}, \bibinfo {author} {\bibfnamefont {D.~T.}\ \bibnamefont {Chen}},
  \bibinfo {author} {\bibfnamefont {S.~J.}\ \bibnamefont {DeCamp}}, \bibinfo
  {author} {\bibfnamefont {M.}~\bibnamefont {Heymann}},\ and\ \bibinfo {author}
  {\bibfnamefont {Z.}~\bibnamefont {Dogic}},\ }\bibfield  {title} {\bibinfo
  {title} {Spontaneous motion in hierarchically assembled active matter},\
  }\href@noop {} {\bibfield  {journal} {\bibinfo  {journal} {Nature}\ }\textbf
  {\bibinfo {volume} {491}},\ \bibinfo {pages} {431} (\bibinfo {year}
  {2012})}\BibitemShut {NoStop}%
\bibitem [{\citenamefont {Zhang}\ \emph {et~al.}(2021)\citenamefont {Zhang},
  \citenamefont {Redford}, \citenamefont {Ruijgrok}, \citenamefont {Kumar},
  \citenamefont {Mozaffari}, \citenamefont {Zemsky}, \citenamefont {Dinner},
  \citenamefont {Vitelli}, \citenamefont {Bryant}, \citenamefont {Gardel},\
  and\ \citenamefont {de~Pablo}}]{zhang2021}%
  \BibitemOpen
  \bibfield  {author} {\bibinfo {author} {\bibfnamefont {R.}~\bibnamefont
  {Zhang}}, \bibinfo {author} {\bibfnamefont {S.~A.}\ \bibnamefont {Redford}},
  \bibinfo {author} {\bibfnamefont {P.~V.}\ \bibnamefont {Ruijgrok}}, \bibinfo
  {author} {\bibfnamefont {N.}~\bibnamefont {Kumar}}, \bibinfo {author}
  {\bibfnamefont {A.}~\bibnamefont {Mozaffari}}, \bibinfo {author}
  {\bibfnamefont {S.}~\bibnamefont {Zemsky}}, \bibinfo {author} {\bibfnamefont
  {A.~R.}\ \bibnamefont {Dinner}}, \bibinfo {author} {\bibfnamefont
  {V.}~\bibnamefont {Vitelli}}, \bibinfo {author} {\bibfnamefont
  {Z.}~\bibnamefont {Bryant}}, \bibinfo {author} {\bibfnamefont {M.~L.}\
  \bibnamefont {Gardel}},\ and\ \bibinfo {author} {\bibfnamefont {J.~J.}\
  \bibnamefont {de~Pablo}},\ }\bibfield  {title} {\bibinfo {title}
  {Spatiotemporal control of liquid crystal structure and dynamics through
  activity patterning},\ }\href@noop {} {\bibfield  {journal} {\bibinfo
  {journal} {Nature Materials}\ }\textbf {\bibinfo {volume} {20}},\ \bibinfo
  {pages} {875} (\bibinfo {year} {2021})}\BibitemShut {NoStop}%
\bibitem [{\citenamefont {Doostmohammadi}\ \emph {et~al.}(2016)\citenamefont
  {Doostmohammadi}, \citenamefont {Adamer}, \citenamefont {Thampi},\ and\
  \citenamefont {Yeomans}}]{Yeomans2016NatComm}%
  \BibitemOpen
  \bibfield  {author} {\bibinfo {author} {\bibfnamefont {A.}~\bibnamefont
  {Doostmohammadi}}, \bibinfo {author} {\bibfnamefont {M.~F.}\ \bibnamefont
  {Adamer}}, \bibinfo {author} {\bibfnamefont {S.~P.}\ \bibnamefont {Thampi}},\
  and\ \bibinfo {author} {\bibfnamefont {J.~M.}\ \bibnamefont {Yeomans}},\
  }\bibfield  {title} {\bibinfo {title} {Stabilization of active matter by
  flow-vortex lattices and defect ordering},\ }\href@noop {} {\bibfield
  {journal} {\bibinfo  {journal} {Nature Communications}\ }\textbf {\bibinfo
  {volume} {7}},\ \bibinfo {pages} {1} (\bibinfo {year} {2016})}\BibitemShut
  {NoStop}%
\bibitem [{\citenamefont {Doostmohammadi}\ \emph {et~al.}(2018)\citenamefont
  {Doostmohammadi}, \citenamefont {Ign{\'e}s-Mullol}, \citenamefont {Yeomans},\
  and\ \citenamefont {Sagu{\'e}s}}]{Doostmohammadi2018-Review}%
  \BibitemOpen
  \bibfield  {author} {\bibinfo {author} {\bibfnamefont {A.}~\bibnamefont
  {Doostmohammadi}}, \bibinfo {author} {\bibfnamefont {J.}~\bibnamefont
  {Ign{\'e}s-Mullol}}, \bibinfo {author} {\bibfnamefont {J.~M.}\ \bibnamefont
  {Yeomans}},\ and\ \bibinfo {author} {\bibfnamefont {F.}~\bibnamefont
  {Sagu{\'e}s}},\ }\bibfield  {title} {\bibinfo {title} {Active nematics},\
  }\href@noop {} {\bibfield  {journal} {\bibinfo  {journal} {Nature
  Communications}\ }\textbf {\bibinfo {volume} {9}},\ \bibinfo {pages} {3246}
  (\bibinfo {year} {2018})}\BibitemShut {NoStop}%
\bibitem [{\citenamefont {Loewe}\ and\ \citenamefont
  {Shendruk}(2021)}]{Loewe2021NJP}%
  \BibitemOpen
  \bibfield  {author} {\bibinfo {author} {\bibfnamefont {B.}~\bibnamefont
  {Loewe}}\ and\ \bibinfo {author} {\bibfnamefont {T.~N.}\ \bibnamefont
  {Shendruk}},\ }\bibfield  {title} {\bibinfo {title} {Passive janus particles
  are self-propelled in active nematics},\ }\href@noop {} {\bibfield  {journal}
  {\bibinfo  {journal} {New Journal of Physics}\ } (\bibinfo {year}
  {2021})}\BibitemShut {NoStop}%
\bibitem [{\citenamefont {Thampi}(2022)}]{Thampi2022}%
  \BibitemOpen
  \bibfield  {author} {\bibinfo {author} {\bibfnamefont {S.~P.}\ \bibnamefont
  {Thampi}},\ }\bibfield  {title} {\bibinfo {title} {Channel confined active
  nematics},\ }\href@noop {} {\bibfield  {journal} {\bibinfo  {journal}
  {Current Opinion in Colloid and Interface Science}\ }\textbf {\bibinfo
  {volume} {61}},\ \bibinfo {pages} {101613} (\bibinfo {year}
  {2022})}\BibitemShut {NoStop}%
\bibitem [{\citenamefont {Ray}\ \emph {et~al.}(2023)\citenamefont {Ray},
  \citenamefont {Zhang},\ and\ \citenamefont {Dogic}}]{Sattvic2023}%
  \BibitemOpen
  \bibfield  {author} {\bibinfo {author} {\bibfnamefont {S.}~\bibnamefont
  {Ray}}, \bibinfo {author} {\bibfnamefont {J.}~\bibnamefont {Zhang}},\ and\
  \bibinfo {author} {\bibfnamefont {Z.}~\bibnamefont {Dogic}},\ }\bibfield
  {title} {\bibinfo {title} {Rectified rotational dynamics of mobile inclusions
  in two-dimensional active nematics},\ }\href@noop {} {\bibfield  {journal}
  {\bibinfo  {journal} {Physical Review Letters}\ }\textbf {\bibinfo {volume}
  {130}},\ \bibinfo {pages} {238301} (\bibinfo {year} {2023})}\BibitemShut
  {NoStop}%
\bibitem [{\citenamefont {Houston}\ and\ \citenamefont
  {Alexander}(2023)}]{Alexander2023}%
  \BibitemOpen
  \bibfield  {author} {\bibinfo {author} {\bibfnamefont {A.~J.~H.}\
  \bibnamefont {Houston}}\ and\ \bibinfo {author} {\bibfnamefont {G.~P.}\
  \bibnamefont {Alexander}},\ }\bibfield  {title} {\bibinfo {title} {Colloids
  in two-dimensional active nematics: conformal cogs and controllable
  spontaneous rotation},\ }\href@noop {} {\bibfield  {journal} {\bibinfo
  {journal} {New Journal of Physics}\ }\textbf {\bibinfo {volume} {25}},\
  \bibinfo {pages} {123006} (\bibinfo {year} {2023})}\BibitemShut {NoStop}%
\bibitem [{\citenamefont {Tang}\ and\ \citenamefont
  {Selinger}(2021)}]{Selinger2021PRE}%
  \BibitemOpen
  \bibfield  {author} {\bibinfo {author} {\bibfnamefont {X.}~\bibnamefont
  {Tang}}\ and\ \bibinfo {author} {\bibfnamefont {J.~V.}\ \bibnamefont
  {Selinger}},\ }\bibfield  {title} {\bibinfo {title} {Alignment of a
  topological defect by an activity gradient},\ }\href@noop {} {\bibfield
  {journal} {\bibinfo  {journal} {Physical Review E}\ }\textbf {\bibinfo
  {volume} {103}},\ \bibinfo {pages} {022703} (\bibinfo {year}
  {2021})}\BibitemShut {NoStop}%
\bibitem [{\citenamefont {Zarei}\ \emph {et~al.}(2023)\citenamefont {Zarei},
  \citenamefont {Berezney}, \citenamefont {Hensley}, \citenamefont {Lemma},
  \citenamefont {Senbil}, \citenamefont {Dogic},\ and\ \citenamefont
  {Fraden}}]{Fraden2023-LightAN}%
  \BibitemOpen
  \bibfield  {author} {\bibinfo {author} {\bibfnamefont {Z.}~\bibnamefont
  {Zarei}}, \bibinfo {author} {\bibfnamefont {J.}~\bibnamefont {Berezney}},
  \bibinfo {author} {\bibfnamefont {A.}~\bibnamefont {Hensley}}, \bibinfo
  {author} {\bibfnamefont {L.}~\bibnamefont {Lemma}}, \bibinfo {author}
  {\bibfnamefont {N.}~\bibnamefont {Senbil}}, \bibinfo {author} {\bibfnamefont
  {Z.}~\bibnamefont {Dogic}},\ and\ \bibinfo {author} {\bibfnamefont
  {S.}~\bibnamefont {Fraden}},\ }\bibfield  {title} {\bibinfo {title}
  {Light-activated microtubule-based two-dimensional active nematic},\
  }\href@noop {} {\bibfield  {journal} {\bibinfo  {journal} {Soft Matter}\
  }\textbf {\bibinfo {volume} {19}},\ \bibinfo {pages} {6691} (\bibinfo {year}
  {2023})}\BibitemShut {NoStop}%
\bibitem [{\citenamefont {Kozhukhov}\ and\ \citenamefont
  {Shendruk}(2022)}]{Kozhukhov2022-ANMPCD}%
  \BibitemOpen
  \bibfield  {author} {\bibinfo {author} {\bibfnamefont {T.}~\bibnamefont
  {Kozhukhov}}\ and\ \bibinfo {author} {\bibfnamefont {T.~N.}\ \bibnamefont
  {Shendruk}},\ }\bibfield  {title} {\bibinfo {title} {Mesoscopic simulations
  of active nematics},\ }\href@noop {} {\bibfield  {journal} {\bibinfo
  {journal} {Science Advances}\ }\textbf {\bibinfo {volume} {8}},\ \bibinfo
  {pages} {eabo5788} (\bibinfo {year} {2022})}\BibitemShut {NoStop}%
\bibitem [{\citenamefont {Shendruk}\ and\ \citenamefont
  {Yeomans}(2015)}]{Shendruk2015SoftMatter-NMPCD}%
  \BibitemOpen
  \bibfield  {author} {\bibinfo {author} {\bibfnamefont {T.~N.}\ \bibnamefont
  {Shendruk}}\ and\ \bibinfo {author} {\bibfnamefont {J.~M.}\ \bibnamefont
  {Yeomans}},\ }\bibfield  {title} {\bibinfo {title} {Multi-particle collision
  dynamics algorithm for nematic fluids},\ }\href@noop {} {\bibfield  {journal}
  {\bibinfo  {journal} {Soft Matter}\ }\textbf {\bibinfo {volume} {11}},\
  \bibinfo {pages} {5101} (\bibinfo {year} {2015})}\BibitemShut {NoStop}%
\bibitem [{\citenamefont {Macías-Durán}\ \emph {et~al.}(2023)\citenamefont
  {Macías-Durán}, \citenamefont {Duarte-Alaniz},\ and\ \citenamefont
  {Híjar}}]{Hijar2023-ANMPCD}%
  \BibitemOpen
  \bibfield  {author} {\bibinfo {author} {\bibfnamefont {J.}~\bibnamefont
  {Macías-Durán}}, \bibinfo {author} {\bibfnamefont {V.}~\bibnamefont
  {Duarte-Alaniz}},\ and\ \bibinfo {author} {\bibfnamefont {H.}~\bibnamefont
  {Híjar}},\ }\bibfield  {title} {\bibinfo {title} {Active nematic liquid
  crystals simulated by particle-based mesoscopic methods},\ }\href@noop {}
  {\bibfield  {journal} {\bibinfo  {journal} {Soft Matter}\ }\textbf {\bibinfo
  {volume} {19}},\ \bibinfo {pages} {8052} (\bibinfo {year}
  {2023})}\BibitemShut {NoStop}%
\bibitem [{\citenamefont {Marenduzzo}\ \emph {et~al.}(2007)\citenamefont
  {Marenduzzo}, \citenamefont {Orlandini}, \citenamefont {Cates},\ and\
  \citenamefont {Yeomans}}]{Marenduzzo2007PRE-ANLB}%
  \BibitemOpen
  \bibfield  {author} {\bibinfo {author} {\bibfnamefont {D.}~\bibnamefont
  {Marenduzzo}}, \bibinfo {author} {\bibfnamefont {E.}~\bibnamefont
  {Orlandini}}, \bibinfo {author} {\bibfnamefont {M.}~\bibnamefont {Cates}},\
  and\ \bibinfo {author} {\bibfnamefont {J.}~\bibnamefont {Yeomans}},\
  }\bibfield  {title} {\bibinfo {title} {Steady-state hydrodynamic
  instabilities of active liquid crystals: Hybrid lattice boltzmann
  simulations},\ }\href@noop {} {\bibfield  {journal} {\bibinfo  {journal}
  {Physical Review E}\ }\textbf {\bibinfo {volume} {76}},\ \bibinfo {pages}
  {031921} (\bibinfo {year} {2007})}\BibitemShut {NoStop}%
\bibitem [{\citenamefont {Keogh}\ \emph {et~al.}(2023)\citenamefont {Keogh},
  \citenamefont {Kozhukhov}, \citenamefont {Thijssen},\ and\ \citenamefont
  {Shendruk}}]{Keogh2023-Darcy}%
  \BibitemOpen
  \bibfield  {author} {\bibinfo {author} {\bibfnamefont {R.~R.}\ \bibnamefont
  {Keogh}}, \bibinfo {author} {\bibfnamefont {T.}~\bibnamefont {Kozhukhov}},
  \bibinfo {author} {\bibfnamefont {K.}~\bibnamefont {Thijssen}},\ and\
  \bibinfo {author} {\bibfnamefont {T.~N.}\ \bibnamefont {Shendruk}},\
  }\bibfield  {title} {\bibinfo {title} {Active darcy's law},\ }\href@noop {}
  {\bibfield  {journal} {\bibinfo  {journal} {arXiv preprint arXiv:2308.05462}\
  } (\bibinfo {year} {2023})}\BibitemShut {NoStop}%
\bibitem [{\citenamefont {Chat{\'e}}\ \emph {et~al.}(2006)\citenamefont
  {Chat{\'e}}, \citenamefont {Ginelli},\ and\ \citenamefont
  {Montagne}}]{Chate2006PRL}%
  \BibitemOpen
  \bibfield  {author} {\bibinfo {author} {\bibfnamefont {H.}~\bibnamefont
  {Chat{\'e}}}, \bibinfo {author} {\bibfnamefont {F.}~\bibnamefont {Ginelli}},\
  and\ \bibinfo {author} {\bibfnamefont {R.}~\bibnamefont {Montagne}},\
  }\bibfield  {title} {\bibinfo {title} {Simple model for active nematics:
  Quasi-long-range order and giant fluctuations},\ }\href@noop {} {\bibfield
  {journal} {\bibinfo  {journal} {Physical Review Letters}\ }\textbf {\bibinfo
  {volume} {96}},\ \bibinfo {pages} {180602} (\bibinfo {year}
  {2006})}\BibitemShut {NoStop}%
\bibitem [{\citenamefont {Ngo}\ \emph {et~al.}(2014)\citenamefont {Ngo},
  \citenamefont {Peshkov}, \citenamefont {Aranson}, \citenamefont {Bertin},
  \citenamefont {Ginelli},\ and\ \citenamefont
  {Chat{\'e}}}]{PeshkovChate2014PRL}%
  \BibitemOpen
  \bibfield  {author} {\bibinfo {author} {\bibfnamefont {S.}~\bibnamefont
  {Ngo}}, \bibinfo {author} {\bibfnamefont {A.}~\bibnamefont {Peshkov}},
  \bibinfo {author} {\bibfnamefont {I.~S.}\ \bibnamefont {Aranson}}, \bibinfo
  {author} {\bibfnamefont {E.}~\bibnamefont {Bertin}}, \bibinfo {author}
  {\bibfnamefont {F.}~\bibnamefont {Ginelli}},\ and\ \bibinfo {author}
  {\bibfnamefont {H.}~\bibnamefont {Chat{\'e}}},\ }\bibfield  {title} {\bibinfo
  {title} {Large-scale chaos and fluctuations in active nematics},\ }\href@noop
  {} {\bibfield  {journal} {\bibinfo  {journal} {Physical Review Letters}\
  }\textbf {\bibinfo {volume} {113}},\ \bibinfo {pages} {038302} (\bibinfo
  {year} {2014})}\BibitemShut {NoStop}%
\bibitem [{\citenamefont {Bertin}\ \emph {et~al.}(2009)\citenamefont {Bertin},
  \citenamefont {Droz},\ and\ \citenamefont {Gr{\'e}goire}}]{Bertin2009JPhysA}%
  \BibitemOpen
  \bibfield  {author} {\bibinfo {author} {\bibfnamefont {E.}~\bibnamefont
  {Bertin}}, \bibinfo {author} {\bibfnamefont {M.}~\bibnamefont {Droz}},\ and\
  \bibinfo {author} {\bibfnamefont {G.}~\bibnamefont {Gr{\'e}goire}},\
  }\bibfield  {title} {\bibinfo {title} {Hydrodynamic equations for
  self-propelled particles: microscopic derivation and stability analysis},\
  }\href@noop {} {\bibfield  {journal} {\bibinfo  {journal} {Journal of Physics
  A: Mathematical and Theoretical}\ }\textbf {\bibinfo {volume} {42}},\
  \bibinfo {pages} {445001} (\bibinfo {year} {2009})}\BibitemShut {NoStop}%
\bibitem [{\citenamefont {B{\"a}r}\ \emph {et~al.}(2020)\citenamefont
  {B{\"a}r}, \citenamefont {Gro{\ss}mann}, \citenamefont {Heidenreich},\ and\
  \citenamefont {Peruani}}]{Bar2020AnnRev}%
  \BibitemOpen
  \bibfield  {author} {\bibinfo {author} {\bibfnamefont {M.}~\bibnamefont
  {B{\"a}r}}, \bibinfo {author} {\bibfnamefont {R.}~\bibnamefont
  {Gro{\ss}mann}}, \bibinfo {author} {\bibfnamefont {S.}~\bibnamefont
  {Heidenreich}},\ and\ \bibinfo {author} {\bibfnamefont {F.}~\bibnamefont
  {Peruani}},\ }\bibfield  {title} {\bibinfo {title} {Self-propelled rods:
  Insights and perspectives for active matter},\ }\href@noop {} {\bibfield
  {journal} {\bibinfo  {journal} {Annual Review of Condensed Matter Physics}\
  }\textbf {\bibinfo {volume} {11}},\ \bibinfo {pages} {441} (\bibinfo {year}
  {2020})}\BibitemShut {NoStop}%
\bibitem [{\citenamefont {Barriuso~Gutiérrez}\ \emph
  {et~al.}(2022)\citenamefont {Barriuso~Gutiérrez}, \citenamefont
  {Martín-Roca}, \citenamefont {Bianco}, \citenamefont {Pagonabarraga},\ and\
  \citenamefont {Valeriani}}]{Valeriani2022}%
  \BibitemOpen
  \bibfield  {author} {\bibinfo {author} {\bibfnamefont {C.~M.}\ \bibnamefont
  {Barriuso~Gutiérrez}}, \bibinfo {author} {\bibfnamefont {J.}~\bibnamefont
  {Martín-Roca}}, \bibinfo {author} {\bibfnamefont {V.}~\bibnamefont
  {Bianco}}, \bibinfo {author} {\bibfnamefont {I.}~\bibnamefont
  {Pagonabarraga}},\ and\ \bibinfo {author} {\bibfnamefont {C.}~\bibnamefont
  {Valeriani}},\ }\bibfield  {title} {\bibinfo {title} {Simulating
  microswimmers under confinement with dissipative particle (hydro) dynamics},\
  }\href@noop {} {\bibfield  {journal} {\bibinfo  {journal} {Frontiers in
  Physics}\ }\textbf {\bibinfo {volume} {10}} (\bibinfo {year}
  {2022})}\BibitemShut {NoStop}%
\bibitem [{\citenamefont {Zantop}\ and\ \citenamefont
  {Stark}(2021)}]{Zantop2021JCP-MPCDState}%
  \BibitemOpen
  \bibfield  {author} {\bibinfo {author} {\bibfnamefont {A.~W.}\ \bibnamefont
  {Zantop}}\ and\ \bibinfo {author} {\bibfnamefont {H.}~\bibnamefont {Stark}},\
  }\bibfield  {title} {\bibinfo {title} {Multi-particle collision dynamics with
  a non-ideal equation of state. i},\ }\href@noop {} {\bibfield  {journal}
  {\bibinfo  {journal} {The Journal of Chemical Physics}\ }\textbf {\bibinfo
  {volume} {154}},\ \bibinfo {pages} {024105} (\bibinfo {year}
  {2021})}\BibitemShut {NoStop}%
\bibitem [{\citenamefont {Cates}\ and\ \citenamefont
  {Tailleur}(2015)}]{Cates2015-MIPS}%
  \BibitemOpen
  \bibfield  {author} {\bibinfo {author} {\bibfnamefont {M.~E.}\ \bibnamefont
  {Cates}}\ and\ \bibinfo {author} {\bibfnamefont {J.}~\bibnamefont
  {Tailleur}},\ }\bibfield  {title} {\bibinfo {title} {Motility-induced phase
  separation},\ }\href@noop {} {\bibfield  {journal} {\bibinfo  {journal}
  {Annual Review of Condensed Matter Physics}\ }\textbf {\bibinfo {volume}
  {6}},\ \bibinfo {pages} {219} (\bibinfo {year} {2015})}\BibitemShut {NoStop}%
\bibitem [{\citenamefont {Bate}\ \emph {et~al.}(2022)\citenamefont {Bate},
  \citenamefont {Varney}, \citenamefont {Taylor}, \citenamefont {Dickie},
  \citenamefont {Chueh}, \citenamefont {Norton},\ and\ \citenamefont
  {Wu}}]{bate2022}%
  \BibitemOpen
  \bibfield  {author} {\bibinfo {author} {\bibfnamefont {T.~E.}\ \bibnamefont
  {Bate}}, \bibinfo {author} {\bibfnamefont {M.~E.}\ \bibnamefont {Varney}},
  \bibinfo {author} {\bibfnamefont {E.~H.}\ \bibnamefont {Taylor}}, \bibinfo
  {author} {\bibfnamefont {J.~H.}\ \bibnamefont {Dickie}}, \bibinfo {author}
  {\bibfnamefont {C.-C.}\ \bibnamefont {Chueh}}, \bibinfo {author}
  {\bibfnamefont {M.~M.}\ \bibnamefont {Norton}},\ and\ \bibinfo {author}
  {\bibfnamefont {K.-T.}\ \bibnamefont {Wu}},\ }\bibfield  {title} {\bibinfo
  {title} {Self-mixing in microtubule-kinesin active fluid from nonuniform to
  uniform distribution of activity},\ }\href@noop {} {\bibfield  {journal}
  {\bibinfo  {journal} {Nature Communications}\ }\textbf {\bibinfo {volume}
  {13}},\ \bibinfo {pages} {6573} (\bibinfo {year} {2022})}\BibitemShut
  {NoStop}%
\bibitem [{\citenamefont {Ruske}\ and\ \citenamefont
  {Yeomans}(2022)}]{ruske2022}%
  \BibitemOpen
  \bibfield  {author} {\bibinfo {author} {\bibfnamefont {L.~J.}\ \bibnamefont
  {Ruske}}\ and\ \bibinfo {author} {\bibfnamefont {J.~M.}\ \bibnamefont
  {Yeomans}},\ }\bibfield  {title} {\bibinfo {title} {Activity gradients in
  two-and three-dimensional active nematics},\ }\href@noop {} {\bibfield
  {journal} {\bibinfo  {journal} {Soft Matter}\ }\textbf {\bibinfo {volume}
  {18}},\ \bibinfo {pages} {5654} (\bibinfo {year} {2022})}\BibitemShut
  {NoStop}%
\bibitem [{\citenamefont {Coelho}\ \emph {et~al.}(2022)\citenamefont {Coelho},
  \citenamefont {Ara{\'u}jo},\ and\ \citenamefont {da~Gama}}]{coelho2022}%
  \BibitemOpen
  \bibfield  {author} {\bibinfo {author} {\bibfnamefont {R.~C.}\ \bibnamefont
  {Coelho}}, \bibinfo {author} {\bibfnamefont {N.~A.}\ \bibnamefont
  {Ara{\'u}jo}},\ and\ \bibinfo {author} {\bibfnamefont {M.~M.~T.}\
  \bibnamefont {da~Gama}},\ }\bibfield  {title} {\bibinfo {title} {Dispersion
  of activity at an active--passive nematic interface},\ }\href@noop {}
  {\bibfield  {journal} {\bibinfo  {journal} {Soft Matter}\ }\textbf {\bibinfo
  {volume} {18}},\ \bibinfo {pages} {7642} (\bibinfo {year}
  {2022})}\BibitemShut {NoStop}%
\bibitem [{\citenamefont {Gompper}\ \emph {et~al.}(2009)\citenamefont
  {Gompper}, \citenamefont {Ihle}, \citenamefont {Kroll},\ and\ \citenamefont
  {Winkler}}]{GompperIhle2009Book-MPCD}%
  \BibitemOpen
  \bibfield  {author} {\bibinfo {author} {\bibfnamefont {G.}~\bibnamefont
  {Gompper}}, \bibinfo {author} {\bibfnamefont {T.}~\bibnamefont {Ihle}},
  \bibinfo {author} {\bibfnamefont {D.}~\bibnamefont {Kroll}},\ and\ \bibinfo
  {author} {\bibfnamefont {R.}~\bibnamefont {Winkler}},\ }\bibfield  {title}
  {\bibinfo {title} {Multi-particle collision dynamics: A particle-based
  mesoscale simulation approach to the hydrodynamics of complex fluids},\
  }\href@noop {} {\bibfield  {journal} {\bibinfo  {journal} {Advanced computer
  simulation approaches for soft matter sciences III}\ ,\ \bibinfo {pages} {1}}
  (\bibinfo {year} {2009})}\BibitemShut {NoStop}%
\bibitem [{\citenamefont {Noguchi}\ \emph {et~al.}(2007)\citenamefont
  {Noguchi}, \citenamefont {Kikuchi},\ and\ \citenamefont
  {Gompper}}]{Gompper2007EPL}%
  \BibitemOpen
  \bibfield  {author} {\bibinfo {author} {\bibfnamefont {H.}~\bibnamefont
  {Noguchi}}, \bibinfo {author} {\bibfnamefont {N.}~\bibnamefont {Kikuchi}},\
  and\ \bibinfo {author} {\bibfnamefont {G.}~\bibnamefont {Gompper}},\
  }\bibfield  {title} {\bibinfo {title} {Particle-based mesoscale hydrodynamic
  techniques},\ }\href@noop {} {\bibfield  {journal} {\bibinfo  {journal} {EPL
  (Europhysics Letters)}\ }\textbf {\bibinfo {volume} {78}},\ \bibinfo {pages}
  {10005} (\bibinfo {year} {2007})}\BibitemShut {NoStop}%
\bibitem [{\citenamefont {Tao}\ \emph {et~al.}(2008)\citenamefont {Tao},
  \citenamefont {Götze},\ and\ \citenamefont {Gompper}}]{tao2008}%
  \BibitemOpen
  \bibfield  {author} {\bibinfo {author} {\bibfnamefont {Y.-G.}\ \bibnamefont
  {Tao}}, \bibinfo {author} {\bibfnamefont {I.~O.}\ \bibnamefont {Götze}},\
  and\ \bibinfo {author} {\bibfnamefont {G.}~\bibnamefont {Gompper}},\
  }\bibfield  {title} {\bibinfo {title} {{Multiparticle collision dynamics
  modeling of viscoelastic fluids}},\ }\href@noop {} {\bibfield  {journal}
  {\bibinfo  {journal} {The Journal of Chemical Physics}\ }\textbf {\bibinfo
  {volume} {128}},\ \bibinfo {pages} {144902} (\bibinfo {year}
  {2008})}\BibitemShut {NoStop}%
\bibitem [{\citenamefont {Kowalik}\ and\ \citenamefont
  {Winkler}(2013)}]{Stark2013-ViscoMPCD}%
  \BibitemOpen
  \bibfield  {author} {\bibinfo {author} {\bibfnamefont {B.}~\bibnamefont
  {Kowalik}}\ and\ \bibinfo {author} {\bibfnamefont {R.~G.}\ \bibnamefont
  {Winkler}},\ }\bibfield  {title} {\bibinfo {title} {{Multiparticle collision
  dynamics simulations of viscoelastic fluids: Shear-thinning Gaussian
  dumbbells}},\ }\href@noop {} {\bibfield  {journal} {\bibinfo  {journal} {The
  Journal of Chemical Physics}\ }\textbf {\bibinfo {volume} {138}},\ \bibinfo
  {pages} {104903} (\bibinfo {year} {2013})}\BibitemShut {NoStop}%
\bibitem [{\citenamefont {Toneian}\ \emph {et~al.}(2019)\citenamefont
  {Toneian}, \citenamefont {Kahl}, \citenamefont {Gompper},\ and\ \citenamefont
  {Winkler}}]{Toneian2019}%
  \BibitemOpen
  \bibfield  {author} {\bibinfo {author} {\bibfnamefont {D.}~\bibnamefont
  {Toneian}}, \bibinfo {author} {\bibfnamefont {G.}~\bibnamefont {Kahl}},
  \bibinfo {author} {\bibfnamefont {G.}~\bibnamefont {Gompper}},\ and\ \bibinfo
  {author} {\bibfnamefont {R.~G.}\ \bibnamefont {Winkler}},\ }\bibfield
  {title} {\bibinfo {title} {Hydrodynamic correlations of viscoelastic fluids
  by multiparticle collision dynamics simulations},\ }\href@noop {} {\bibfield
  {journal} {\bibinfo  {journal} {The Journal of Chemical Physics}\ }\textbf
  {\bibinfo {volume} {151}},\ \bibinfo {pages} {194110} (\bibinfo {year}
  {2019})}\BibitemShut {NoStop}%
\bibitem [{\citenamefont {Simha}\ and\ \citenamefont
  {Ramaswamy}(2002)}]{SimhaRamaswamy2002PhysicaA}%
  \BibitemOpen
  \bibfield  {author} {\bibinfo {author} {\bibfnamefont {R.~A.}\ \bibnamefont
  {Simha}}\ and\ \bibinfo {author} {\bibfnamefont {S.}~\bibnamefont
  {Ramaswamy}},\ }\bibfield  {title} {\bibinfo {title} {Statistical
  hydrodynamics of ordered suspensions of self-propelled particles: waves,
  giant number fluctuations and instabilities},\ }\href@noop {} {\bibfield
  {journal} {\bibinfo  {journal} {Physica A: Statistical Mechanics and its
  Applications}\ }\textbf {\bibinfo {volume} {306}},\ \bibinfo {pages} {262}
  (\bibinfo {year} {2002})}\BibitemShut {NoStop}%
\bibitem [{\citenamefont {Thampi}\ \emph {et~al.}(2014)\citenamefont {Thampi},
  \citenamefont {Golestanian},\ and\ \citenamefont {Yeomans}}]{Thampi2014EPL}%
  \BibitemOpen
  \bibfield  {author} {\bibinfo {author} {\bibfnamefont {S.~P.}\ \bibnamefont
  {Thampi}}, \bibinfo {author} {\bibfnamefont {R.}~\bibnamefont
  {Golestanian}},\ and\ \bibinfo {author} {\bibfnamefont {J.~M.}\ \bibnamefont
  {Yeomans}},\ }\bibfield  {title} {\bibinfo {title} {Instabilities and
  topological defects in active nematics},\ }\href@noop {} {\bibfield
  {journal} {\bibinfo  {journal} {EPL (Europhysics Letters)}\ }\textbf
  {\bibinfo {volume} {105}},\ \bibinfo {pages} {18001} (\bibinfo {year}
  {2014})}\BibitemShut {NoStop}%
\bibitem [{\citenamefont {Marchetti}\ \emph {et~al.}(2013)\citenamefont
  {Marchetti}, \citenamefont {Joanny}, \citenamefont {Ramaswamy}, \citenamefont
  {Liverpool}, \citenamefont {Prost}, \citenamefont {Rao},\ and\ \citenamefont
  {Simha}}]{Marchetti2013RevModPhys-Hydrodynamics}%
  \BibitemOpen
  \bibfield  {author} {\bibinfo {author} {\bibfnamefont {M.~C.}\ \bibnamefont
  {Marchetti}}, \bibinfo {author} {\bibfnamefont {J.-F.}\ \bibnamefont
  {Joanny}}, \bibinfo {author} {\bibfnamefont {S.}~\bibnamefont {Ramaswamy}},
  \bibinfo {author} {\bibfnamefont {T.~B.}\ \bibnamefont {Liverpool}}, \bibinfo
  {author} {\bibfnamefont {J.}~\bibnamefont {Prost}}, \bibinfo {author}
  {\bibfnamefont {M.}~\bibnamefont {Rao}},\ and\ \bibinfo {author}
  {\bibfnamefont {R.~A.}\ \bibnamefont {Simha}},\ }\bibfield  {title} {\bibinfo
  {title} {Hydrodynamics of soft active matter},\ }\href@noop {} {\bibfield
  {journal} {\bibinfo  {journal} {Reviews of Modern Physics}\ }\textbf
  {\bibinfo {volume} {85}},\ \bibinfo {pages} {1143} (\bibinfo {year}
  {2013})}\BibitemShut {NoStop}%
\bibitem [{\citenamefont {Hemingway}\ \emph {et~al.}(2016)\citenamefont
  {Hemingway}, \citenamefont {Mishra}, \citenamefont {Marchetti},\ and\
  \citenamefont {Fielding}}]{Hemingway2016SoftMatter}%
  \BibitemOpen
  \bibfield  {author} {\bibinfo {author} {\bibfnamefont {E.~J.}\ \bibnamefont
  {Hemingway}}, \bibinfo {author} {\bibfnamefont {P.}~\bibnamefont {Mishra}},
  \bibinfo {author} {\bibfnamefont {M.~C.}\ \bibnamefont {Marchetti}},\ and\
  \bibinfo {author} {\bibfnamefont {S.~M.}\ \bibnamefont {Fielding}},\
  }\bibfield  {title} {\bibinfo {title} {Correlation lengths in hydrodynamic
  models of active nematics},\ }\href@noop {} {\bibfield  {journal} {\bibinfo
  {journal} {Soft Matter}\ }\textbf {\bibinfo {volume} {12}},\ \bibinfo {pages}
  {7943} (\bibinfo {year} {2016})}\BibitemShut {NoStop}%
\bibitem [{\citenamefont {Keogh}\ \emph {et~al.}(2022)\citenamefont {Keogh},
  \citenamefont {Chandragiri}, \citenamefont {Loewe}, \citenamefont
  {Ala-Nissila}, \citenamefont {Thampi},\ and\ \citenamefont
  {Shendruk}}]{Keogh2022}%
  \BibitemOpen
  \bibfield  {author} {\bibinfo {author} {\bibfnamefont {R.~R.}\ \bibnamefont
  {Keogh}}, \bibinfo {author} {\bibfnamefont {S.}~\bibnamefont {Chandragiri}},
  \bibinfo {author} {\bibfnamefont {B.}~\bibnamefont {Loewe}}, \bibinfo
  {author} {\bibfnamefont {T.}~\bibnamefont {Ala-Nissila}}, \bibinfo {author}
  {\bibfnamefont {S.~P.}\ \bibnamefont {Thampi}},\ and\ \bibinfo {author}
  {\bibfnamefont {T.~N.}\ \bibnamefont {Shendruk}},\ }\bibfield  {title}
  {\bibinfo {title} {Helical flow states in active nematics},\ }\href@noop {}
  {\bibfield  {journal} {\bibinfo  {journal} {Physical Review E}\ }\textbf
  {\bibinfo {volume} {106}},\ \bibinfo {pages} {L012602} (\bibinfo {year}
  {2022})}\BibitemShut {NoStop}%
\bibitem [{\citenamefont {Head}\ \emph {et~al.}(2024)\citenamefont {Head},
  \citenamefont {Dor{\'e}}, \citenamefont {Keogh}, \citenamefont {Bonn},
  \citenamefont {Negro}, \citenamefont {Marenduzzo}, \citenamefont
  {Doostmohammadi}, \citenamefont {Thijssen}, \citenamefont
  {L{\'o}pez-Le{\'o}n},\ and\ \citenamefont {Shendruk}}]{Head2024-DTensor}%
  \BibitemOpen
  \bibfield  {author} {\bibinfo {author} {\bibfnamefont {L.~C.}\ \bibnamefont
  {Head}}, \bibinfo {author} {\bibfnamefont {C.}~\bibnamefont {Dor{\'e}}},
  \bibinfo {author} {\bibfnamefont {R.~R.}\ \bibnamefont {Keogh}}, \bibinfo
  {author} {\bibfnamefont {L.}~\bibnamefont {Bonn}}, \bibinfo {author}
  {\bibfnamefont {G.}~\bibnamefont {Negro}}, \bibinfo {author} {\bibfnamefont
  {D.}~\bibnamefont {Marenduzzo}}, \bibinfo {author} {\bibfnamefont
  {A.}~\bibnamefont {Doostmohammadi}}, \bibinfo {author} {\bibfnamefont
  {K.}~\bibnamefont {Thijssen}}, \bibinfo {author} {\bibfnamefont
  {T.}~\bibnamefont {L{\'o}pez-Le{\'o}n}},\ and\ \bibinfo {author}
  {\bibfnamefont {T.~N.}\ \bibnamefont {Shendruk}},\ }\bibfield  {title}
  {\bibinfo {title} {Spontaneous self-constraint in active nematic flows},\
  }\href@noop {} {\bibfield  {journal} {\bibinfo  {journal} {Nature Physics}\ }
  (\bibinfo {year} {2024})}\BibitemShut {NoStop}%
\bibitem [{\citenamefont {Blaschke}\ \emph {et~al.}(2016)\citenamefont
  {Blaschke}, \citenamefont {Maurer}, \citenamefont {Menon}, \citenamefont
  {Z{\"o}ttl},\ and\ \citenamefont {Stark}}]{Zottl2016}%
  \BibitemOpen
  \bibfield  {author} {\bibinfo {author} {\bibfnamefont {J.}~\bibnamefont
  {Blaschke}}, \bibinfo {author} {\bibfnamefont {M.}~\bibnamefont {Maurer}},
  \bibinfo {author} {\bibfnamefont {K.}~\bibnamefont {Menon}}, \bibinfo
  {author} {\bibfnamefont {A.}~\bibnamefont {Z{\"o}ttl}},\ and\ \bibinfo
  {author} {\bibfnamefont {H.}~\bibnamefont {Stark}},\ }\bibfield  {title}
  {\bibinfo {title} {Phase separation and coexistence of hydrodynamically
  interacting microswimmers},\ }\href@noop {} {\bibfield  {journal} {\bibinfo
  {journal} {Soft Matter}\ }\textbf {\bibinfo {volume} {12}},\ \bibinfo {pages}
  {9821} (\bibinfo {year} {2016})}\BibitemShut {NoStop}%
\bibitem [{\citenamefont {Wani}\ \emph {et~al.}(2022)\citenamefont {Wani},
  \citenamefont {Kovakas}, \citenamefont {Nikoubashman},\ and\ \citenamefont
  {Howard}}]{Wani2022}%
  \BibitemOpen
  \bibfield  {author} {\bibinfo {author} {\bibfnamefont {Y.~M.}\ \bibnamefont
  {Wani}}, \bibinfo {author} {\bibfnamefont {P.~G.}\ \bibnamefont {Kovakas}},
  \bibinfo {author} {\bibfnamefont {A.}~\bibnamefont {Nikoubashman}},\ and\
  \bibinfo {author} {\bibfnamefont {M.~P.}\ \bibnamefont {Howard}},\ }\bibfield
   {title} {\bibinfo {title} {{Diffusion and sedimentation in colloidal
  suspensions using multiparticle collision dynamics with a discrete particle
  model}},\ }\href@noop {} {\bibfield  {journal} {\bibinfo  {journal} {The
  Journal of Chemical Physics}\ }\textbf {\bibinfo {volume} {156}},\ \bibinfo
  {pages} {024901} (\bibinfo {year} {2022})}\BibitemShut {NoStop}%
\bibitem [{\citenamefont {Jain}\ and\ \citenamefont {Thakur}(2022)}]{Jain2022}%
  \BibitemOpen
  \bibfield  {author} {\bibinfo {author} {\bibfnamefont {N.}~\bibnamefont
  {Jain}}\ and\ \bibinfo {author} {\bibfnamefont {S.}~\bibnamefont {Thakur}},\
  }\bibfield  {title} {\bibinfo {title} {Collapse dynamics of chemically active
  flexible polymer},\ }\href@noop {} {\bibfield  {journal} {\bibinfo  {journal}
  {Macromolecules}\ }\textbf {\bibinfo {volume} {55}},\ \bibinfo {pages} {2375}
  (\bibinfo {year} {2022})}\BibitemShut {NoStop}%
\bibitem [{\citenamefont {Chen}\ \emph {et~al.}(2019)\citenamefont {Chen},
  \citenamefont {Poling-Skutvik}, \citenamefont {Howard}, \citenamefont
  {Nikoubashman}, \citenamefont {Egorov}, \citenamefont {Conrad},\ and\
  \citenamefont {Palmer}}]{chen2019}%
  \BibitemOpen
  \bibfield  {author} {\bibinfo {author} {\bibfnamefont {R.}~\bibnamefont
  {Chen}}, \bibinfo {author} {\bibfnamefont {R.}~\bibnamefont
  {Poling-Skutvik}}, \bibinfo {author} {\bibfnamefont {M.~P.}\ \bibnamefont
  {Howard}}, \bibinfo {author} {\bibfnamefont {A.}~\bibnamefont
  {Nikoubashman}}, \bibinfo {author} {\bibfnamefont {S.~A.}\ \bibnamefont
  {Egorov}}, \bibinfo {author} {\bibfnamefont {J.~C.}\ \bibnamefont {Conrad}},\
  and\ \bibinfo {author} {\bibfnamefont {J.~C.}\ \bibnamefont {Palmer}},\
  }\bibfield  {title} {\bibinfo {title} {Influence of polymer flexibility on
  nanoparticle dynamics in semidilute solutions},\ }\href@noop {} {\bibfield
  {journal} {\bibinfo  {journal} {Soft Matter}\ }\textbf {\bibinfo {volume}
  {15}},\ \bibinfo {pages} {1260} (\bibinfo {year} {2019})}\BibitemShut
  {NoStop}%
\bibitem [{\citenamefont {Narayan}\ \emph {et~al.}(2007)\citenamefont
  {Narayan}, \citenamefont {Ramaswamy},\ and\ \citenamefont
  {Menon}}]{Ramaswamy2007Science}%
  \BibitemOpen
  \bibfield  {author} {\bibinfo {author} {\bibfnamefont {V.}~\bibnamefont
  {Narayan}}, \bibinfo {author} {\bibfnamefont {S.}~\bibnamefont {Ramaswamy}},\
  and\ \bibinfo {author} {\bibfnamefont {N.}~\bibnamefont {Menon}},\ }\bibfield
   {title} {\bibinfo {title} {Long-lived giant number fluctuations in a
  swarming granular nematic},\ }\href@noop {} {\bibfield  {journal} {\bibinfo
  {journal} {Science}\ }\textbf {\bibinfo {volume} {317}},\ \bibinfo {pages}
  {105} (\bibinfo {year} {2007})}\BibitemShut {NoStop}%
\bibitem [{\citenamefont {Patelli}\ \emph {et~al.}(2019)\citenamefont
  {Patelli}, \citenamefont {Djafer-Cherif}, \citenamefont {Aranson},
  \citenamefont {Bertin},\ and\ \citenamefont {Chat{\'e}}}]{Chate2019PRL}%
  \BibitemOpen
  \bibfield  {author} {\bibinfo {author} {\bibfnamefont {A.}~\bibnamefont
  {Patelli}}, \bibinfo {author} {\bibfnamefont {I.}~\bibnamefont
  {Djafer-Cherif}}, \bibinfo {author} {\bibfnamefont {I.~S.}\ \bibnamefont
  {Aranson}}, \bibinfo {author} {\bibfnamefont {E.}~\bibnamefont {Bertin}},\
  and\ \bibinfo {author} {\bibfnamefont {H.}~\bibnamefont {Chat{\'e}}},\
  }\bibfield  {title} {\bibinfo {title} {Understanding dense active nematics
  from microscopic models},\ }\href@noop {} {\bibfield  {journal} {\bibinfo
  {journal} {Physical Review Letters}\ }\textbf {\bibinfo {volume} {123}},\
  \bibinfo {pages} {258001} (\bibinfo {year} {2019})}\BibitemShut {NoStop}%
\bibitem [{\citenamefont {Kumar}\ \emph {et~al.}(2022)\citenamefont {Kumar},
  \citenamefont {Zhang}, \citenamefont {Redford}, \citenamefont {de~Pablo},\
  and\ \citenamefont {Gardel}}]{Kumar2022SoftMatter}%
  \BibitemOpen
  \bibfield  {author} {\bibinfo {author} {\bibfnamefont {N.}~\bibnamefont
  {Kumar}}, \bibinfo {author} {\bibfnamefont {R.}~\bibnamefont {Zhang}},
  \bibinfo {author} {\bibfnamefont {S.~A.}\ \bibnamefont {Redford}}, \bibinfo
  {author} {\bibfnamefont {J.~J.}\ \bibnamefont {de~Pablo}},\ and\ \bibinfo
  {author} {\bibfnamefont {M.~L.}\ \bibnamefont {Gardel}},\ }\bibfield  {title}
  {\bibinfo {title} {Catapulting of topological defects through elasticity
  bands in active nematics},\ }\href@noop {} {\bibfield  {journal} {\bibinfo
  {journal} {Soft Matter}\ }\textbf {\bibinfo {volume} {18}},\ \bibinfo {pages}
  {5271} (\bibinfo {year} {2022})}\BibitemShut {NoStop}%
\bibitem [{\citenamefont {DeCamp}\ \emph {et~al.}(2015)\citenamefont {DeCamp},
  \citenamefont {Redner}, \citenamefont {Baskaran}, \citenamefont {Hagan},\
  and\ \citenamefont {Dogic}}]{Dogic2015NatureMat}%
  \BibitemOpen
  \bibfield  {author} {\bibinfo {author} {\bibfnamefont {S.~J.}\ \bibnamefont
  {DeCamp}}, \bibinfo {author} {\bibfnamefont {G.~S.}\ \bibnamefont {Redner}},
  \bibinfo {author} {\bibfnamefont {A.}~\bibnamefont {Baskaran}}, \bibinfo
  {author} {\bibfnamefont {M.~F.}\ \bibnamefont {Hagan}},\ and\ \bibinfo
  {author} {\bibfnamefont {Z.}~\bibnamefont {Dogic}},\ }\bibfield  {title}
  {\bibinfo {title} {Orientational order of motile defects in active
  nematics},\ }\href@noop {} {\bibfield  {journal} {\bibinfo  {journal} {Nature
  materials}\ }\textbf {\bibinfo {volume} {14}},\ \bibinfo {pages} {1110}
  (\bibinfo {year} {2015})}\BibitemShut {NoStop}%
\bibitem [{\citenamefont {Tang}\ and\ \citenamefont
  {Selinger}(2019)}]{Selinger2018SoftMatter}%
  \BibitemOpen
  \bibfield  {author} {\bibinfo {author} {\bibfnamefont {X.}~\bibnamefont
  {Tang}}\ and\ \bibinfo {author} {\bibfnamefont {J.~V.}\ \bibnamefont
  {Selinger}},\ }\bibfield  {title} {\bibinfo {title} {Theory of defect motion
  in 2d passive and active nematic liquid crystals},\ }\href@noop {} {\bibfield
   {journal} {\bibinfo  {journal} {Soft Matter}\ }\textbf {\bibinfo {volume}
  {15}},\ \bibinfo {pages} {587} (\bibinfo {year} {2019})}\BibitemShut
  {NoStop}%
\bibitem [{\citenamefont {Giomi}\ \emph {et~al.}(2014)\citenamefont {Giomi},
  \citenamefont {Bowick}, \citenamefont {Mishra}, \citenamefont {Sknepnek},\
  and\ \citenamefont {Cristina~Marchetti}}]{Giomi2014PhilTransactionsA}%
  \BibitemOpen
  \bibfield  {author} {\bibinfo {author} {\bibfnamefont {L.}~\bibnamefont
  {Giomi}}, \bibinfo {author} {\bibfnamefont {M.~J.}\ \bibnamefont {Bowick}},
  \bibinfo {author} {\bibfnamefont {P.}~\bibnamefont {Mishra}}, \bibinfo
  {author} {\bibfnamefont {R.}~\bibnamefont {Sknepnek}},\ and\ \bibinfo
  {author} {\bibfnamefont {M.}~\bibnamefont {Cristina~Marchetti}},\ }\bibfield
  {title} {\bibinfo {title} {Defect dynamics in active nematics},\ }\href@noop
  {} {\bibfield  {journal} {\bibinfo  {journal} {Philosophical Transactions of
  the Royal Society A: Mathematical, Physical and Engineering Sciences}\
  }\textbf {\bibinfo {volume} {372}},\ \bibinfo {pages} {20130365} (\bibinfo
  {year} {2014})}\BibitemShut {NoStop}%
\bibitem [{\citenamefont {Theers}\ and\ \citenamefont
  {Winkler}(2014)}]{theers2014}%
  \BibitemOpen
  \bibfield  {author} {\bibinfo {author} {\bibfnamefont {M.}~\bibnamefont
  {Theers}}\ and\ \bibinfo {author} {\bibfnamefont {R.~G.}\ \bibnamefont
  {Winkler}},\ }\bibfield  {title} {\bibinfo {title} {Effects of thermal
  fluctuations and fluid compressibility on hydrodynamic synchronization of
  microrotors at finite oscillatory reynolds number: a multiparticle collision
  dynamics simulation study},\ }\href@noop {} {\bibfield  {journal} {\bibinfo
  {journal} {Soft Matter}\ }\textbf {\bibinfo {volume} {10}},\ \bibinfo {pages}
  {5894} (\bibinfo {year} {2014})}\BibitemShut {NoStop}%
\bibitem [{\citenamefont {Akhter}\ and\ \citenamefont
  {Rohlf}(2014)}]{Akhter2014}%
  \BibitemOpen
  \bibfield  {author} {\bibinfo {author} {\bibfnamefont {T.}~\bibnamefont
  {Akhter}}\ and\ \bibinfo {author} {\bibfnamefont {K.}~\bibnamefont {Rohlf}},\
  }\bibfield  {title} {\bibinfo {title} {Quantifying compressibility and slip
  in multiparticle collision (mpc) flow through a local constriction},\
  }\href@noop {} {\bibfield  {journal} {\bibinfo  {journal} {Entropy}\ }\textbf
  {\bibinfo {volume} {16}},\ \bibinfo {pages} {418} (\bibinfo {year}
  {2014})}\BibitemShut {NoStop}%
\bibitem [{\citenamefont {Chat{\'e}}\ and\ \citenamefont
  {Mahault}(2019)}]{Chate2019}%
  \BibitemOpen
  \bibfield  {author} {\bibinfo {author} {\bibfnamefont {H.}~\bibnamefont
  {Chat{\'e}}}\ and\ \bibinfo {author} {\bibfnamefont {B.}~\bibnamefont
  {Mahault}},\ }\bibfield  {title} {\bibinfo {title} {Dry, aligning, dilute,
  active matter: A synthetic and self-contained overview},\ }\href@noop {}
  {\bibfield  {journal} {\bibinfo  {journal} {arXiv preprint arXiv:1906.05542}\
  } (\bibinfo {year} {2019})}\BibitemShut {NoStop}%
\bibitem [{\citenamefont {Ramaswamy}\ \emph {et~al.}(2003)\citenamefont
  {Ramaswamy}, \citenamefont {Simha},\ and\ \citenamefont
  {Toner}}]{Ramaswamy2003EPL}%
  \BibitemOpen
  \bibfield  {author} {\bibinfo {author} {\bibfnamefont {S.}~\bibnamefont
  {Ramaswamy}}, \bibinfo {author} {\bibfnamefont {R.~A.}\ \bibnamefont
  {Simha}},\ and\ \bibinfo {author} {\bibfnamefont {J.}~\bibnamefont {Toner}},\
  }\bibfield  {title} {\bibinfo {title} {Active nematics on a substrate: Giant
  number fluctuations and long-time tails},\ }\href@noop {} {\bibfield
  {journal} {\bibinfo  {journal} {Europhysics Letters ({EPL})}\ }\textbf
  {\bibinfo {volume} {62}},\ \bibinfo {pages} {196} (\bibinfo {year}
  {2003})}\BibitemShut {NoStop}%
\bibitem [{\citenamefont {Shi}\ and\ \citenamefont
  {Ma}(2013)}]{Shi2013NatComm}%
  \BibitemOpen
  \bibfield  {author} {\bibinfo {author} {\bibfnamefont {X.-q.}\ \bibnamefont
  {Shi}}\ and\ \bibinfo {author} {\bibfnamefont {Y.-q.}\ \bibnamefont {Ma}},\
  }\bibfield  {title} {\bibinfo {title} {Topological structure dynamics
  revealing collective evolution in active nematics},\ }\href@noop {}
  {\bibfield  {journal} {\bibinfo  {journal} {Nature Communications}\ }\textbf
  {\bibinfo {volume} {4}},\ \bibinfo {pages} {1} (\bibinfo {year}
  {2013})}\BibitemShut {NoStop}%
\bibitem [{\citenamefont {Henkes}\ \emph {et~al.}(2011)\citenamefont {Henkes},
  \citenamefont {Fily},\ and\ \citenamefont {Marchetti}}]{Henkes2011-Jamming}%
  \BibitemOpen
  \bibfield  {author} {\bibinfo {author} {\bibfnamefont {S.}~\bibnamefont
  {Henkes}}, \bibinfo {author} {\bibfnamefont {Y.}~\bibnamefont {Fily}},\ and\
  \bibinfo {author} {\bibfnamefont {M.~C.}\ \bibnamefont {Marchetti}},\
  }\bibfield  {title} {\bibinfo {title} {Active jamming: Self-propelled soft
  particles at high density},\ }\href@noop {} {\bibfield  {journal} {\bibinfo
  {journal} {Physical Review E}\ }\textbf {\bibinfo {volume} {84}},\ \bibinfo
  {pages} {040301} (\bibinfo {year} {2011})}\BibitemShut {NoStop}%
\bibitem [{\citenamefont {Henkes}\ \emph {et~al.}(2018)\citenamefont {Henkes},
  \citenamefont {Marchetti},\ and\ \citenamefont
  {Sknepnek}}]{Henkes2018-NematicSphere}%
  \BibitemOpen
  \bibfield  {author} {\bibinfo {author} {\bibfnamefont {S.}~\bibnamefont
  {Henkes}}, \bibinfo {author} {\bibfnamefont {M.~C.}\ \bibnamefont
  {Marchetti}},\ and\ \bibinfo {author} {\bibfnamefont {R.}~\bibnamefont
  {Sknepnek}},\ }\bibfield  {title} {\bibinfo {title} {Dynamical patterns in
  nematic active matter on a sphere},\ }\href@noop {} {\bibfield  {journal}
  {\bibinfo  {journal} {Physical Review E}\ }\textbf {\bibinfo {volume} {97}},\
  \bibinfo {pages} {042605} (\bibinfo {year} {2018})}\BibitemShut {NoStop}%
\bibitem [{\citenamefont {Toner}(2019)}]{Toner2019}%
  \BibitemOpen
  \bibfield  {author} {\bibinfo {author} {\bibfnamefont {J.}~\bibnamefont
  {Toner}},\ }\bibfield  {title} {\bibinfo {title} {{Giant number fluctuations
  in dry active polar fluids: A shocking analogy with lightning rods}},\
  }\href@noop {} {\bibfield  {journal} {\bibinfo  {journal} {The Journal of
  Chemical Physics}\ }\textbf {\bibinfo {volume} {150}},\ \bibinfo {pages}
  {154120} (\bibinfo {year} {2019})}\BibitemShut {NoStop}%
\bibitem [{\citenamefont {Moussa{\"\i}d}\ \emph {et~al.}(2011)\citenamefont
  {Moussa{\"\i}d}, \citenamefont {Helbing},\ and\ \citenamefont
  {Theraulaz}}]{Helbing2011}%
  \BibitemOpen
  \bibfield  {author} {\bibinfo {author} {\bibfnamefont {M.}~\bibnamefont
  {Moussa{\"\i}d}}, \bibinfo {author} {\bibfnamefont {D.}~\bibnamefont
  {Helbing}},\ and\ \bibinfo {author} {\bibfnamefont {G.}~\bibnamefont
  {Theraulaz}},\ }\bibfield  {title} {\bibinfo {title} {How simple rules
  determine pedestrian behavior and crowd disasters},\ }\href@noop {}
  {\bibfield  {journal} {\bibinfo  {journal} {Proceedings of the National
  Academy of Sciences}\ }\textbf {\bibinfo {volume} {108}},\ \bibinfo {pages}
  {6884} (\bibinfo {year} {2011})}\BibitemShut {NoStop}%
\bibitem [{\citenamefont {Liu}\ \emph {et~al.}(2011)\citenamefont {Liu},
  \citenamefont {Fu}, \citenamefont {Liu}, \citenamefont {Ren}, \citenamefont
  {Chau}, \citenamefont {Li}, \citenamefont {Xiang}, \citenamefont {Zeng},
  \citenamefont {Chen}, \citenamefont {Tang}, \citenamefont {Lenz},
  \citenamefont {Cui}, \citenamefont {Huang}, \citenamefont {Hwa},\ and\
  \citenamefont {Huang}}]{Huang2011}%
  \BibitemOpen
  \bibfield  {author} {\bibinfo {author} {\bibfnamefont {C.}~\bibnamefont
  {Liu}}, \bibinfo {author} {\bibfnamefont {X.}~\bibnamefont {Fu}}, \bibinfo
  {author} {\bibfnamefont {L.}~\bibnamefont {Liu}}, \bibinfo {author}
  {\bibfnamefont {X.}~\bibnamefont {Ren}}, \bibinfo {author} {\bibfnamefont
  {C.~K.}\ \bibnamefont {Chau}}, \bibinfo {author} {\bibfnamefont
  {S.}~\bibnamefont {Li}}, \bibinfo {author} {\bibfnamefont {L.}~\bibnamefont
  {Xiang}}, \bibinfo {author} {\bibfnamefont {H.}~\bibnamefont {Zeng}},
  \bibinfo {author} {\bibfnamefont {G.}~\bibnamefont {Chen}}, \bibinfo {author}
  {\bibfnamefont {L.-H.}\ \bibnamefont {Tang}}, \bibinfo {author}
  {\bibfnamefont {P.}~\bibnamefont {Lenz}}, \bibinfo {author} {\bibfnamefont
  {X.}~\bibnamefont {Cui}}, \bibinfo {author} {\bibfnamefont {W.}~\bibnamefont
  {Huang}}, \bibinfo {author} {\bibfnamefont {T.}~\bibnamefont {Hwa}},\ and\
  \bibinfo {author} {\bibfnamefont {J.-D.}\ \bibnamefont {Huang}},\ }\bibfield
  {title} {\bibinfo {title} {Sequential establishment of stripe patterns in an
  expanding cell population},\ }\href@noop {} {\bibfield  {journal} {\bibinfo
  {journal} {Science}\ }\textbf {\bibinfo {volume} {334}},\ \bibinfo {pages}
  {238} (\bibinfo {year} {2011})}\BibitemShut {NoStop}%
\bibitem [{\citenamefont {Fu}\ \emph {et~al.}(2012)\citenamefont {Fu},
  \citenamefont {Tang}, \citenamefont {Liu}, \citenamefont {Huang},
  \citenamefont {Hwa},\ and\ \citenamefont {Lenz}}]{Lenz2012}%
  \BibitemOpen
  \bibfield  {author} {\bibinfo {author} {\bibfnamefont {X.}~\bibnamefont
  {Fu}}, \bibinfo {author} {\bibfnamefont {L.-H.}\ \bibnamefont {Tang}},
  \bibinfo {author} {\bibfnamefont {C.}~\bibnamefont {Liu}}, \bibinfo {author}
  {\bibfnamefont {J.-D.}\ \bibnamefont {Huang}}, \bibinfo {author}
  {\bibfnamefont {T.}~\bibnamefont {Hwa}},\ and\ \bibinfo {author}
  {\bibfnamefont {P.}~\bibnamefont {Lenz}},\ }\bibfield  {title} {\bibinfo
  {title} {Stripe formation in bacterial systems with density-suppressed
  motility},\ }\href@noop {} {\bibfield  {journal} {\bibinfo  {journal}
  {Physical Review Letters}\ }\textbf {\bibinfo {volume} {108}},\ \bibinfo
  {pages} {198102} (\bibinfo {year} {2012})}\BibitemShut {NoStop}%
\bibitem [{\citenamefont {Dogic}\ \emph {et~al.}(2014)\citenamefont {Dogic},
  \citenamefont {Sharma},\ and\ \citenamefont {Zakhary}}]{Dogic2014}%
  \BibitemOpen
  \bibfield  {author} {\bibinfo {author} {\bibfnamefont {Z.}~\bibnamefont
  {Dogic}}, \bibinfo {author} {\bibfnamefont {P.}~\bibnamefont {Sharma}},\ and\
  \bibinfo {author} {\bibfnamefont {M.~J.}\ \bibnamefont {Zakhary}},\
  }\bibfield  {title} {\bibinfo {title} {Hypercomplex liquid crystals},\
  }\href@noop {} {\bibfield  {journal} {\bibinfo  {journal} {Annual Review of
  Condensed Matter Physics}\ }\textbf {\bibinfo {volume} {5}},\ \bibinfo
  {pages} {137} (\bibinfo {year} {2014})}\BibitemShut {NoStop}%
\end{thebibliography}%


%

\clearpage
\onecolumngrid
\appendix

\section{Supplementary Tables}
\begin{table*}[h]
    \centering
    \begin{tabular}{|c||c|c||p{0.5\linewidth}|}
        \hline
        \textbf{Parameter} & \textbf{Units} & \textbf{Typical Value} & \textbf{Description}
        \\ \hline \hline
        MPCD cell size, $a$ & 
        [Length] & 
        $1$ & 
        Unit length 
        \\ \hline
        MPCD thermal energy unit, $\kbt$ & 
        [Energy] & 
        $1$ & 
        Unit energy 
        \\ \hline
        Particle mass, $m$ & 
        [Mass] & 
        $1$ & 
        Unit mass 
        \\ \hline
        Unit time, $\tau$ & 
        $a\sqrt{m/\kbt}$ & 
        $1$ & 
        Derived unit of time 
        \\ \hline
        Simulation box length, $\lenSys$ & 
        $a$ & 
        $200$ & 
        System width and length 
        \\ \hline
        Local number density, $\rho_C$ & 
        $a^{-2}$ & 
        N/A & 
        Instantaneous number density of cell $C$ 
        \\ \hline
        Particle number per cell, $\NCell$ & 
        N/A & 
        $20$ & 
        System-averaged particle number per cell (equivalent to $\av{\rho_C}$)
        \\ \hline
        Timestep size, $\delta t$ & 
        $\tau$ & 
        0.1 & 
        Duration of ballistic streaming step 
        \\ \hline
        Warmup time & 
        $\delta t$ & 
        $10000$ & 
        Time before data collection begins 
        \\ \hline
        Simulation time & 
        $\delta t$ & 
        $50000$ & 
         Data collection time 
        \\ \hline
        Mean field potential, $U$ & 
        $\kbt$ & 
        10 & 
        Cell-based inter-molecular orientation interaction 
        \\ \hline
        Rotational friction, $\gamma_\mathrm{R}$ & 
        $\kbt \tau$ & 
        0.01 & 
        Rotational friction during orientation collision step 
        \\ \hline
        Hydrodynamic susceptibility, $\chi$ & 
        N/A & 
        0.5 & 
        Heuristic shear coupling factor controlling alignment relaxation relative to $\delta t$ in Jeffery's equation
        \\ \hline
        Tumbling parameter, $\lambda$ & 
        N/A & 
        2 & 
        Bare tumbling parameter used in Jeffery equation 
        \\ \hline
        Particle activity, $\act$ & 
        $ma/\tau^{2}$ & 
        $[10^{-4}, 1]$ & 
        Single particle contribution towards cell's active dipole  
        \\ \hline
        Sigmoidal modulation position, $\Sigpos$ & 
        N/A & 
        0.4 & 
        Midpoint of the sigmoidal function $\SigF$ relative to $\av{\rho_C}$  
        \\ \hline
        Sigmoidal modulation width, $\Sigwid$ & 
        N/A & 
        0.5 & 
        Width of the sigmoidal function $\SigF$ relative to $\av{\rho_C}$ 
        \\ \hline
    \end{tabular}
    \caption{
        \textbf{Reference table of active-nematic multiparticle collision dynamics (AN-MPCD) simulation parameters. }
        Table of all simulation parameters used in this work.
        Units, typical values and a brief description of each are stated.
    }
    \label{sitab:UnitsRef}
\end{table*}

\begin{table*}[h]
    \centering
    \begin{tabular}{|c|c|c||c||c|c|c|c|c|}
        \hline
        \multirow{2}{*}{\textbf{Scaling Quantity}} & \multirow{2}{*}{\textbf{Variable}} & \multirow{2}{*}{\textbf{Figure}} & \multirow{2}{*}{\textbf{Theory}} & 
        \multicolumn{4}{c|}{\textbf{Scaling}} & \multirow{2}{*}{\textbf{Measurement Regime}} \\
        \cline{5-8}
        & & & & $\ActSum$ & $\ActAv$ & $\SigSum$ & $\SigAv$&  \\
        \hline \hline
        Defect separation & $\lenDef$ & \fig{fig:Turb-Defects} &
        -1/2 & 
        $-0.53\pm 0.02$ & 
        $-0.45\pm 0.01$ &
        $-0.49\pm 0.04$ &
        $-0.52\pm 0.04$ &
        [0.04, 0.5] 
        \\ \hline
        Bend-wall speed & $v_\mathrm{av}$ &  &
        - & 
        $1.28\pm 0.03$ & 
        $1.21\pm 0.03$ &
        $1.22\pm 0.02$ &
        $1.13\pm 0.04$ &
        [0.001, 0.04] 



        \\ \hline
        Turbulence speed & $v_\mathrm{av}$ & \fig{fig:Turb-FlowSpeed}a &
        +1/2 & 
        $0.86\pm 0.03$ & 
        $0.50\pm 0.02$ &
        $0.42\pm 0.03$ &
        $0.45\pm 0.03$ &
        [0.04, 0.3] 



        \\ \hline
        Velocity length & $\ell_v$ & \fig{fig:Turb-FlowSpeed}b &
        -1/2 & 
        $-0.55\pm 0.04$ & 
        $-0.47\pm 0.07$ &
        $-0.42\pm 0.06$ &
        $-0.45\pm 0.07$ &
        [0.03, 0.5] 
        \\ \hline
        Density length & $\ell_\rho$ & \fig{fig:Dens-Ficks}c &
        - & 
        $-1.41\pm 0.16$ & 
        $-0.89\pm 0.17$ &
        $-0.92\pm 0.06$ &
        $-0.66\pm 0.09$ &
        [0.03, 0.5] 



        \\ \hline
    \end{tabular}
    \caption{
        \textbf{Fitted scalings. } 
        Computed fits for the defect separation, bend-wall speed, turbulence speed, velocity length scale, and density length scale.
        Fits are performed for each activity formulation within the same range of activities. 
    }
    \label{sitab:ScalingRef}
\end{table*}

\begin{table}[h]
    \centering
    \begin{tabular}{|c|c||c|c|c|}
        \hline
        \textbf{Activity Formulation} & \textbf{Variable} & $\actMIN$ & $\actMin$ & $\actMax$ \\
        \hline \hline
        Particle-carried & $\ActSum$ & 0.0035 & 0.0263 & 0.0950 \\
        \hline
        Cell-carried & $\ActAv$ & 0.0010 & 0.0288 & 0.141 \\
        \hline
        Modulated particle-carried & $\SigSum$ & 0.0031 & 0.0380 & 0.287 \\
        \hline
        Modulated cell-carried & $\SigAv$ & 0.0040 & 0.0365 & 0.278 \\
        \hline
    \end{tabular}
    \caption{
        \textbf{Critical activities for each formulation of cellular activity. }
        Critical activities are determined by the giant number fluctuation prefactor values (\fig{fig:Dens-GNF}b, c). 
        Activity overcomes the thermostat at $\actMIN$, above which bend walls occur. 
        At $\actMin$, the bend instability can fit within the system size, giving rise to bulk active turbulence.
        At $\actMax$, density fluctuations are so substantial that the continuum assumption no longer holds at large scales. 
    }
    \label{sitab:OperationalRegimeRef}
\end{table}

\section{Supplementary Movie Captions}
\begin{enumerate}
    \item \label{simov:DirSnapshots}
        \textbf{\underline{M1.mp4:}
        Bulk active turbulence is maintained for all four formulations of cellular activity.}
        Each line segment corresponds to the director $\vec{n}_C$ in one MPCD cell, with no smoothing applied to each frame, colored by scalar order parameter $S_C$. 
        Parameters in each are $\act = 0.15$, $\lenSys=50$, simulation length is $2000 \delta t$ following a $10000 \delta t$ warmup, and each frame is $5\delta t$ apart.
    \item \label{simov:BendInstability}
        \textbf{\underline{M2.mp4:}
        Transient dynamics leading to active turbulence are maintained upon cellular activity modulation. }
        Same simulation parameters as \movie{simov:DirSnapshots} but without a warmup. 
    \item \label{simov:DensSnapshots}
        \textbf{\underline{M3.mp4:}
        Density fluctuations in bulk active turbulence are mitigated by cellular activity modulation.} 
        Particle density fields $\rho_C$ corresponding to the same simulations as \movie{simov:DirSnapshots}. 
\end{enumerate}

\section{Supplementary Figures}

\begin{figure*}[tb]
    \centering
    \begin{subfigure}
        \centering
        \includegraphics[width=0.49\linewidth]{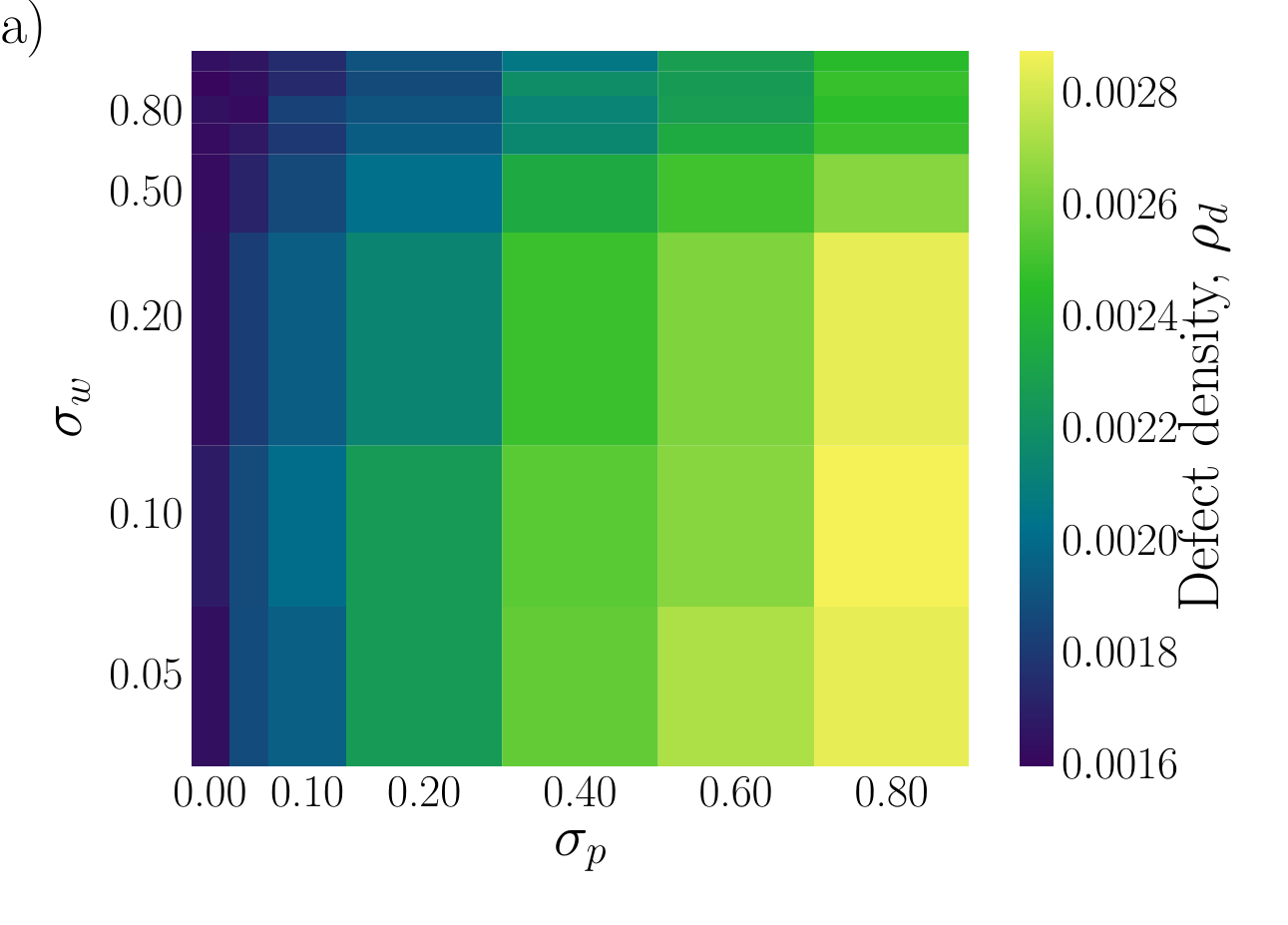}
    \end{subfigure}
    \begin{subfigure}
        \centering
        \includegraphics[width=0.49\linewidth]{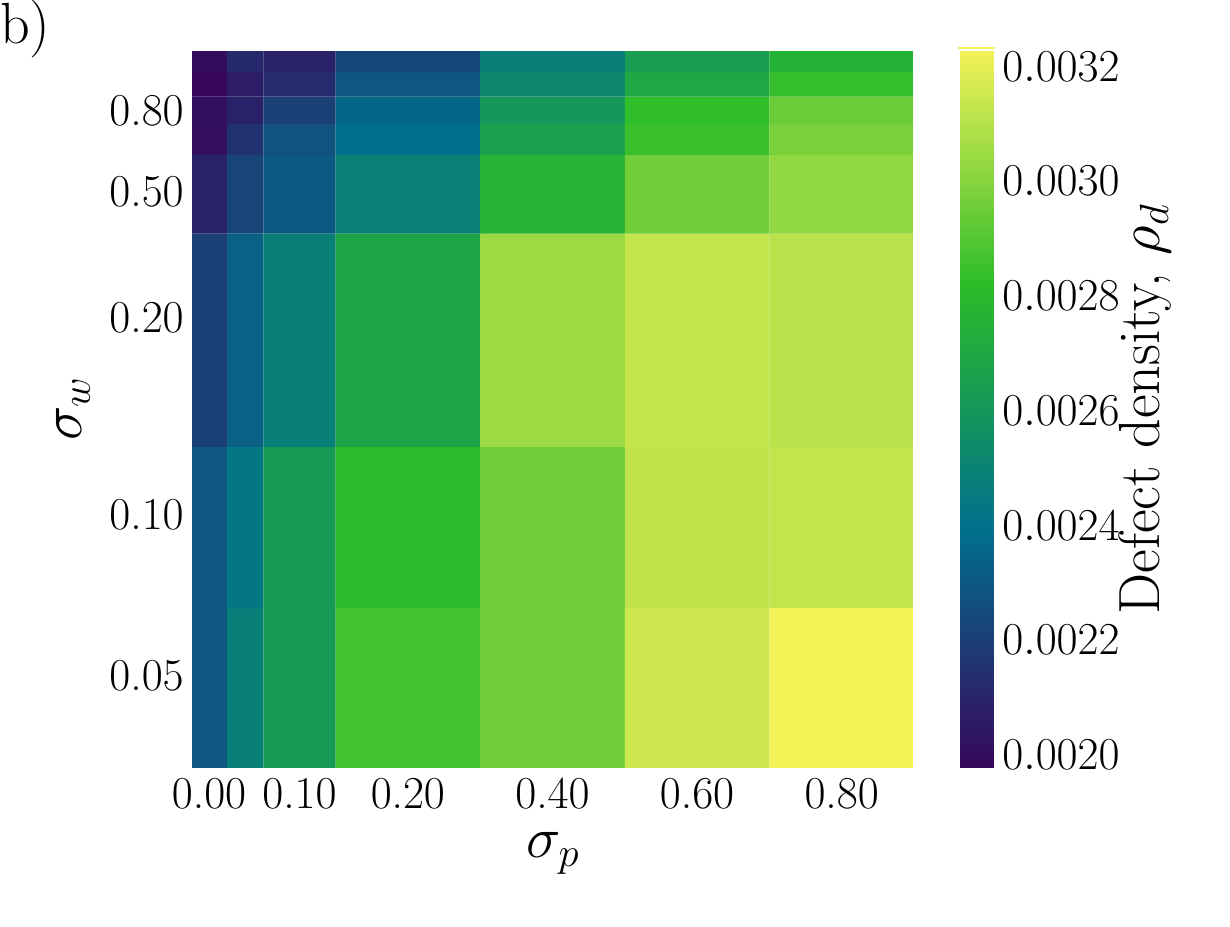}
    \end{subfigure}
    \caption{
        \textbf{Active turbulence is maintained only for a limited range of modulation parameters $\Sigpos$ and $\Sigwid$. } 
        Parameter space depicting the total defect number density as a function of the modulation parameters of $\SigF$ (\eq{eq:SigFnDef}), for a system of activity $\act=0.1$. 
        \textbf{(a)} Parameter space for \emph{modulated particle-carried} activity, $\SigSum$.
        \textbf{(b)} Parameter space for \emph{modulated cell-carried} activity, $\SigAv$.
        The choice of modulation parameters controls active turbulence akin to lowering of an effective activity, which motivates our choice of $\Sigpos=0.4$ and $\Sigwid=0.5$ (main text and \tabsi{sitab:UnitsRef}). 
    }
    \label{sifig:PhaseDiagrams}
\end{figure*}

\begin{figure*}[tb]
    \centering
    \begin{subfigure}
        \centering
        \includegraphics[width=0.45\linewidth]{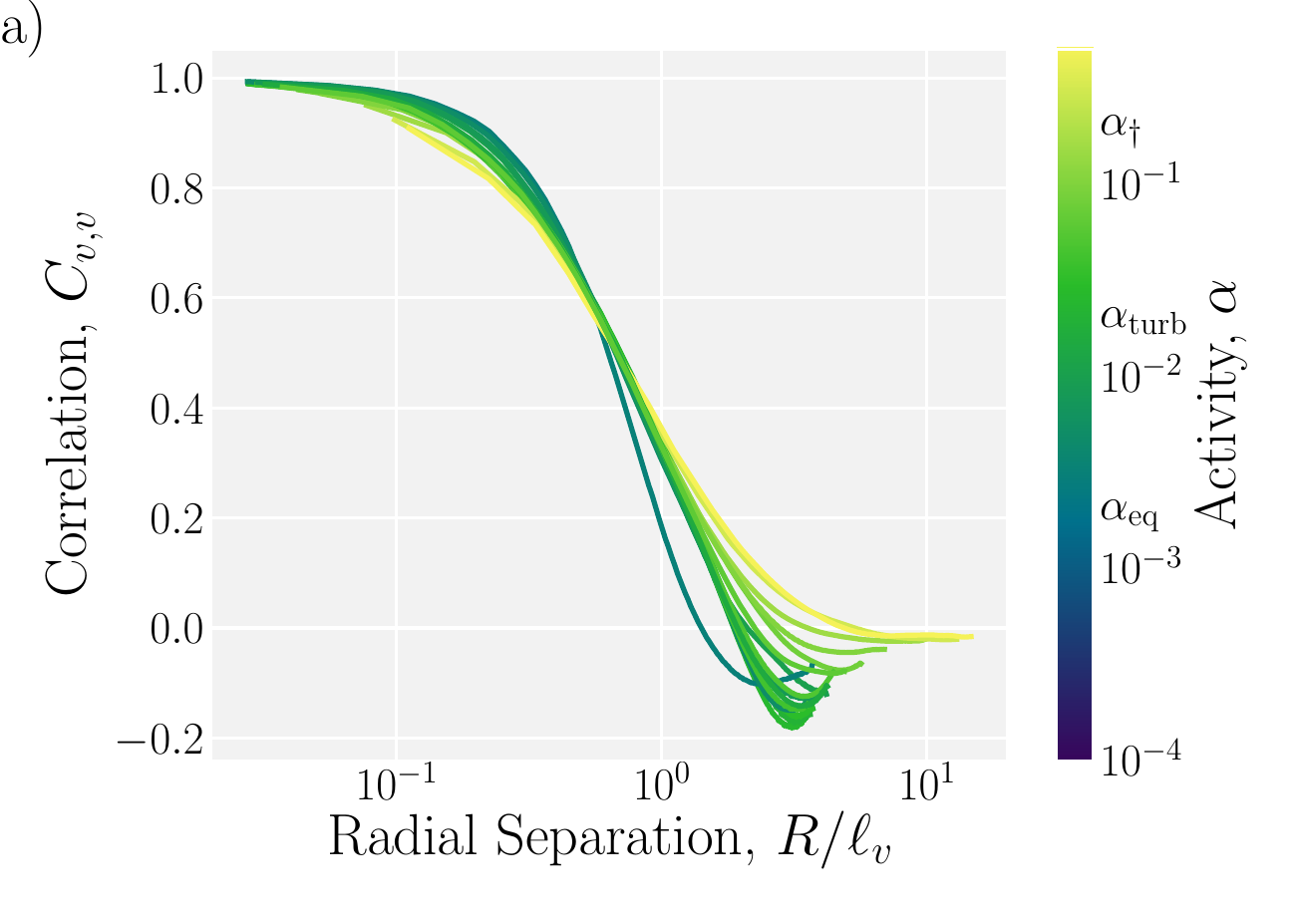}
    \end{subfigure}
    \begin{subfigure}
        \centering
        \includegraphics[width=0.45\linewidth]{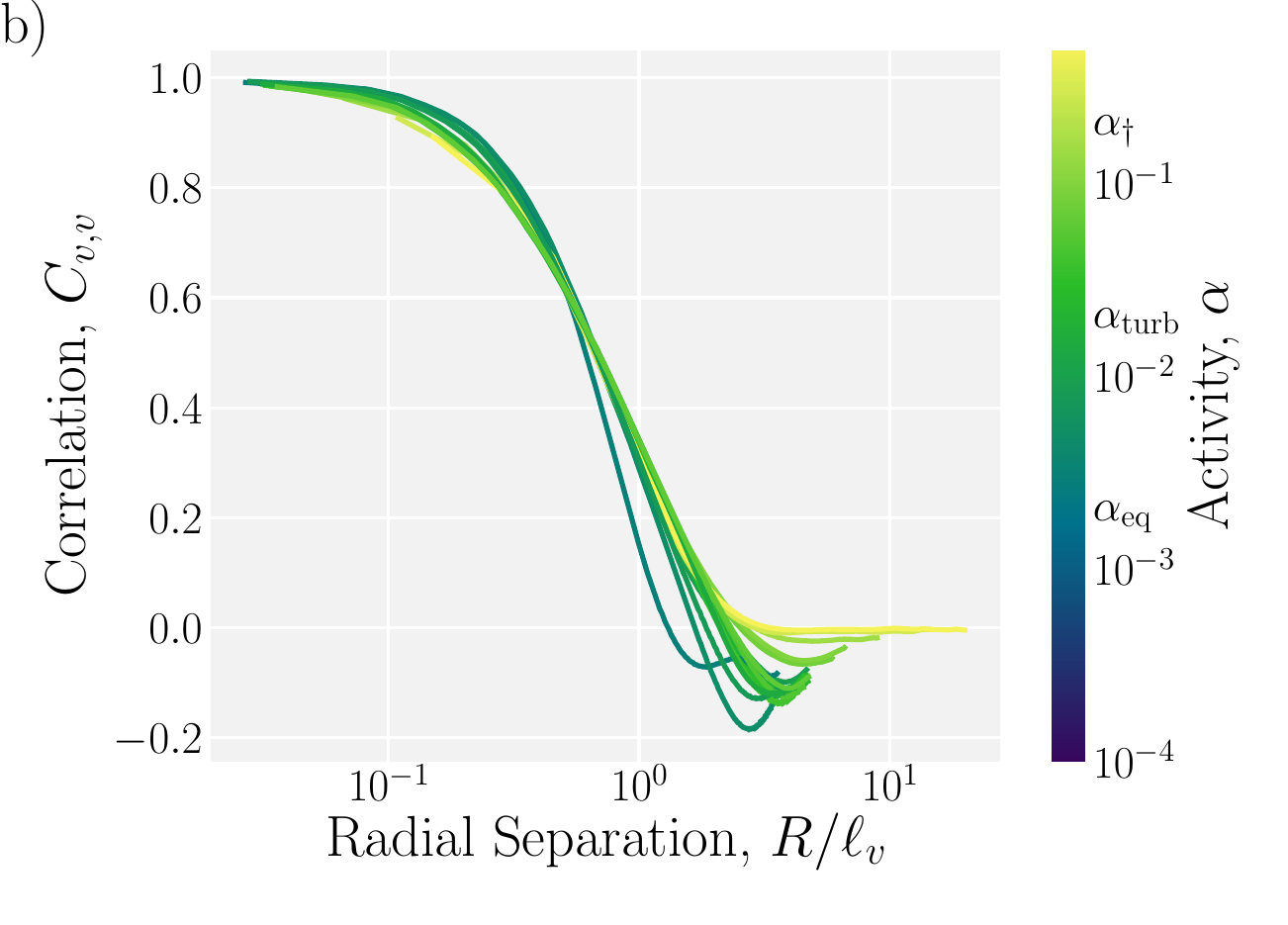}
    \end{subfigure}
    \begin{subfigure}
        \centering
        \includegraphics[width=0.45\linewidth]{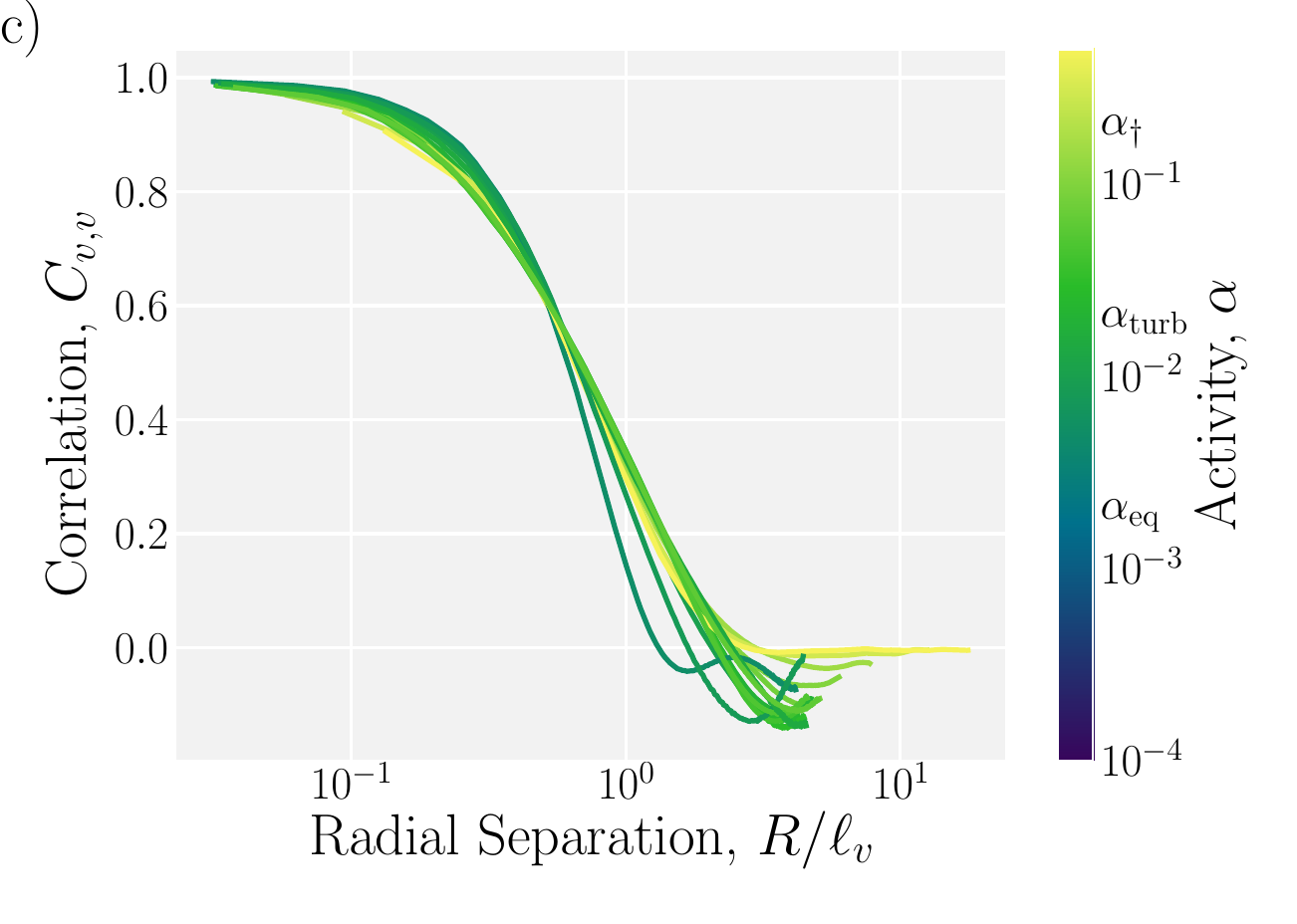}
    \end{subfigure}
    \begin{subfigure}
        \centering
        \includegraphics[width=0.45\linewidth]{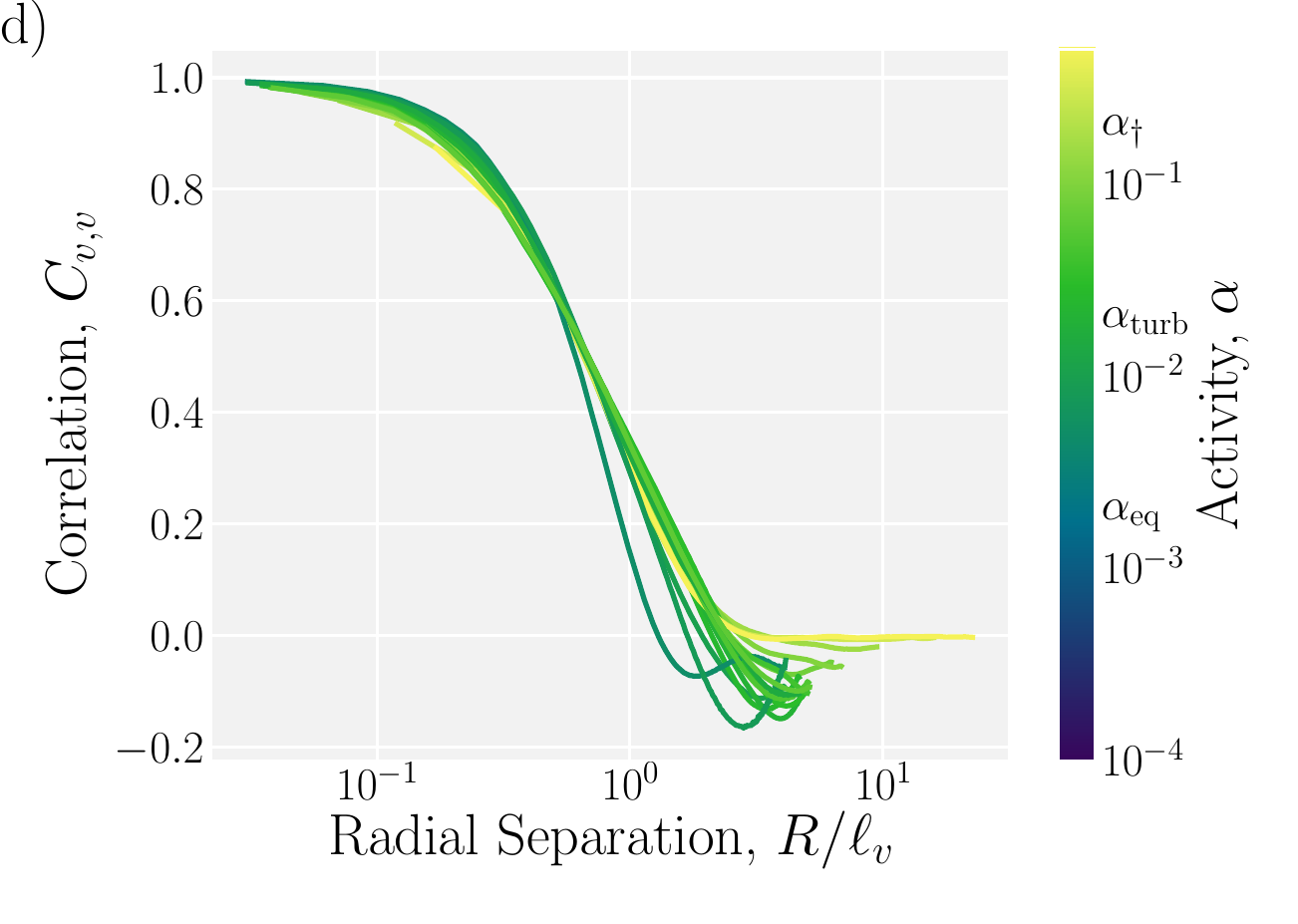}
    \end{subfigure}
    \caption{\textbf{Radial velocity auto-correlation functions $\corr{v}{v}$ collapse when re-scaled by the velocity length scale $\lenVel$.} 
    Correlation functions for 
    \textbf{(a)} \emph{particle-carried},  
    \textbf{(b)} \emph{cell-carried}, 
    \textbf{(c)} \emph{modulated particle-carried}, and
    \textbf{(d)} \emph{modulated cell-carried} activity.
    }
    \label{sifig:Vel-Corr}
\end{figure*}

\begin{figure*}[tb]
    \centering
    \begin{subfigure}
        \centering
        \includegraphics[width=0.45\linewidth]{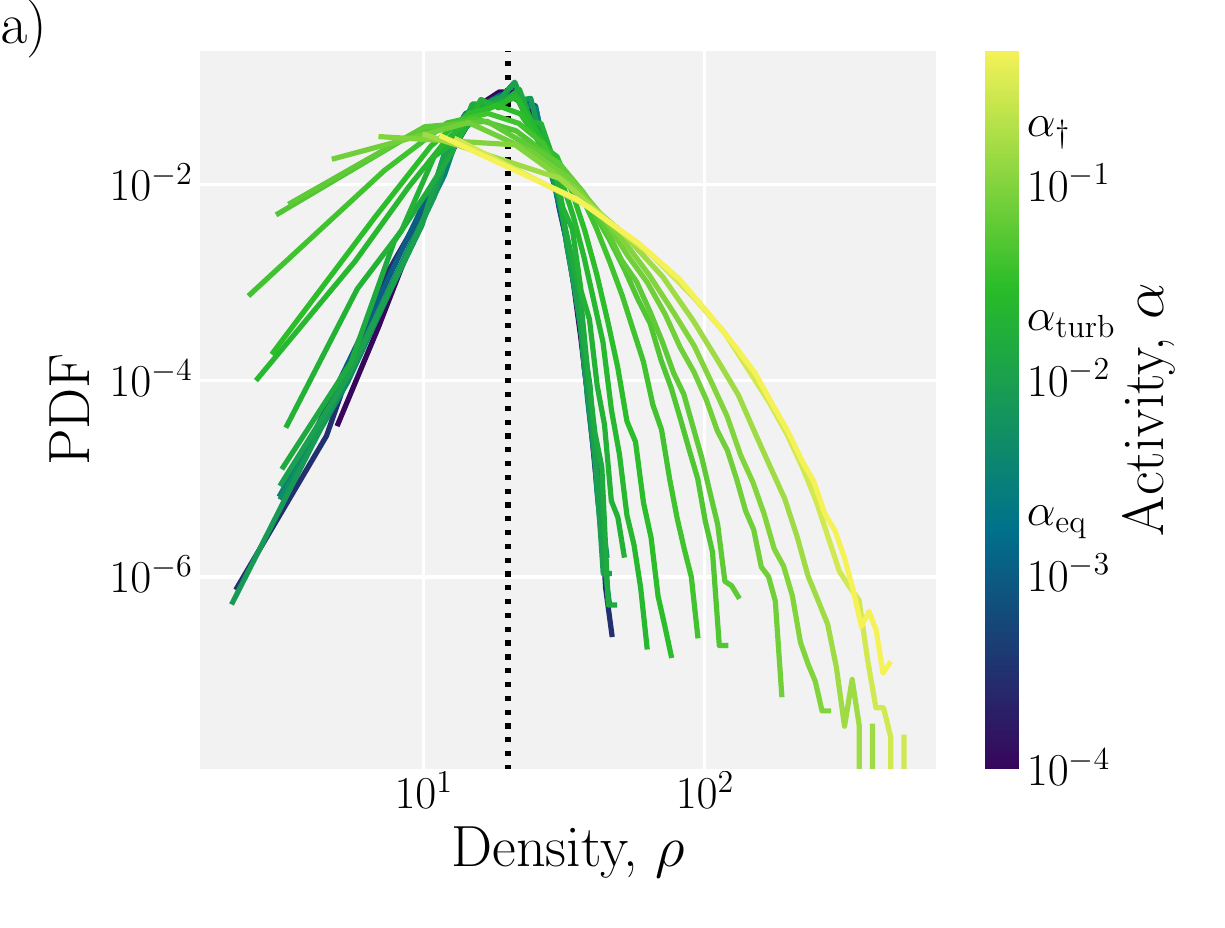}
    \end{subfigure}
    \begin{subfigure}
        \centering
        \includegraphics[width=0.45\linewidth]{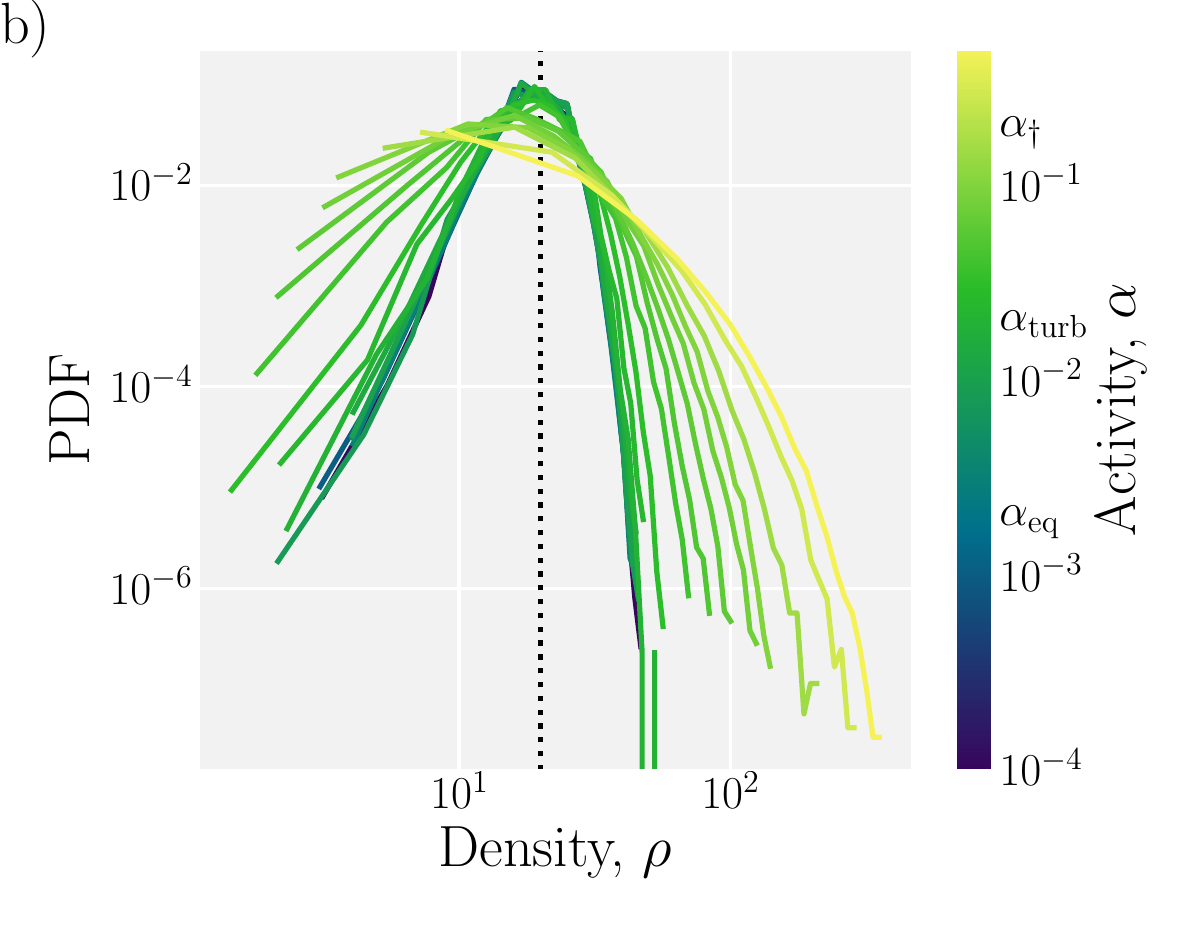}
    \end{subfigure}
    \begin{subfigure}
        \centering
        \includegraphics[width=0.45\linewidth]{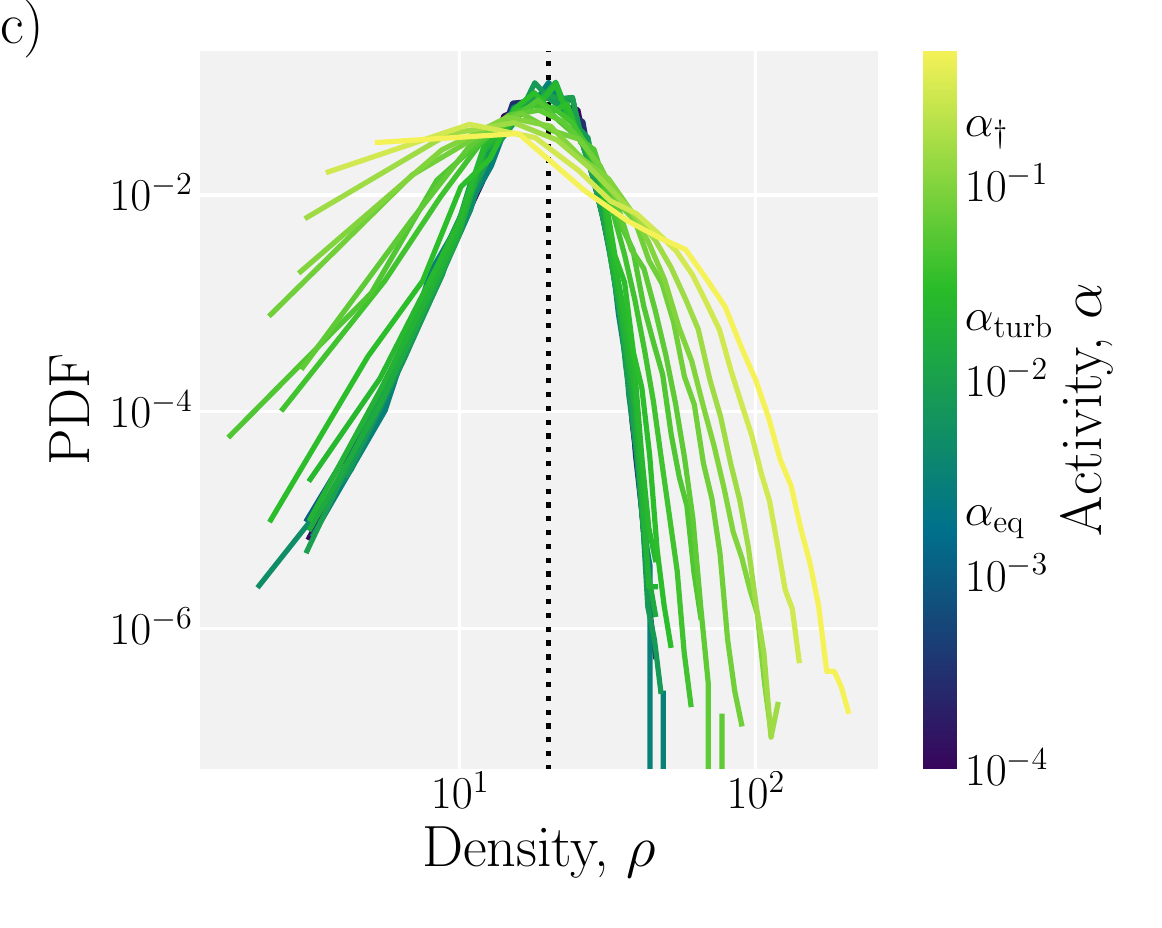}
    \end{subfigure}
    \begin{subfigure}
        \centering
        \includegraphics[width=0.45\linewidth]{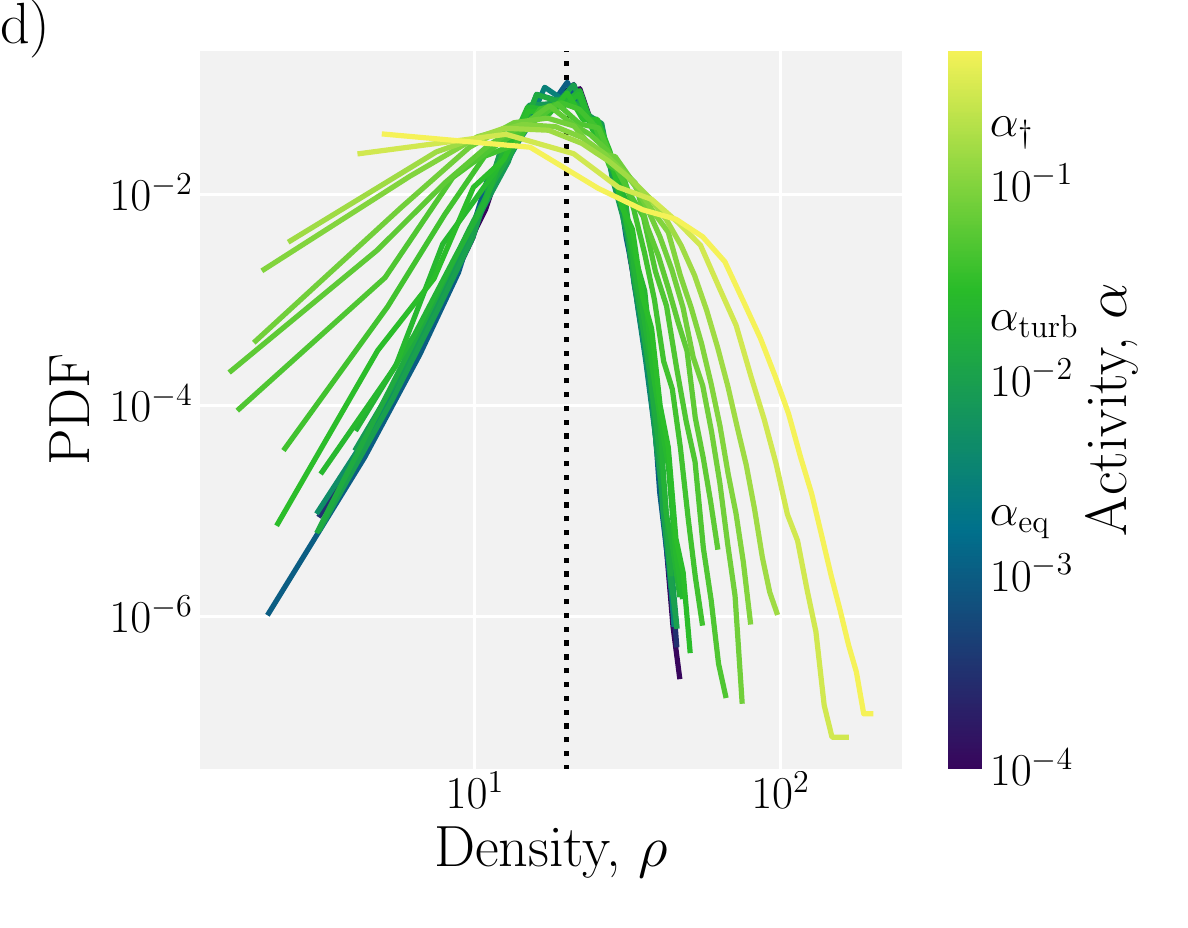}
    \end{subfigure}
    \caption{\textbf{Probability distributions of density widen as activity is increased for all activity formulations. } 
    The dotted line indicates the mean density of $\av{\rho}=20$. 
    Probability density function for 
    \textbf{(a)} \emph{particle-carried},  
    \textbf{(b)} \emph{cell-carried}, 
    \textbf{(c)} \emph{modulated particle-carried}, and
    \textbf{(d)} \emph{modulated cell-carried} activity.
    }
    \label{sifig:Dens-Prob}
\end{figure*}

\begin{figure*}[tb]
    \centering
    \begin{subfigure}
        \centering
        \includegraphics[width=0.45\linewidth]{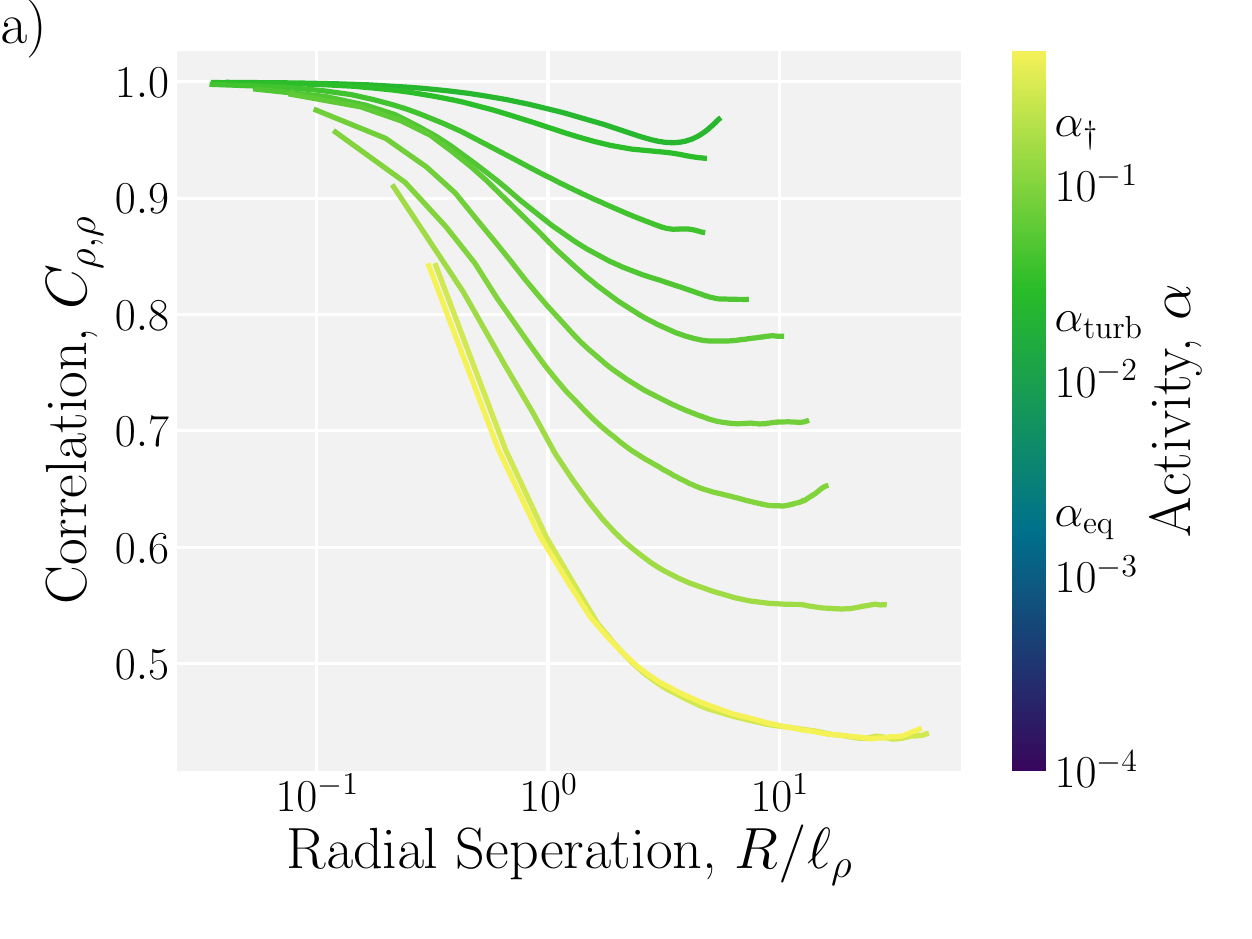}
    \end{subfigure}
    \begin{subfigure}
        \centering
        \includegraphics[width=0.45\linewidth]{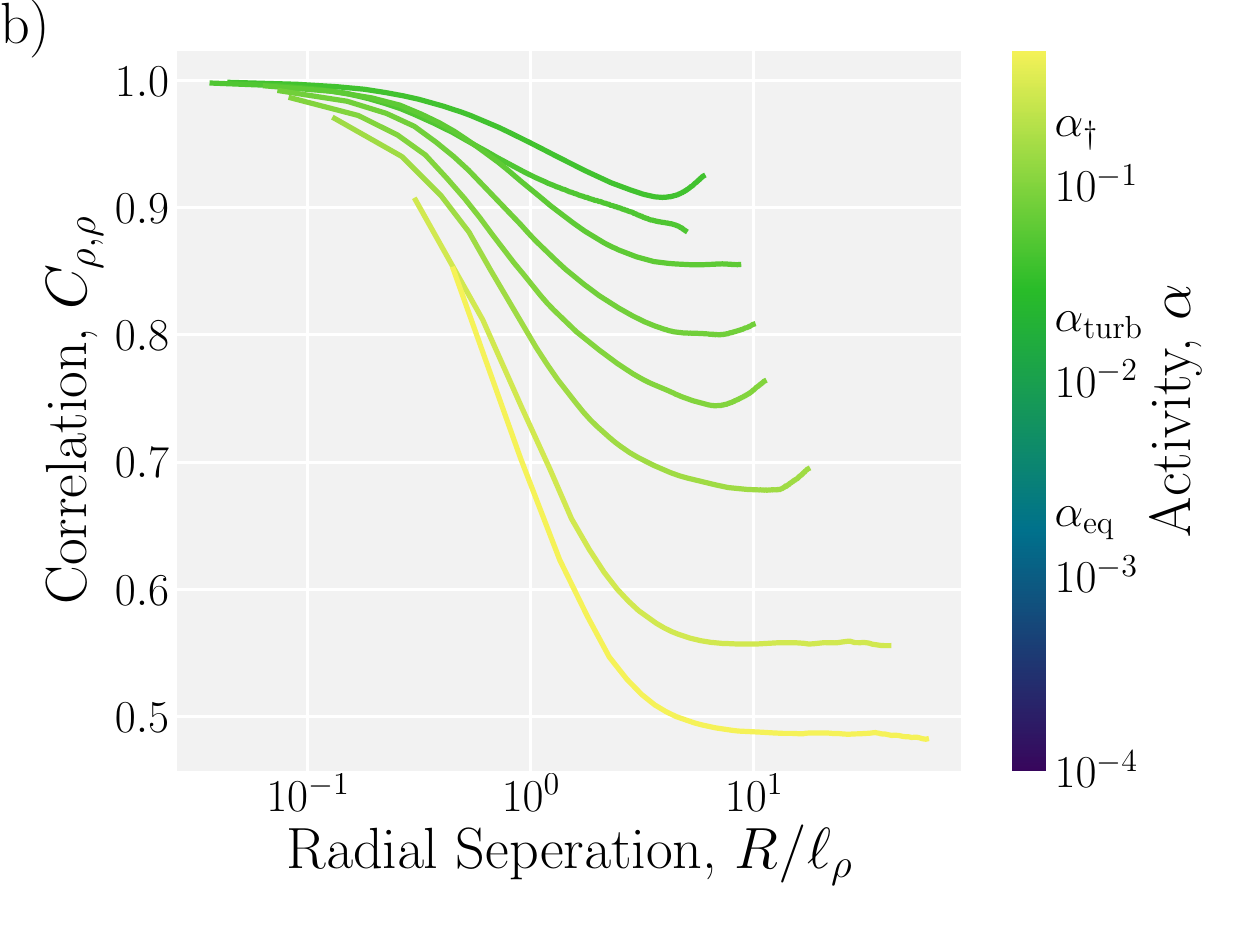}
    \end{subfigure}
    \begin{subfigure}
        \centering
        \includegraphics[width=0.45\linewidth]{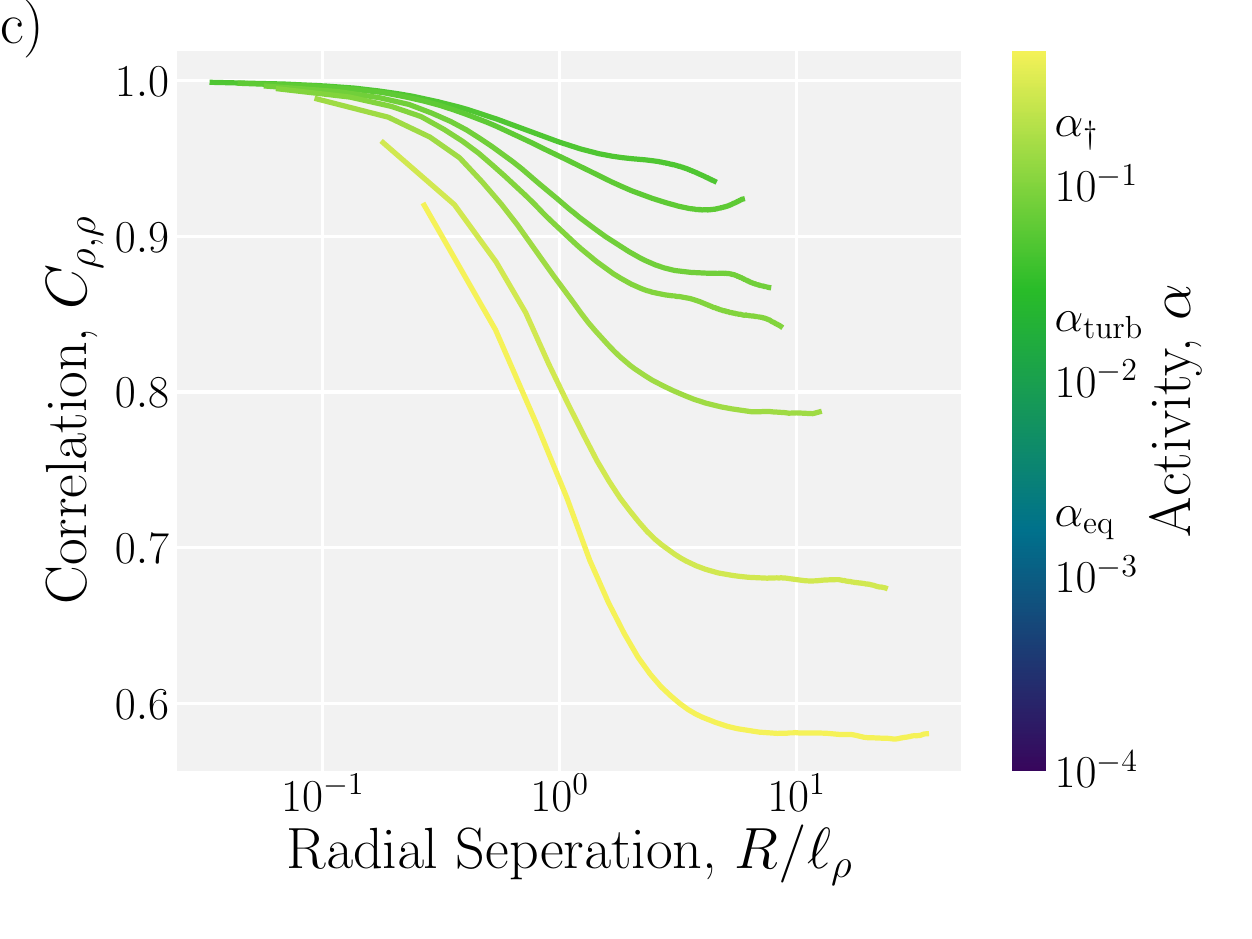}
    \end{subfigure}
    \begin{subfigure}
        \centering
        \includegraphics[width=0.45\linewidth]{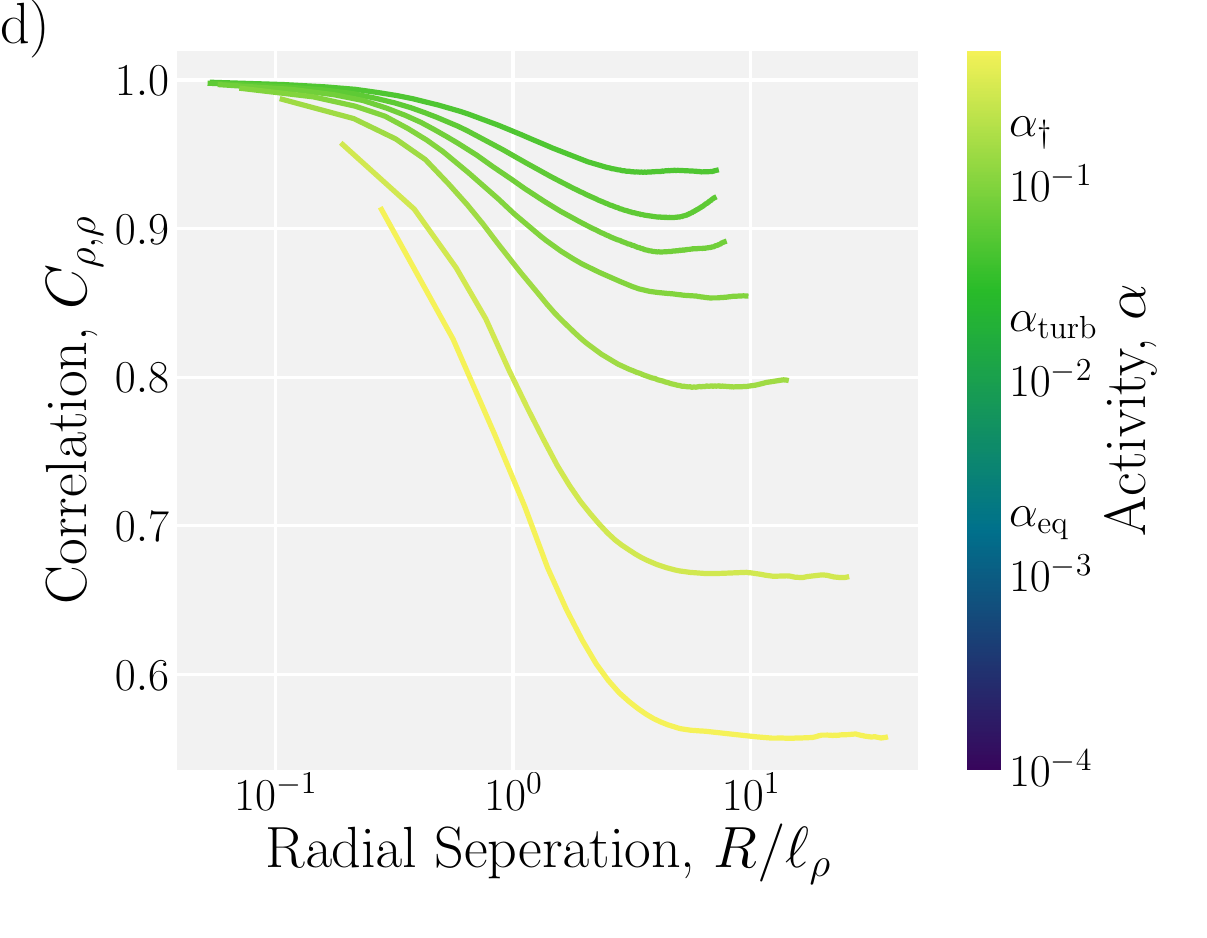}
    \end{subfigure}
    \caption{\textbf{Radial density auto-correlation function $\corr{\rho}{\rho}$ exhibit different far field values when rescaled by density correlation length $\ell_\rho$.} 
    Correlation functions for 
    \textbf{(a)} \emph{particle-carried},  
    \textbf{(b)} \emph{cell-carried}, 
    \textbf{(c)} \emph{modulated particle-carried}, and
    \textbf{(d)} \emph{modulated cell-carried} activity.
    }
    \label{sifig:Dens-Corr}
\end{figure*}

\begin{figure*}[tb]
    \centering
    \begin{subfigure}
        \centering
        \includegraphics[width=0.49\linewidth]{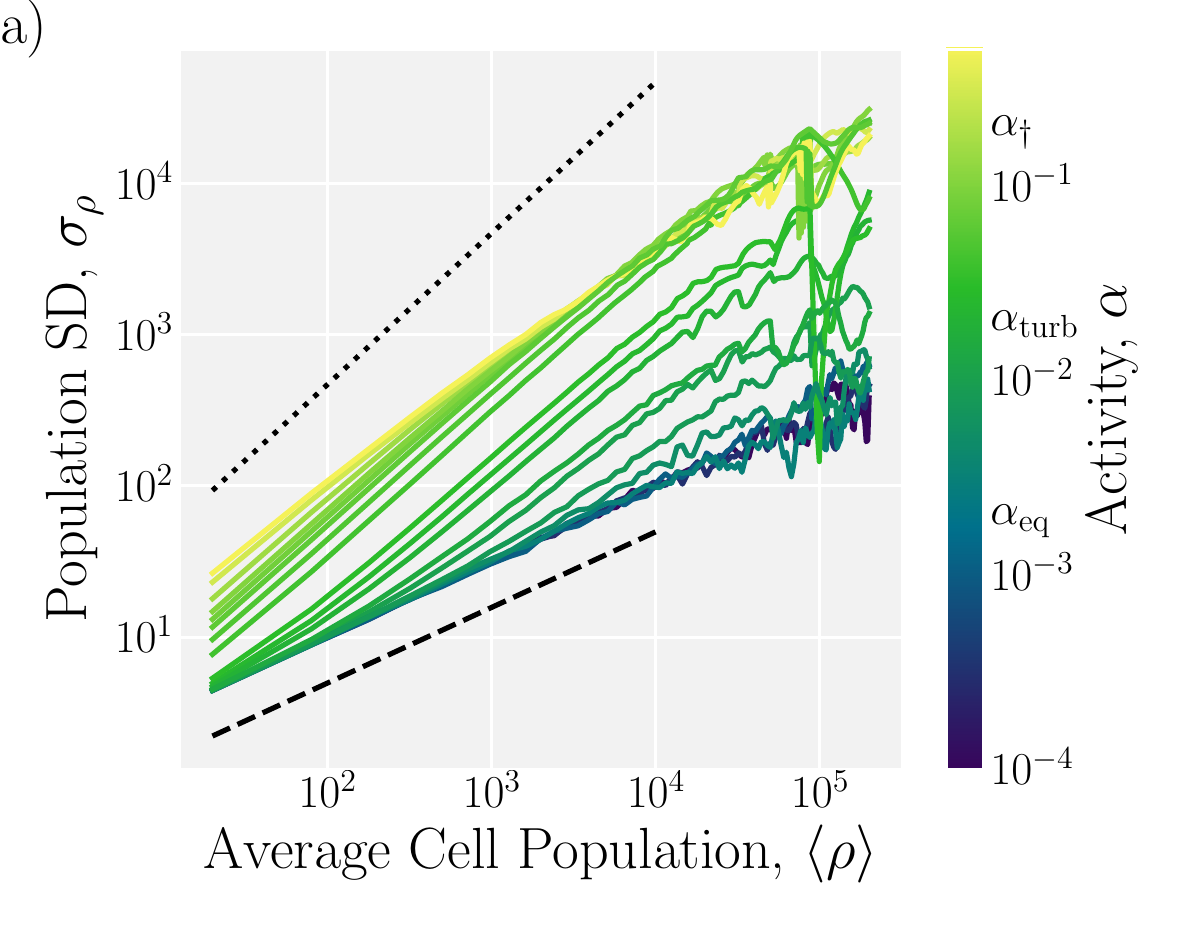}
    \end{subfigure}
    \begin{subfigure}
        \centering
        \includegraphics[width=0.49\linewidth]{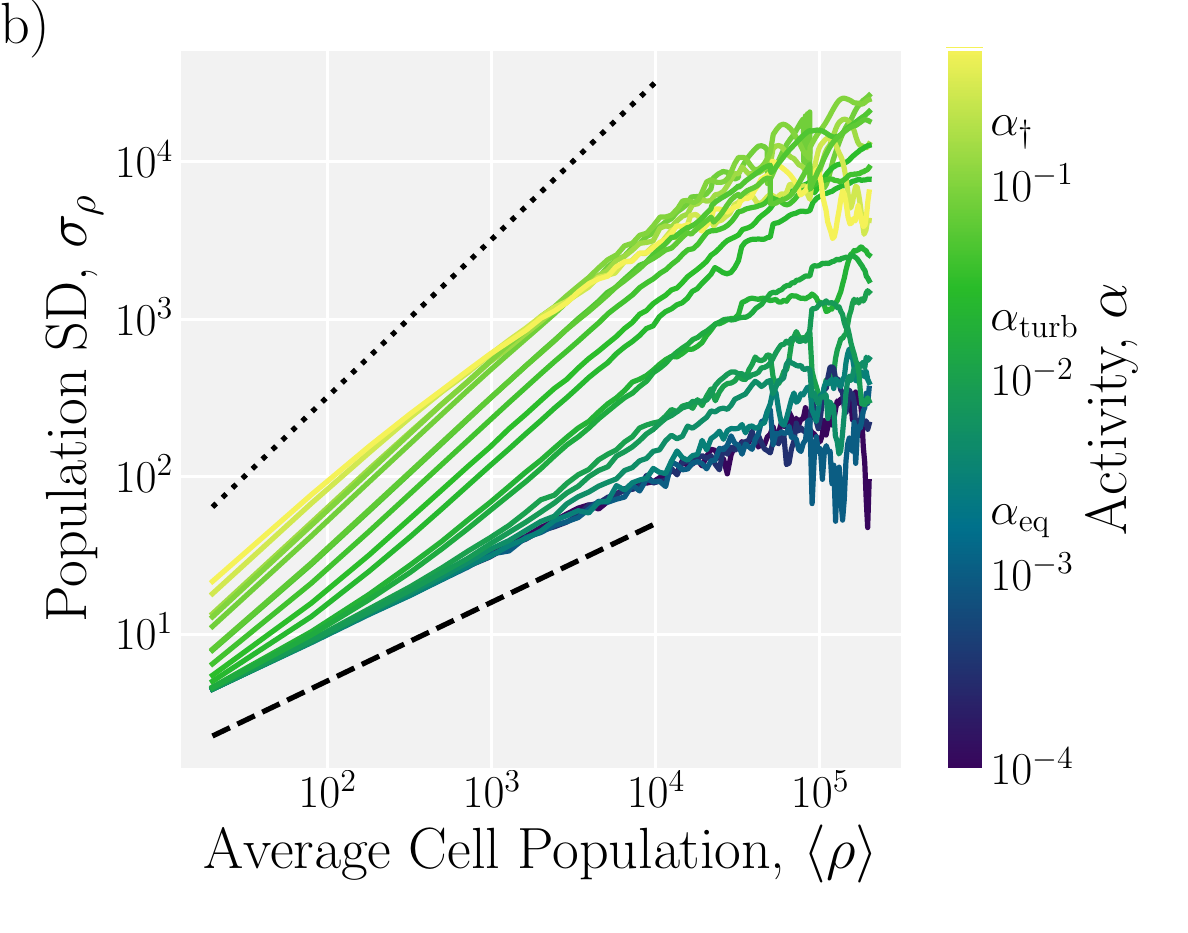}
    \end{subfigure}
    \begin{subfigure}
        \centering
        \includegraphics[width=0.49\linewidth]{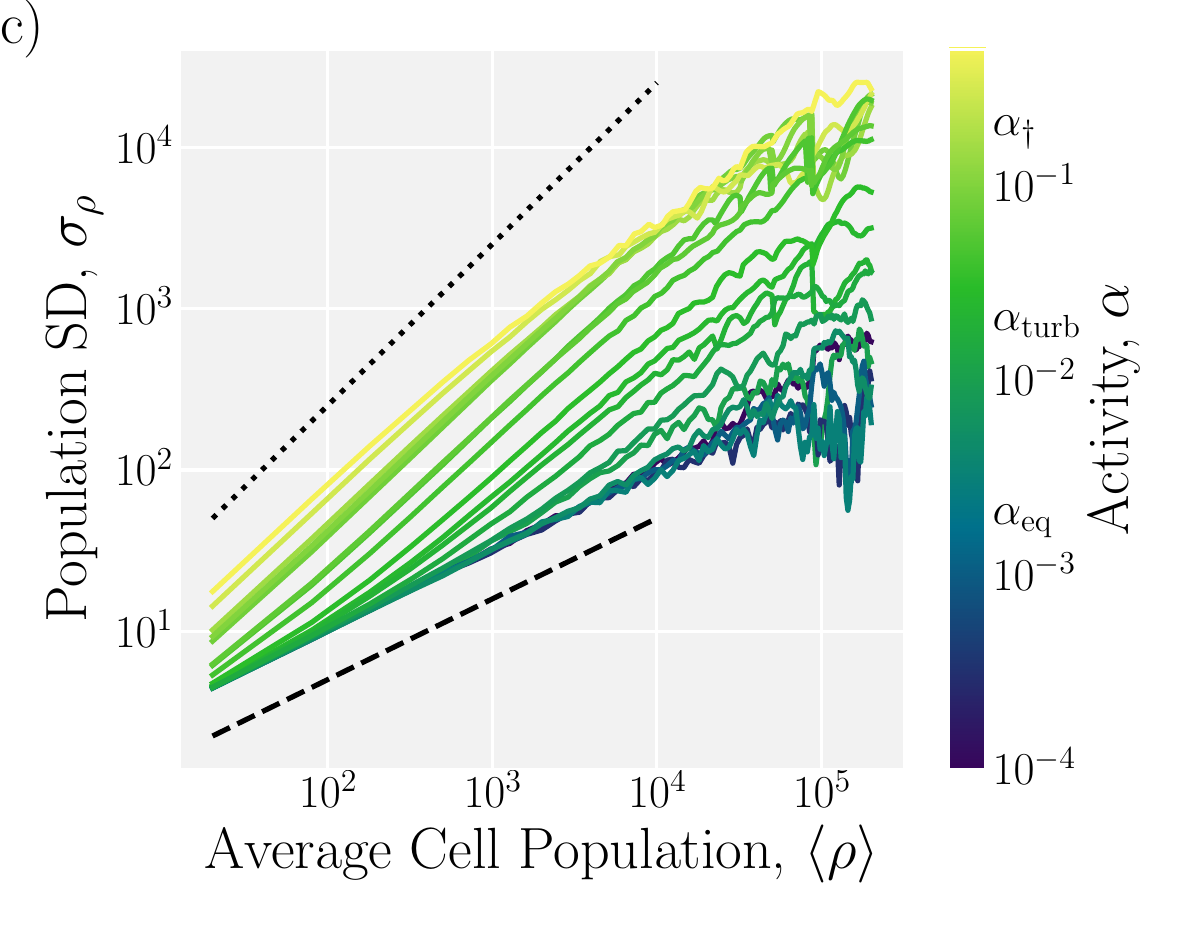}
    \end{subfigure}
    \begin{subfigure}
        \centering
        \includegraphics[width=0.49\linewidth]{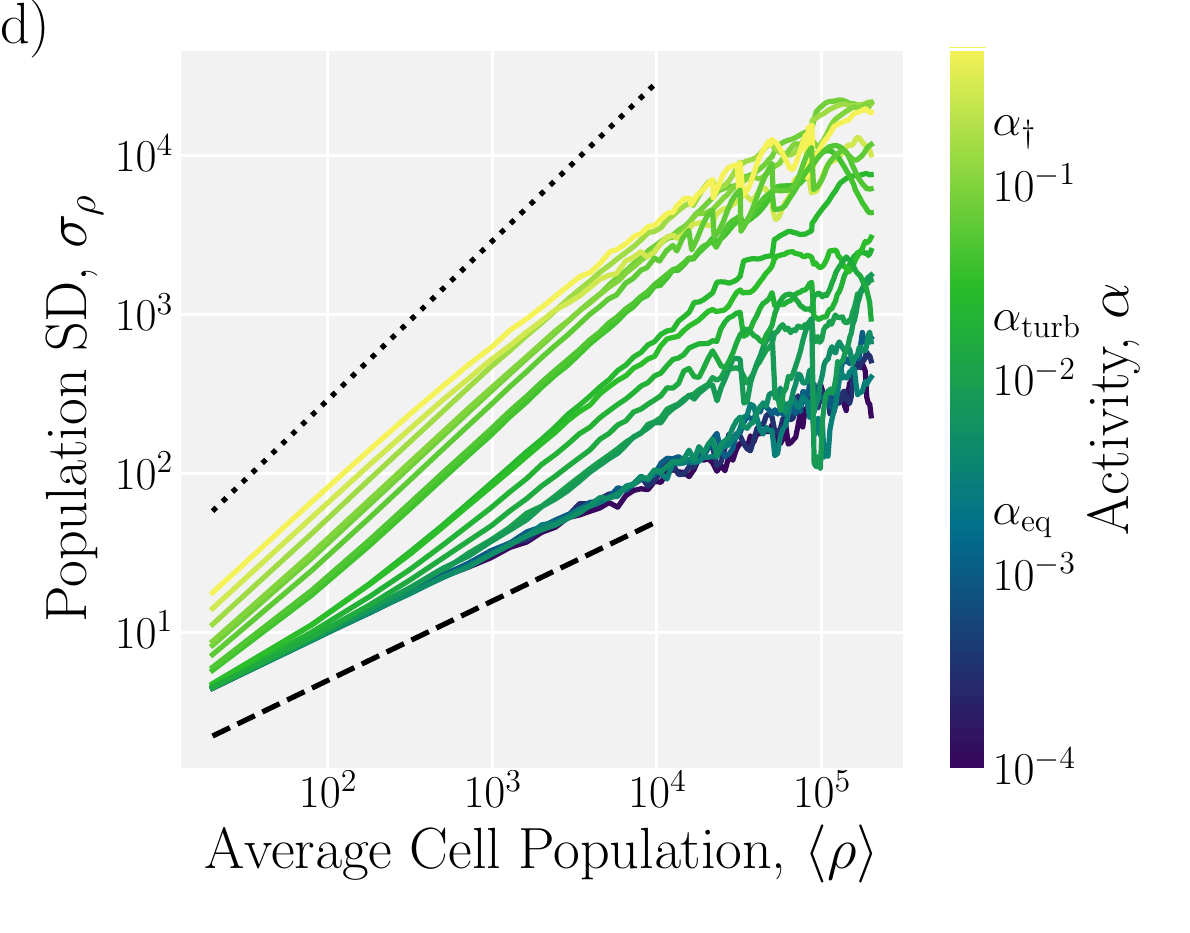}
    \end{subfigure}
    \caption{\textbf{Giant number fluctuations are present in the turbulence regime for all activity formulations.} 
    The dashed line indicates a scaling of $\nu=0.5$, whereas the dotted line indicates a scaling of $\nu=1$.
    Population standard deviation (SD) $\sigma_\rho$ as function of average population $\av{\rho}$ within an area for 
    \textbf{(a)} \emph{particle-carried},  
    \textbf{(b)} \emph{cell-carried}, 
    \textbf{(c)} \emph{modulated particle-carried}, and
    \textbf{(d)} \emph{modulated cell-carried} activity.
    }
    \label{sifig:GNF}
\end{figure*}

\begin{figure*}[tb]
    \centering
    \begin{subfigure}
        \centering
        \includegraphics[width=0.49\linewidth]{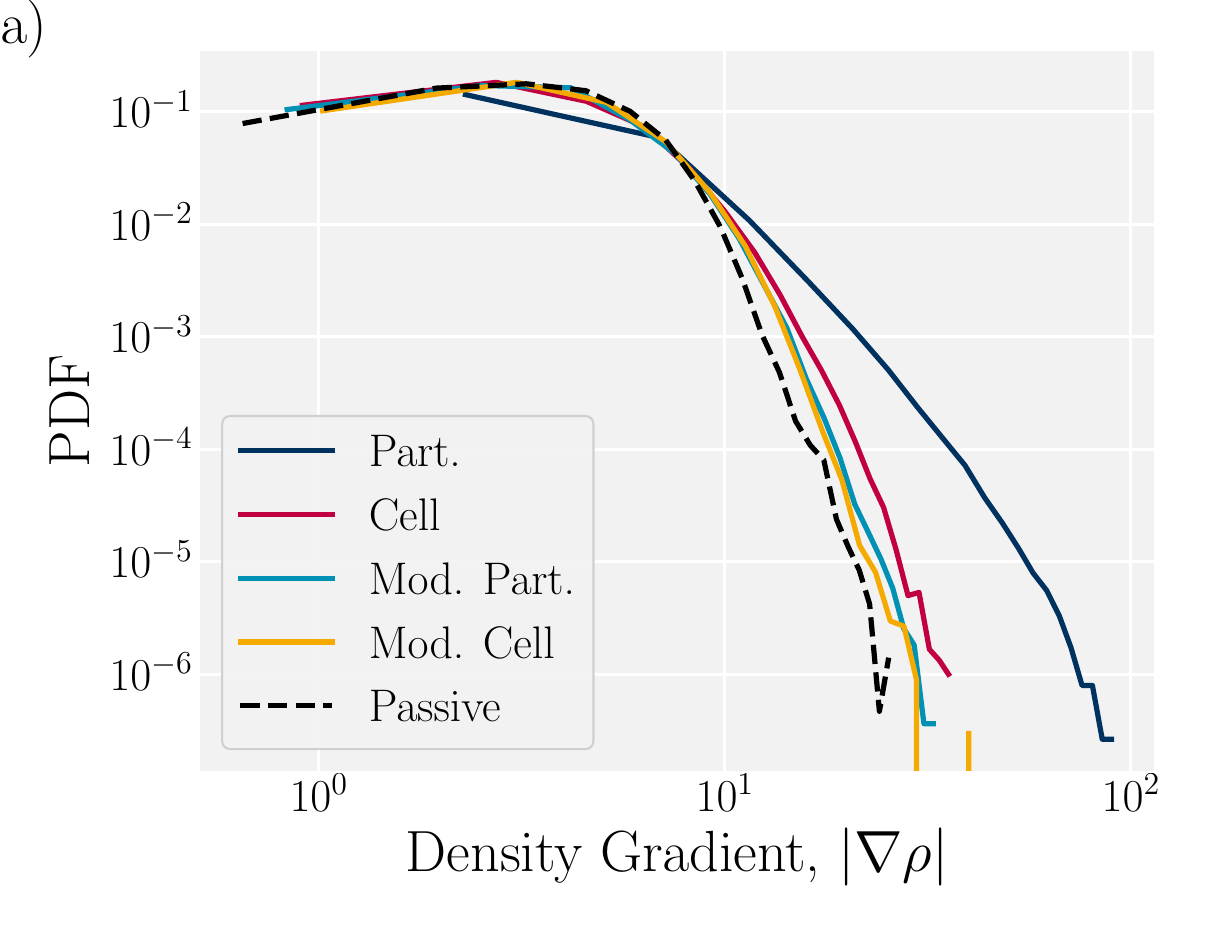}
    \end{subfigure}
    \begin{subfigure}
        \centering
        \includegraphics[width=0.49\linewidth]{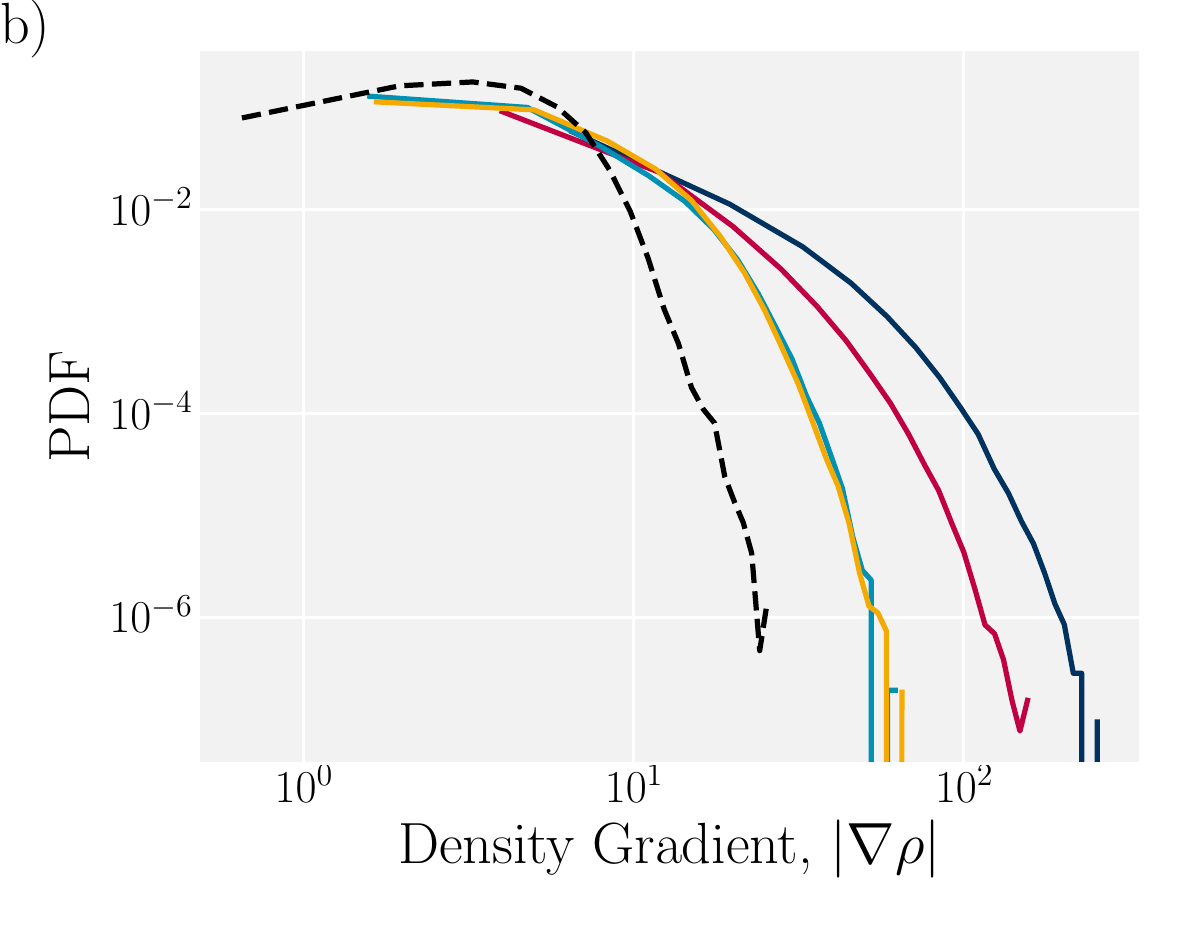}
    \end{subfigure}
    \caption{\textbf{Distributions of density gradients $\abs{\nabla \rho}$ (proportional to Fickian fluxes)}
    Fickian flux $\abs{\nabla \rho}$ for 
    \textbf{(a)} $\act=0.08$ and
    \textbf{(b) } $\act=0.3$. 
    Curves correspond to \emph{particle-carried} (Part.), \emph{cell-carried} (Cell), \emph{modulated particle-carried} (Mod. Part), and \emph{modulated cell-carried} (Mod. Cell) activity, as well as the passive case. 
    }
    \label{sifig:Ficks-Prob}
\end{figure*}

\begin{figure*}[tb]
    \centering
    \begin{subfigure}
        \centering
        \includegraphics[width=0.49\linewidth]{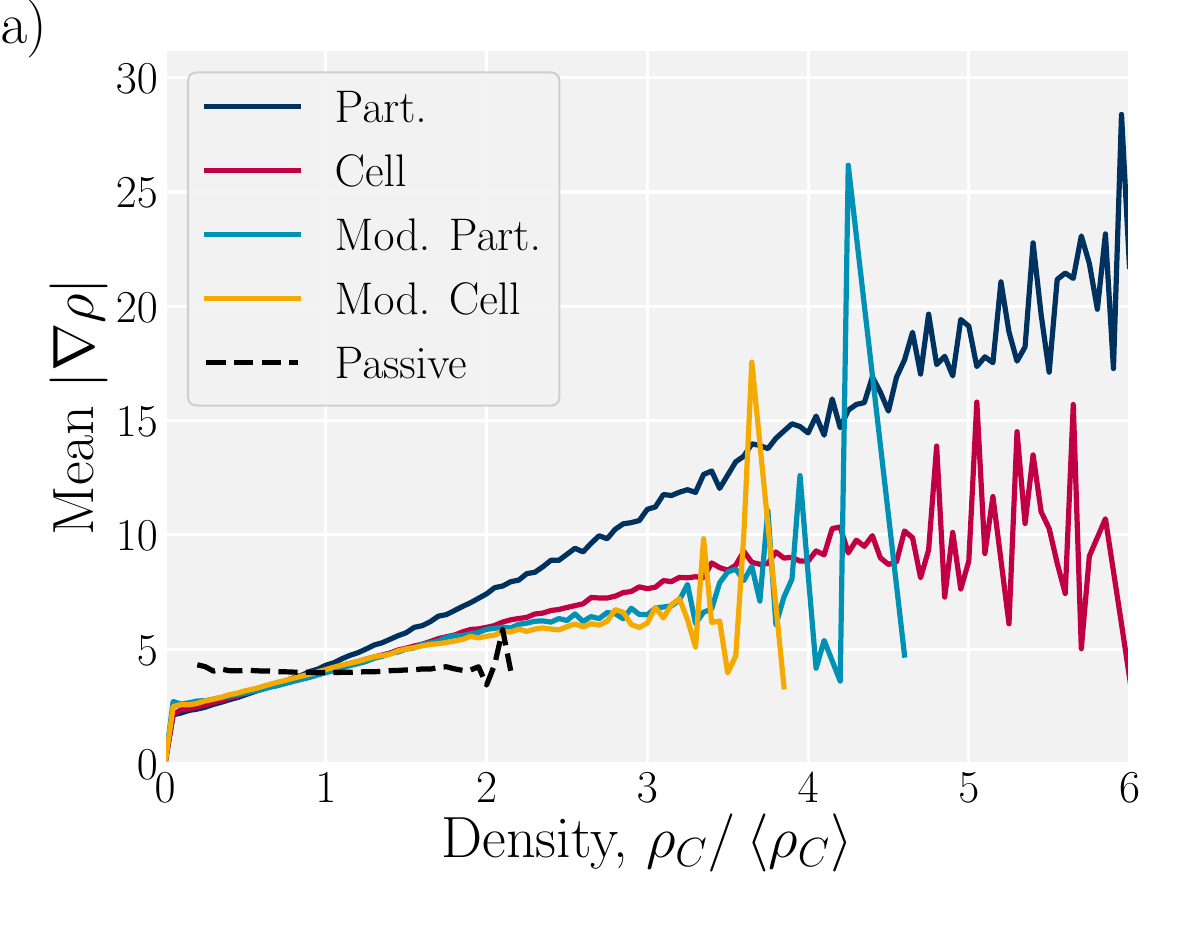}
    \end{subfigure}
    \begin{subfigure}
        \centering
        \includegraphics[width=0.49\linewidth]{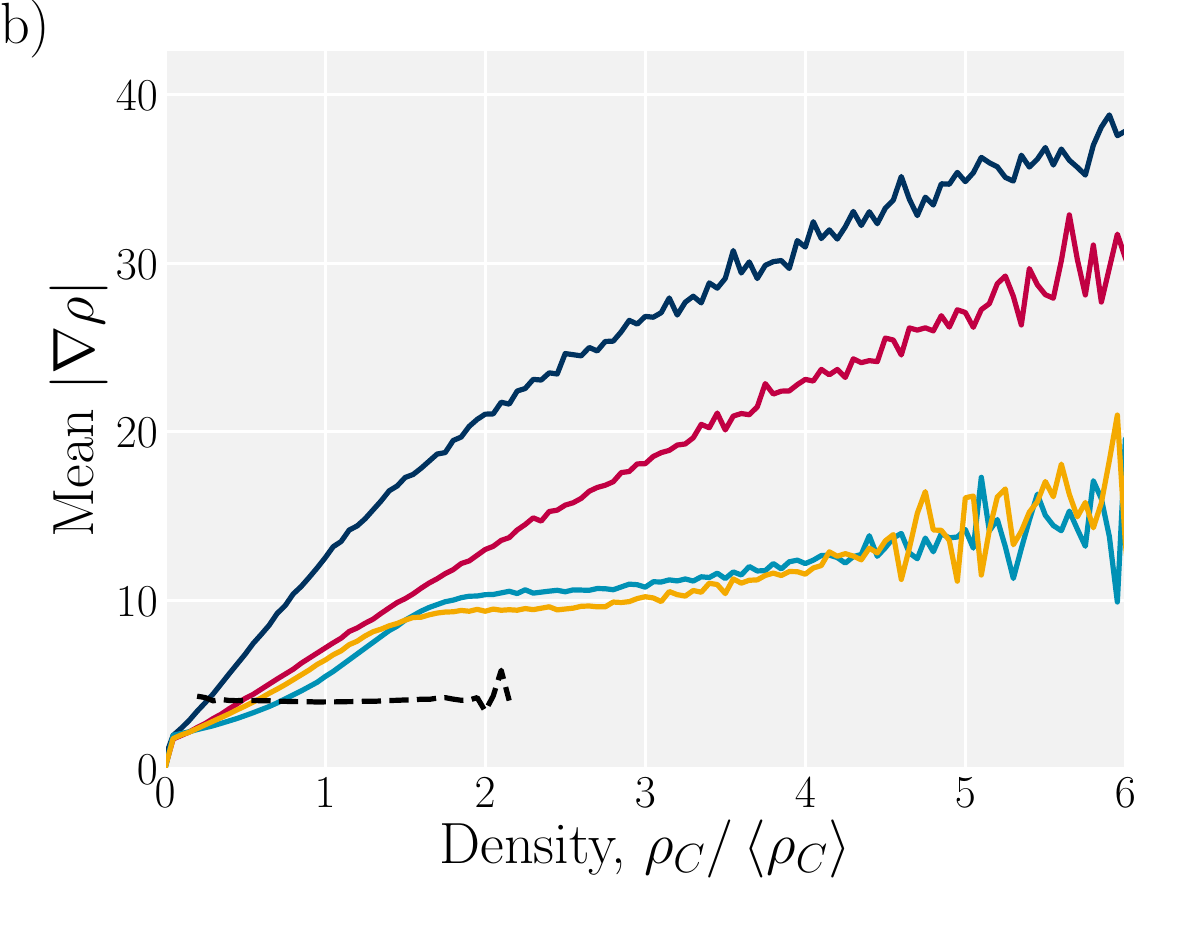}
    \end{subfigure}
    \caption{\textbf{Density gradient magnitudes saturate at high densities for modulated activity formulations.} 
    Fickian flux $\abs{\nabla \rho}$ as a function of local density $\rho_C$ for 
    \textbf{(a)} $\act=0.08$ and \textbf{(b)} $\act=0.3$. 
    Colors are the same as in \figsi{sifig:Ficks-Prob}. 
    }
    \label{sifig:Ficks-Dens}
\end{figure*}

\begin{figure*}[tb]
    \centering
    \begin{subfigure}
        \centering
        \includegraphics[width=0.49\linewidth]{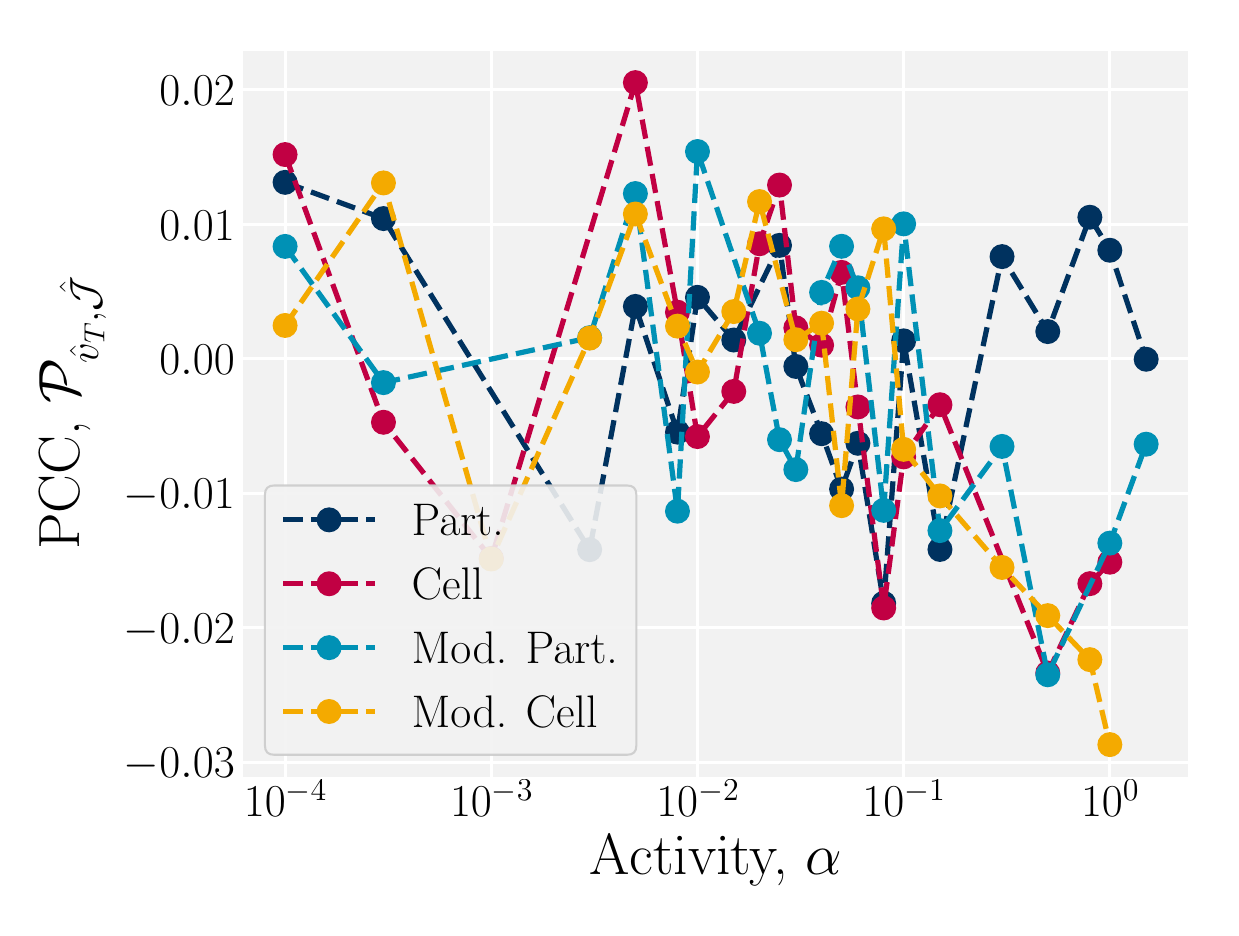}
    \end{subfigure}
    \caption{\textbf{Direction of tracer particles is de-correlated from the local direction of density gradients for all cellular activity formulations. } 
    Pearson correlation coefficient (PCC) of tracer particle direction $\hat{v}_{T}$ compared to the direction of pressure induced drift $\hat{\mathcal{J}} = \grad \rho /\abs{\mathcal{J}}$. 
    Colors are the same as in \figsi{sifig:Ficks-Prob}. 
    }
    \label{sifig:PCC-Angular}
\end{figure*}

\end{document}